\documentclass[aps,prd,superscriptaddress,nofootinbib,11pt]{revtex4}
\usepackage[english]{babel}
\usepackage[utf8]{inputenc}
\usepackage{graphicx}   
\usepackage{slashed}
\usepackage{epstopdf}
\usepackage{verbatim}   
\usepackage{color}      
\usepackage{subfigure}  
\usepackage{multirow}
\usepackage[colorlinks=true, linkcolor=blue,
            citecolor=blue, urlcolor=blue]{hyperref}   
\usepackage{float}
\usepackage{epsfig,rotating}
\usepackage{amsmath,amssymb}
\usepackage{dsfont}
\usepackage{braket}
\restylefloat{table}
\numberwithin{equation}{section}

\newcommand{\Ok}{\ensuremath{\omega_k}}
\newcommand{\Okr}{\ensuremath{\Omega^R_k}}
\newcommand{\vx}{\ensuremath{\vec{x}}}
\newcommand{\vy}{\ensuremath{\vec{y}}}
\newcommand{\vk}{\ensuremath{\vec{k}}}
\newcommand{\vq}{\ensuremath{\vec{q}}}

\newcommand{\be}{\begin{equation}}
\newcommand{\ee}{\end{equation}}
\newcommand{\bea}{\begin{eqnarray}}
\newcommand{\eea}{\end{eqnarray}}
\newcommand{\du}{d_{\mathcal{U}}}

\begin{document}
\title{Quantum Open Effective Field Theory: \\   anomalous decoherence.}
\author{Gabriela Barenboim}
\email{Gabriela.Barenboim@uv.es} \affiliation{Instituto de Fısica Corpuscular, CSIC Universitat de Valencia, Paterna 46980, Spain}\affiliation{Departament de Fısica Teorica, Universitat de Valencia, Burjassot 46100, Spain}

\author{Daniel Boyanovsky}
\email{boyan@pitt.edu} \affiliation{Department
of Physics and Astronomy, University of Pittsburgh,Pittsburgh,
Pennsylvania 15260, USA}
\date{\today}
\begin{abstract}

We study environment-induced decoherence   of a scalar  field probe  coupled to ``anomalous'' environments featuring critical behavior with complementary formulations of quantum open effective field theory: Schwinger-Keldysh including the  influence functional,  and quantum master equation (QME) in Lindblad form. Although we obtain the influence functional to all orders, we assume   weak system-environment coupling and consider both up to second order   in this coupling. The Schwinger-Keldysh formulation yields a Langevin equation and the Lindblad (QME) includes   \emph{time dependent rates} allowing to study transient dynamics. Their comparison clarifies and justifies various   approximations. We focus on three   examples featuring non-Markov dynamics that cannot be described with S-matrix/Fermi's golden rule rates \textbf{i) Unparticle physics:} a conformal sector beyond the Standard Model at a strongly coupled infrared fixed point featuring Lorentz invariant correlation functions with anomalous scaling dimensions, \textbf{ii) Threshold and infrared enhancement:} when the mass shell of the probe field coincides with or is near to a  kinematic threshold of the spectral density of the bath, \textbf{iii) Critical environments:} the bath is at a critical point of a second order phase transition and the probe field couples to the order parameter whose correlation functions scale with anomalous   dimensions and long-wavelength fluctuations feature critical slowing down. In all cases the reduced density matrix undergoes decoherence in the energy basis of the probe (pointer basis), and off-diagonal matrix elements decay as $e^{-\Phi(t)}$ with $\Phi(t)>0 $ the \emph{decoherence function} and $\Gamma(t) = \dot{\Phi}(t)$ the decoherence rate. In all cases we find that either transient or asymptotically $\Phi(t) \propto t^{-\mu}~;~ \mu>0$ with $\mu$ being in general fractional, determined by the behavior of the spectral density at threshold and related to the anomalous scaling dimensions of the bath correlation functions, or $\Phi(t) \propto \ln(t)$ for the infrared case, strikingly different from a constant or vanishing rate $\Gamma$ predicted by  the S-matrix/Fermi's golden rule. For a critical environment a new dynamical scaling exponent-$z$-  emerges. A probe of frequency $\omega_q$ is resonant with a \emph{critically slow} quasiparticle for $q_*\propto \omega^{1/z}_q$ at which decoherence is strongly \emph{enhanced}. The decoherence rate of a probe at rest features anomalous scaling with the probe mass with the new dynamical critical exponent, a direct imprint of dynamical scaling on a physical observable. Possible implications in cosmology are discussed.

\end{abstract}


\maketitle

\section{Introduction}\label{sec:intro}

Effective field theories provide a powerful organizational principle to describe phenomena below some energy scale, or large spatio-temporal scales that emerges upon coarse graining degrees of freedom on short wavelengths, or high energy scales, thereby embodying  the \emph{Wilsonian} paradigm of modern field theory. This approach  has become a pillar of fundamental understanding in diverse areas, an historically important example is   the Landau-Ginzburg theory of phase transitions and the classification of universality classes in critical phenomena\cite{ma,fradkin,gold}. In high energy physics, effective quantum field theory provides a systematic program to describe phenomena below some higher energy scale exploiting  symmetry principles and renormalization group concepts to include information from the higher scales\cite{eft1,eft2,eft3}.

A similar conceptual framework for non-equilibrium phenomena was developed in foundational studies of \emph{quantum Brownian motion}\cite{schwinger,feyver,leggett,ford} in which a quantum particle is ``immersed'' in a bath of quantum oscillators, which are ``traced out'' yielding an effective description for the Brownian particle. This pioneering work spawned the new paradigm of quantum open systems, in which a set of degrees of freedom, namely the  system,  is coupled to other degrees of freedom, a reservoir or bath,  which are ``integrated out'', or ``traced over'' leading to a typically dissipative and stochastic effective description of the dynamics of the system\cite{breuer,zoeller,carmichael,weiss}. This formulation begins with the time evolution of an initial density matrix, which implies evolution forward and backward in time leading to the Schwinger-Keldysh\cite{schwinger,keldysh} or ``in-in'' formulation   yielding causal and retarded equations of motion\cite{maha}.  Integrating out or tracing over the reservoir degrees of freedom leads to an \emph{influence functional}\cite{feyver,leggett} in the  effective dynamics of the system which is built out of correlations of the reservoir fields.  It is obtained exactly in the case when the bath is a collection of independent oscillators\cite{feyver,leggett,carmichael,grabert} but otherwise typically requires a perturbative expansion in terms of free field correlations\cite{hupaz}, this program has been extended to quantum field theory out of equilibrium\cite{stoca,calhu,calhubuk,calhuthreshold,boyabrow}.

 An alternative framework within the quantum open system paradigm is the quantum master equation (QME)\cite{lin,gori,pearle,carmichael,breuer,zoeller,linme} which, while beginning with the time evolution of an initial density matrix, similarly to the ``in-in'' formulation,  develops along different but complementary pathways, and a direct link between the two has been provided in some specific cases\cite{hupaz,calhu}.

The quantum open system paradigm is the fundamental anchor in the understanding of the emergence of classical behavior from a quantum theory in presence of environments\cite{zeh,zurek,pazurek}, it  plays a fundamental role in quantum information\cite{breuer,zoeller,nielsen}, driven quantum systems\cite{diehl}, and has since been implemented to study decoherence   \cite{calhuuniv,lombardo,martin,holman,burg,colas,barenoise},  effective field theory in cosmology\cite{green}, and   misalignment dynamics\cite{axionbrow}.

\textbf{Motivation and objectives}

 Upon tracing over the environmental degrees of freedom  the dynamics of the system is described by the reduced density matrix in which  the influence of the environment enters via correlation functions of the reservoir fields. One of the most ubiquitous assumptions is that  the environment features short range spatio-temporal correlations, leading to a \emph{Markovian}, in other words, memoryless description of the effective dynamics of the system\cite{breuer,zoeller,carmichael}.

The Markov approximation   requires a large separation between the system and environment time scales and  is often invoked to justify a phenomenological  dissipative term in a Lindblad\cite{lin,gori,pearle} form of the (QME), the validity of this along with other approximations have been the subject of recent critical scrutiny\cite{linme}. Non-Markovian dynamics has been argued\cite{nonmarkov,vega} to be of  fundamental importance in various relevant settings where structured environments with  spectral densities featuring thresholds,   entanglement and ``slow'' environmental degrees of freedom    are in general sources of ``memory'', suggesting that a quantum system  coupled to the environment is  a probe of these particular environmental features.

The ubiquity of quantum open effective field theory across fields and its relevance in cosmology motivate us to study important   cases where the Markov approximation, and more generally within quantum field theory, the applicability of S-matrix/Fermi's golden rule  interaction rates, are demonstrably not valid as a consequence of ``slow modes'' or memory of the system.

\textbf{Objectives:} We are   focused on studying decoherence, properly defined, and transient dynamics  in cases in which the bath features aspects of critical behavior\cite{unparticle,banksZaks,bare} either as a consequence of a (second order) phase transition or from infrared or threshold behavior\cite{boyir}. We also seek a fundamental and  more overarching understanding in cases  where collective long wavelength fluctuations  result in  anomalous scaling dimensions of correlation functions of the  bath, which typically result from resummations of infrared physics. These generically imply long range temporal correlations suggesting a breakdown of the Markov (memoryless) approximation, along with the S-matrix/Fermi's golden rule- inspired phenomenological constant rates.

 By relaxing the Markov approximation and including memory in the time evolution, we open a window into hitherto unexplored transient   and   long time phenomena that transcend the usual damping of correlations with constant rates determined by S-matrix theory, or equivalently Fermi's golden rule (FGR). We refer to these  set of phenomena as \emph{anomalous decoherence}.

  To this aim we implement two complementary formulations of quantum open effective field theory: the Schwinger-Keldysh or ``in-in'' effective action including the influence functional and the quantum master equation in Lindblad form but allowing for \emph{time dependent rates} to study environment-induced decoherence of a   scalar field $\phi$ coupled to an environment. We assume that initially the environment is in thermal equilibrium and the total density matrix factorizes,  and also that the system-environment interaction factorizes as $g\,S[\phi]\,\mathcal{O}[\chi]$ with $g$ the coupling and $S[\phi]$, $\mathcal{O}[\chi]$ depend solely on system (S) and environmental ($\mathcal{O}$) fields. These are ubiquitous approximations in quantum open systems and are natural within the context of quantum field theory of non-equilibrium systems, they are tantamount to neglecting initial correlations.

   We study three cases that feature ``critical behavior'', with slow degrees of freedom  which are  expected to lead to non-Markovian dynamics of decoherence:

  \vspace{1mm}

  \textbf{i) Unparticle physics\cite{unparticle,bare}: } Refers to a hypothetical conformal sector beyond the Standard Model described by a strongly interacting (Banks-Zaks) infrared fixed point of the renormalization group\cite{banksZaks}. The probe field $\phi$ couples to scalar unparticle operators $\mathcal{O}$  whose correlation functions are   Lorentz invariant and scale covariant     featuring  anomalous scaling dimensions. The Lehmann representation of the two point function is described by a spectral density that exhibits these anomalous dimensions and a kinematic threshold at which it vanishes with a power law.

  \vspace{1mm}

  \textbf{ii) Threshold and infrared phenomena:} The $\phi$ field couples to a two-particle continuum with a spectral density featuring a threshold at which it vanishes with a power law, the mass shell of the probe $\phi$ coincides with or is near to the threshold. As discussed in ref.\cite{boyir} the behavior near threshold is very similar to that of a critical system, and infrared resummations based on the renormalization group yield anomalous scaling, similar to the unparticle case.

  \vspace{1mm}

  \textbf{iii) An environment at a critical point of a second order phase transition.} The scalar field $\phi$ is coupled to the order parameter $\mathcal{O}$ featuring scale covariant correlation functions with anomalous dimensions, critical slowing down and a new dynamical exponent\cite{ma,gold,hohenberg,tauber,hotboy} that describes the long relaxation time of long wavelength fluctuations.

\vspace{1mm}

\textbf{Brief summary of results:} Within the Schwinger-Keldysh (``in-in'') framework we obtain the influence functional formally to \emph{all orders} in the coupling $g$  and also to all orders in the interaction of the environmental fields with themselves. It may be expressed as a cumulant expansion in $g$ and the higher order correlation functions of $\mathcal{O}$ in the initial density matrix. While formally exact, we focus on weak system-environment coupling with $\phi$ a free scalar field linearly coupled to $\mathcal{O}$ as a \emph{probe} of the environment, to clearly identify environment-induced decoherence unencumbered by self-interactions of the system.  We obtain the ``in-in'' effective action, which determines the time evolution of the reduced density matrix, up to $\mathcal{O}(g^2)$, yielding to this order a description of the dynamics via a Langevin equation with dissipative and noise terms obeying the fluctuation dissipation relation\cite{kms}. We also obtain the (QME) up to this order, which is of the Lindblad form\cite{lin,gori,pearle,carmichael,breuer,zoeller,linme}  but with \emph{time dependent rates} in the dissipator. Their comparison clarifies and justifies various approximations in the (QME). In the three cases considered we find that the reduced density matrix tends to become diagonal in the energy basis of the probe field $\phi$, the off-diagonal matrix elements in this basis, \emph{decohere} and are suppressed as $\propto e^{-\Phi(t)}$ with $\Phi(t) >0$ the \emph{decoherence function} with $\Gamma \equiv \dot{\Phi}(t)$ the decoherence rate. The   time dependence of the populations, namely diagonal elements, describes thermalization and \emph{particle production}. In all cases we find that transient or long time asymptotic behavior is described by  $\Phi(t) \propto t^{-\mu} ~;~ \mu \geq 0$ with $\mu$ being in general fractional and depends on the anomalous dimensions, with the infrared case ($\mu =0$) yielding $\Phi(t) \propto \ln(t)$ in clear contrast to the predictions from Markovian dynamics with S-matrix/Fermi's golden rule constant rates. The critical case stands out in that a probe of frequency $\omega_q$ is \emph{resonant} with a critically slow quasiparticle for wavelength $q_* \propto \omega^{1/z}_q$, with $z$ the new dynamical exponent, with enhanced decoherence. The decoherence rate   of a probe at rest features anomalous scaling dimension with the mass of the probe with the new dynamical critical exponent.

\section{Quantum open effective field theory: }\label{sec:twoapps}

\subsection{Schwinger-Keldysh   and influence functional}\label{subsec:inin}

 We study  the non equilibrium effective action for the case of interacting
 scalar fields, spinor and vector fields may be included, while technically more involved their treatment follows without major conceptual difficulties. We consider a bosonic field $\phi$ of mass $m_\phi$ referred to as the ``system'' and environmental fields, collectively denoted by   $\chi$, described by the Lagrangian density   given by
\begin{equation}
{\cal L}[\phi(x),\chi(x)]= {\cal L}_{ S}[\phi(x)]+{\cal
L}_{ E}[\chi(x)]-g~S[\phi(x)]~{\cal O}[\chi(x)]\,\label{lagra}
\end{equation}
where the system and environment composite operators  $S[\phi]$ and ${\cal O}[\chi]$ are in general  non-linear function(als) of the respective fields, and ${\mathcal L}_{S,E}$ are the Lagrangian densities for the fields $\phi$ and $\chi$ respectively. These include self-interactions of the system ($\phi$) fields and interactions among the environmental ($\chi$)  fields, but the coupling between the system and environmental fields is described by the last term.  We will obtain the reduced density matrix  for the field system  $\phi$ by tracing over or ``integrating out'' the environmental field $\chi$. The tracing or integrating out the environmental field leads to   an effective real-time
functional   of the field $\phi$, known as  the influence functional\cite{feyver,calhubuk} in the
theory of quantum brownian motion. The time evolution of this reduced density matrix is
represented by a path integral in terms of the non-equilibrium
effective action that includes the influence functional.  This
method has been used previously to study quantum brownian
motion\cite{feyver,leggett,grabert,hupaz}, and quantum fields out of equilibrium\cite{stoca,calhu,calhubuk,calhuthreshold,boyabrow}.

Consider the initial density matrix at a time
$t=0$ to be of the form
\begin{equation}
{\rho}(0) = {\rho}_{S}(0) \otimes
{\rho}_{E}(0) \,.\label{inidensmtx}
\end{equation}

The initial density matrix of the $\chi$ fields will be taken to
describe a statistical ensemble in thermal equilibrium at a temperature
$T=1/\beta$, namely

\begin{equation}
{\rho}_{E}(0) = \frac{e^{-\beta\,H_E}}{\underset{\{E\}}{\mathrm{Tr}} e^{-\beta\,H_E}} \,, \label{rhochi}
\end{equation}

\noindent where $H_{E}$ is the full exact Hamiltonian for the environmental
field $\chi$. The zero temperature limit $ {\rho}_{E}(0)   \rightarrow \ket{0_E}\bra{0_E}$, with $\ket{0_E}$ the \emph{exact} ground state of the environment will also be considered.

The factorization of the initial density matrix is an assumption often explicitly or implicitly made in the literature, it can be relaxed by including initial correlations, we will not consider here this important case, relegating it to future study. The initial factorization entails that
the time evolution of the system can be described as ``switching-on'' the coupling between fields at the initial time resulting  in transient dynamics which, along with the long time evolution towards a stationary state will be the focus of our study.

In the field basis the matrix elements of ${\rho}_{S}(0)$
are given by
\begin{equation}
\langle \phi |\rho_S(0) | \phi'\rangle \equiv
\rho_{S}(\phi ,\phi';0)~~;~~\langle \chi | {\rho}_{E}(0) | \chi'\rangle \equiv
\rho_{E}(\chi ,\chi';0)\,,
\end{equation} we emphasize that this is a \emph{functional} density matrix since the fields feature spatial arguments.
The density matrix $\rho_S(0)$   represents an initial out of
equilibrium state or ensemble, describing the physical situation where the
    (system) field $\phi$ which is coupled at $t=0^+$ to a heat bath at a temperature
  $T=1/\beta$.

   Once the system  and bath are put in contact their mutual interaction will
  evolve the initial state out of equilibrium because the initial density matrix does not commute with the total Hamiltonian.

To obtain the effective quantum field theory    out of equilibrium for the light field $\phi$  we will evolve the initial density matrix in time and trace over the ``bath'' degrees of freedom, leading to a reduced density matrix for $\phi$.    Once we obtain the reduced density matrix for the field
$\phi$ we can compute expectation values or correlation functions
of this field.

In the Schroedinger picture the time evolution of the initial density matrix is given by

\begin{equation}
 {\rho}(t)=  e^{-iHt} \, {\rho}(0) \,   e^{iHt}\,, \label{rhooft}
\end{equation}
where the total Hamiltonian $H$ is given by
\begin{equation}
H=H_{S} + H_{E}+H_{SE}(\phi,\chi)~~;~~H_{SE}(\phi,\chi)=g\int d^3x  ~S[\phi(x)]~{\cal O}[\chi(x)]\,, \label{hami}
\end{equation} and $H_{S}~,~H_E$  are the  Hamiltonians for the respective fields including their self-interactions, we assume that $H_S,H_E$ commute.

The reduced density matrix of the system is defined by tracing over the environmental degrees of freedom,

\be \rho^r_S(t) = \underset{\{E\}}{\mathrm{Tr}}\,\rho(t) \,. \label{reddm}\ee

 Introducing a complete set of field eigenstates, the density matrix in field space is given by

\bea  & & \rho(\phi_f,\chi_f;\phi'_f,\chi'_f;t)   =       \langle \phi_f;\chi_f|U(t) {\rho}(0)U^{-1}(t)|\phi'_f;\chi'_f\rangle \nonumber \\
& = & \int D\phi_i D\chi_i D\phi'_i D\chi'_i ~ \langle \phi_f;\chi_f|U(t)|\phi_i;\chi_i\rangle\,\rho_{S}(\phi_i;\phi'_i;0)\,\rho_{E}(\chi_i;\chi'_i;0)\,
 \langle \phi'_i;\chi'_i|U^{-1}(t)|\phi'_f;\chi'_f\rangle \label{evolrhot}\eea The $\int D\phi \int D\chi \cdots $   are functional integrals where the spatial argument has been suppressed. The matrix elements of the time evolution forward and backward can be written as path integrals along forward ($+$)  and backward ($-$)  paths, namely
 \bea   \langle \phi_f;\chi_f|U(t)|\phi_i;\chi_i\rangle  & = &    \int \mathcal{D}\phi^+ \mathcal{D}\chi^+\, e^{i \int d^4 x \mathcal{L}[\phi^+,\chi^+]}\label{piforward}\\
 \langle \phi'_i;\chi'_i|U^{-1}(t)|\phi'_f;\chi'_f\rangle &  =  &   \int \mathcal{D}\phi^- \mathcal{D}\chi^-\, e^{-i \int d^3 x \mathcal{L}[\phi^-,\chi^-]}\label{piback}
 \eea where we use the shorthand notation
 \be \int d^4 x \equiv \int_0^t dt \int d^3 x \,,\label{d4xdef}\ee
 $ \mathcal{L}[\phi,\chi] $ is given by (\ref{lagra})   and
 the boundary conditions on the path integrals are
  \bea     \phi^+(\vec{x},t=0)=\phi_i(\vec{x})~;~
 \phi^+(\vec{x},t)  &  =  &   \phi_f(\vec{x})\,,\nonumber \\   \chi^+(\vec{x},t=0)=\chi_i(\vec{x})~;~
 \chi^+(\vec{x},t) & = & \chi_f(\vec{x}) \,,\label{piforwardbc}\\
     \phi^-(\vec{x},t=0)=\phi'_i(\vec{x})~;~
 \phi^-(\vec{x},t) &  = &    \phi'_f(\vec{x})\,,\nonumber \\   \chi^-(\vec{x},t=0)=\chi'_i(\vec{x})~;~
 \chi^-(\vec{x},t) & = & \chi'_f(\vec{x}) \,.\label{pibackbc}
 \eea

The field variables $\phi^\pm, \chi^\pm$ along the forward ($+$) and backward ($-$) evolution branches are recognized as those necessary for the Schwinger-Keldysh\cite{schwinger,keldysh,maha,calhubuk} or ``in-in''  closed time path approach to the time evolution of a density matrix.

 The reduced density matrix for the system  field $\phi$ (\ref{reddm}) is obtained by tracing over the environmental fields ($\chi$), namely

\be \rho^{r}_S(\phi_f,\phi'_f;t) = \int D\chi_f \,\rho(\phi_f,\chi_f;\phi'_f,\chi_f;t) \,,\label{rhored2} \ee we find
\be \rho^{r}_S(\phi_f,\phi'_f;t) = \int D\phi_i   D\phi'_i  \,  \mathcal{T}[\phi_f,\phi'_f;\phi_i,\phi'_i;t] \,\rho_S(\phi_i,\phi'_i;0)\,, \ee
where the time propagating kernel $\mathcal{T}$ is given by
\be \mathcal{T}[\phi_f,\phi_i;\phi'_f,\phi'_i;t] = {\int} \mathcal{D}\phi^+ \, \int \mathcal{D}\phi^- \, e^{i  \int d^4x \left[\mathcal{L}_S[\phi^+]-\mathcal{L}_S[\phi^-]\right]}\,e^{i\mathcal{F}[S^+;S^-]} \ee with
the following boundary conditions on the forward ($\phi^+$) and backward  ($\phi^-$) path integrals
\bea &  &   \phi^+(\vec{x},t=0)=\phi_i(\vec{x})~;~
 \phi^+(\vec{x},t)  =   \phi_f(\vec{x}) \nonumber \\
&  &   \phi^-(\vec{x},t=0)=\phi'_i(\vec{x})~;~
 \phi^-(\vec{x},t)  =   \phi'_f(\vec{x}) \,.\label{bcfipm}\eea   $\mathcal{F}[S^+;S^-]$ is the \emph{influence action} where $S^\pm \equiv S[\phi^\pm]$, it is  given by
 \bea  e^{i\mathcal{F}[S^+;S^-]}  &  =  & \int D\chi_i \,D\chi'_i \int   D\chi_f  \int \mathcal{D}\chi^+ \int \mathcal{D}\chi^- \, e^{i  \int d^4x \Big[\mathcal{L}_E[\chi^+]-g S[\phi^+]~\mathcal{O}[\chi^+]\Big]}~   e^{-i  \int d^4x \Big[\mathcal{L}_E[\chi^-]-g S[\phi^-]~\mathcal{O}[\chi^-]\Big]}\, \nonumber \\ & \times  & \rho_{E}(\chi_i,\chi'_i;0)\,, \label{inffunc}\eea where the boundary conditions on the path integrals are
 \be \chi^+(\vec{x},t=0)=\chi_i(\vec{x})~;~\chi^-(\vec{x},t=0)=\chi'_i(\vec{x}) ~~;~~
 \chi^+(\vec{x},t)= \chi^-(\vec{x},t)=\chi_f(\vec{x}) \,, \label{bcchis} \ee the last equality being a consequence of taking the trace.

In the above path integral defining the influence action, $S[\phi^\pm]$ acts as an \emph{external source} coupled to the composite operator $\mathcal{O}(\chi)$, therefore, it is straightforward to conclude that (\ref{inffunc}) is a path integral representation of
\be e^{i\mathcal{F}[S^+;S^-]} = \underset{\{\chi\}}{\mathrm{Tr}}  \Big[ \mathcal{U}(t;S^+)\,\rho_E(0)\,  \mathcal{U}^{-1}(t;S^-) \Big]\,, \label{trasa}\ee where  $\mathcal{U}(t;S^\pm)$ is the   time evolution operator acting solely in the $\chi$ sector but in presence of \emph{external sources} $S^\pm$,  namely \be \mathcal{U}(t;S^+) = T\Big( e^{-i \int_0^t H_E[\chi;S^+(t')]dt'}\Big) ~~;~~
\mathcal{U}^{-1}(t;S^-) = \widetilde{T}\Big( e^{i \int_0^t H_E[\chi;S^-(t')]dt'}\Big)  \label{timevchi}\ee where
\be H _E[\chi;S^\pm(t)] = H_E+g\,\int d^3x S[\phi^\pm ]\, \mathcal{O}[\chi]\,,\label{HS}   \ee
and $\widetilde{T}$ is the \emph{anti-time evolution operator} corresponding to    evolution backwards in time. The calculation of the influence action in an expansion in the system-environment coupling $g$ is facilitated by passing to the interaction picture of the environmental fields,   defining
\be  \mathcal{U}(t;S^+) =  e^{i H_{E}\,t}\,\mathcal{U}_{ip}(t;S^+)~;~  \mathcal{U}^{-1}(t;S^-)= \mathcal{U}^{-1}_{ip}(t;S^-)\,e^{-iH_Et} \,, \label{ipicture} \ee where the $e^{\pm i H_{E}\,t}$ cancel out in the trace in (\ref{trasa}) . The influence functional is obtained from the expression
\be \mathcal{F}[S^+;S^-;t] = -i \ln \Bigg\{ \underset{\{E\}}{\mathrm{Tr}}  \Big[ \mathcal{U}_{ip}(t;S^+)\,\rho_E(0)\,  \mathcal{U}^{-1}_{ip}(t;S^-) \Big] \Bigg\} \,.\label{Fln} \ee This is the \emph{exact} form of  the influence functional valid to all orders of the self interactions of environmental fields and in the system environment coupling, unitarity implies that $\mathcal{F}[S,S;t]=0$. It is worth emphasizing that this form of the influence functional is the real time  equivalent of the usual radiative (loop) corrections to an effective action within the background field method.

 The trace in (\ref{Fln}) is in the initial environmental density matrix, this is  a consequence of the assumption of factorization of the initial density
matrix into (uncoupled) system-environment.

 With the expansion
\bea  && \mathcal{U}_{ip}(t;S) =  1-i g \int^t_0 dt_1 \int d^3 x_1 \mathcal{O}[\chi(\vx_1,t_1]\,S[\phi(\vx_1,t_1)]   \\ & + &  i^2 \, g^2 \int^t_0 dt_1 \int^{t_1}_0 dt_2 \int d^3x_1 \int d^3 x_2  \mathcal{O}[\chi(\vx_1,t_1)]\,S[\phi(\vx_1,t_1)]~\mathcal{O}[\chi(\vx_2,t_2)]\,S[\phi(\vx_2,t_2)] + \cdots \nonumber \label{Uip}\eea and a similar expansion for $\mathcal{U}^{-1}_{ip}(t;S)$ where the composite operators $\mathcal{O}$ are in the Heisenberg picture of the environmental Hamiltonian $H_E$, the influence functional can be obtained as a series in the coupling $g$ from a cumulant expansion of the result (\ref{Fln}), in terms of the correlation functions of the operators $\mathcal{O}$ in the initial density matrix of the environment $\rho_E(0)$. A power $g^n$ in $\mathcal{F}$ corresponds to the connected $n-point$ correlation functions of $\mathcal{O}$ in $\rho_E(0)$.

The ``in-in'' \emph{effective action}  describing the non-equilibrium evolution of the \emph{reduced density matrix}  is given by
\be  {S}_{eff}[\phi^+,\phi^-] = \int^t_0 dt_1 \int d^3 x \Big\{ \mathcal{L}_S[\phi^+]-\mathcal{L}_S[\phi^-]\Big\} +\mathcal{F}[S[\phi^+],S[\phi^-]] \,.\label{Leff} \ee We emphasize that this effective action is formally to \emph{all orders} in self-interactions of the system field $\phi$ and the environmental correlation functions are \emph{to all orders} of any interactions of the environmental fields with themselves.

Although the expression (\ref{Fln}) is exact, and (\ref{Leff}) is the full effective action to all orders in all couplings, we will assume that the system and the bath are weakly coupled, and  we obtain $\mathcal{F}$ up to $\mathcal{O}(g^2)$ in the cumulant expansion by  expanding the logarithm in powers of $g$ up  this order.

We find
\bea \mathcal{F}[S^+,S^-] & = &    - g \int d^4x \Big( S^+[\phi(x)]-S^-[\phi(x)]\Big)\,\langle \mathcal{O}(x)\rangle \nonumber + \\ & & \frac{i g^2 }{2} \int d^4x_1 \int d^4x_2 \Bigg\{ S^+[\phi(x_1)]\,S^+[\phi(x_2)]\,C_c^{++}(x_1-x_2)+ S^-[\phi(x_1)]\,S^-[\phi(x_2)]\,C_c^{--}(x_1-x_2) \nonumber \\
 & - & S^+[\phi(x_1)]\,S^-[\phi(x_2)]\,C_c^{+-}(x_1-x_2)- S^-[\phi(x_1)]\,S^+[\phi(x_2)]\,C_c^{-+}(x_1-x_2)\Bigg\}\,. \label{finF}\eea In this expression  the \emph{connected} correlation functions are given by
\begin{eqnarray}
&& C_c^{-+}(x_1-x_2) =   \langle
{\cal O}(x_1) {\cal O}(x_2)\rangle - \langle \mathcal{O}(x_1)\rangle  \langle \mathcal{O}(x_2)\rangle =   {C}_c^>(x_1-x_2) \,,\label{ggreat} \\&&  C_c^{+-}(x_1-x_2) =   \langle
{\cal O}(x_2) {\cal O}(x_1)\rangle - \langle \mathcal{O}(x_2)\rangle  \langle \mathcal{O}(x_1)\rangle =   {C}_c^<(x_1-x_2)\,,\label{lesser} \\&& C_c^{++}(x_1-x_2)
  =
{ C}_c^>(x_1-x_2)\Theta(t_1-t_2)+ {  C}_c^<(x_1-x_2)\Theta(t_2-t_1) \,,\label{timeordered} \\&& C_c^{--}(x_1-x_2)
  =
{ C}_c^>(x_1-x_2)\Theta(t_2-t_1)+ {  C}_c^<(x_1-x_2)\Theta(t_1-t_2)\,,\label{antitimeordered}
\end{eqnarray}   where
\be \langle (\cdots) \rangle = \mathrm{Tr}(\cdots)\rho_E(0)\,. \label{expec}\ee
Furthermore, for the case of hermitian operators $\mathcal{O}$ as considered here it follows that
\be C_c^>(x_1-x_2) = C_c^<(x_2-x_1)\,. \label{iden}\ee

We emphasize that these are \emph{exact} correlation functions of the bath, to all orders in the couplings of the environmental fields with themselves. This is important because as stated in the objectives, we focus on the case where these correlation functions feature scaling behavior that emerges at a critical point and result from a (renormalization group) resummation of infrared modes, beyond perturbation theory.

Since   $\rho_E(0)$ is in equilibrium and is translational invariant, the expectation value $\langle \mathcal{O}(x_1)\rangle$ is independent of space-time coordinates. Therefore the first term linear in the coupling $g$ on the right hand side of (\ref{finF}) can be cancelled by a normal ordering prescription of the composite operator $\mathcal{O}(\chi)$ in the initial density matrix $\rho_E(0)$, namely
\be \mathcal{O}(\chi) \rightarrow \mathcal{O}(\chi) - \langle \mathcal{O}(\chi)\rangle \,.\label{NO}\ee    This normal ordered operator features vanishing expectation value in $\rho_E(0)$.  This is tantamount to introducing a counterterm in the Lagrangian density of the system  that cancels   the first term on the right hand side of (\ref{finF}) and the terms $\langle \mathcal{O}(x_{1,2})\rangle$ in the connected correlation functions (\ref{ggreat}-\ref{antitimeordered}), hence in what follows we suppress the subscript ``c'' in the correlation functions.

With the ``in-in'' path integral representation of the time evolved density matrix for the system, we can obtain multi-time correlation functions of the system field by introducing sources $J^{\pm}(\vx,t)$ linearly coupled to the system field variables $\phi^{\pm}$, namely
\be \mathcal{L}_S[\phi^\pm] \rightarrow \mathcal{L}_S[\phi^\pm] + J^{\pm}(\vx,t)\,\phi^{\pm}(\vx,t) \,,\label{sources}\ee yielding
\be S_{eff}[ \phi^\pm;J^\pm]= S_{eff}[\phi^+,\phi^-]+i\,\int d^3 x \int dt \Big(J^+(\vx,t)\phi^+(\vx,t)-J^-(\vx,t)\phi^-(\vx,t)\Big)\,,\label{Seffjs}\ee  and taking the trace of the reduced density matrix with these sources defining the generating functional
\be \mathcal{Z}[J^+;J^-] \equiv \int D\phi_i D\phi'_i D\phi_f \mathcal{D}\phi^+ \mathcal{D}\phi^- \,e^{i S_{eff}[\phi^\pm;J^\pm] }\,\rho_S(\phi,\phi'_i;0) \,,\label{genfun}\ee where the boundary conditions (\ref{bcfipm}) are modified to $\phi'_f=\phi_f$ to account for the trace.   Multi-time correlation functions with appropriate time orderings can now be obtained  by taking variational derivatives of this generating functional with respect to the sources $J^\pm$.

We now re-write the influence action (\ref{finF}) solely in terms of the two correlation functions $C^\lessgtr$ by implementing the following steps:

\begin{itemize}
\item{In the term with $S^+(x_1)S^+(x_2)$: in the contribution $C^<(x_1-x_2)\Theta(t_2-t_1)$ (see eqn. (\ref{timeordered})) relabel $t_1 \leftrightarrow t_2$ and use the property (\ref{iden}).  }
\item{ In the term with $S^-(x_1)S^-(x_2)$: in the contribution $C^>(x_1-x_2)\Theta(t_2-t_1)$ (see eqn. (\ref{antitimeordered})) relabel $t_1 \leftrightarrow t_2$ and use the property (\ref{iden}). }
    \item{ In the term with $S^+(x_1)S^-(x_2)$: multiply $C^<(x_1-x_2)$ by $\Theta(t_1-t_2)+\Theta(t_2-t_1)=1$, and in the term with $\Theta(t_2-t_1)$ relabel $ t_1 \leftrightarrow t_2$ and use the property (\ref{iden}). }
 \item{ In the term with $S^-(x_1)S^+(x_2)$: multiply $C^>(x_1-x_2)$ by $\Theta(t_1-t_2)+\Theta(t_2-t_1)=1$, and in the term with $\Theta(t_2-t_1)$ relabel $t_1 \leftrightarrow t_2$ and use the property (\ref{iden})\,, }
\end{itemize} yielding up to second order in the system-environment coupling

\bea   \mathcal{F}[S^+, S^-] &     =  &   i\,g^2\int d^3x_1 d^3x_2 \int^t_0 dt_1\,\int^{t_1}_0 dt_2\,\Bigg\{ \Bigg[ S^+(\vx_1,t_1)S^+(\vx_2,t_2) - S^-(\vx_1,t_1)S^+(\vx_2,t_2)\Bigg]\,C^>(x_1-x_2)   \nonumber \\ & +  & \bigg[S^-(\vx_1,t_1)S^-(\vx_2,t_2)-  S^+(\vx_1,t_1)S^-(\vx_2,t_2)\Bigg]\,C^<(x_1-x_2)
 \Bigg\} \label{Funravel}\eea where $C^{\lessgtr}$ are given by eqns. (\ref{ggreat},\ref{lesser}).  This is the general form of the influence function up to second order in the system-environment coupling.

 Since the environment is a bath in thermal equilibrium the two point correlation functions $C^{\lessgtr}(x_1-x_2)$ admit an exact Lehmann representation which is analyzed in detain in appendix (\ref{app:fd}).

We recognize the first two terms in the effective action (\ref{Leff}) as the representation of the time evolution of the reduced density matrix in absence of coupling to the environment, namely
\be \rho^r_S(t) = e^{-iH_S t}\,\rho_S(0) e^{i H_S t} \Rightarrow \dot{\rho}^r_S(t) = -i [H_S,\rho^r_S(t)] \,,\label{frevol}\ee
and we  note that
\bea \frac{d}{dt}\Bigg\{e^{i\mathcal{F}[S^+, S^-]}\Bigg\} & = &  -g^2\int d^3x_1 d^3x_2  \,\int^{t}_0 dt_2\,\Bigg\{ S^+(\vx_1,t)S^+(\vx_2,t_2)\,C^>(\vx_1-\vx_2,t-t_2) + \nonumber \\ && S^-(\vx_1,t)S^-(\vx_2,t_2)\,C^<(\vx_1-\vx_2,t-t_2)- S^+(\vx_1,t)S^-(\vx_2,t_2)\,C^<(\vx_1-\vx_2,t-t_2)\nonumber \\ & - & S^-(\vx_1,t)S^+(\vx_2,t_2)\,C^>(\vx_1-\vx_2,t-t_2)\Bigg\} ~\Bigg\{e^{i\mathcal{F}[S^+, S^-]}\Bigg\}\,.  \label{dtF}\eea Because the time evolution of the density matrix is determined by $U(t)\rho(0)U^{-1}(t)$ with $U(t)$ the forward time evolution operator (time ordered) and $U^{-1}(t)$ the backward time evolution operator (anti-time ordered),  operator insertions with $(+)$ go \emph{before the density matrix}, and operator insertions with $(-)$ go \emph{after} the density matrix. Therefore when multiplying the density matrix the equivalence becomes
\be S^+ ~ S^+ \rightarrow S~S\rho ~~;~~S^- ~ S^- \rightarrow \rho S~S ~~;~~ S^+ ~ S^- \rightarrow  S \rho S~~;~~ S^- ~ S^+ \rightarrow S \rho S\,, \label{equivalence}\ee
this result will be of paramount importance when  comparing to the quantum master equation in section (\ref{subsec:QME}).

\subsubsection{Free scalar fields coupled to the environment: Schwinger-Keldysh.}\label{subsub:free}

The results above are general and valid to \emph{all orders} in the self-interactions of the system and environmental fields, but up to second order in the system-environment coupling $g$. As emphasized in our objectives, we seek to probe environmental correlations by coupling the environment to a free scalar field, but considering the \emph{exact} correlation functions of the environment, including the self-interaction of the environmental degrees of freedom  in absence of its coupling to the system.

 Analyzing the dynamics of this probe, namely the system field $\phi$,  and comparing it to free field dynamics allows us to understand the influence of the environmental correlations in the simplest manner and to extract cleanly the decoherence effects from the coupling to the environment, namely environment induced decoherence, unencumbered by self-interactions of the system field. This case is also relevant as
an example of ``misalignment dynamics'' in presence of environmental fields\cite{axionbrow}, an aspect relevant in cosmology.

We now consider the system field $\phi$ as a free field, with

\be  \mathcal{L}_S[\phi] = \frac{1}{2}\Big[ \partial_{\mu}\phi~ \partial^\mu \phi - m^2_\phi~ \phi^2(x) \Big]~~;~~ S[\phi(x)] \equiv \phi(x)\,, \label{free}\ee  but consider the \emph{exact} correlation functions of the environmental degrees of freedom including their self-interactions. Including external sources $J^\pm$ coupled to the scalar field as (\ref{sources}) in the generating functional (\ref{genfun}) one can obtain arbitrary multi-time correlation functions, however, as in the case of Brownian motion, we will focus on equal time correlation functions\cite{feyver,schwinger,leggett,calhubuk}. For this purpose, it is useful to introduce  the Keldysh   center of mass and
relative variables
\be \Psi(\vec x,t)   =   \frac{1}{2}
\left(\phi^+(\vec x,t) + \phi^-(\vec x,t) \right) \; \; ; \; \;
R(\vec x,t) = \left(\phi^+(\vec x,t) - \phi^-(\vec x,t) \right)\,
\label{wigvars} \ee which are particularly suited to study the dynamics of correlation functions of the scalar field as probes of the environment as can be seen as follows:
for the expectation value in the reduced density matrix, using the trace property and recalling that the Keldysh labels $(+),(-)$ corresponds to insertions before and after of the density matrix respectively (time evolution forward $(+)$ and backward $(-)$),  it follows that
\be \mathrm{Tr} \, \phi \,\rho^r_S = \mathrm{Tr} \, \phi^+ \,\rho^r_S = \mathrm{Tr}  \,  \rho^r_S \, \phi = \mathrm{Tr} \,   \rho^r_S\,  \phi^-  \equiv \mathrm{Tr}\, \Psi \, \rho^r_S \,,\label{exvalPhi}\ee similarly, the \emph{equal time} two point correlation function

\be \mathrm{Tr} \, \phi^2(t)\,\rho^r_S = \mathrm{Tr} \, \phi^+(t)\,\phi^+(t)  \,\rho^r_S = \mathrm{Tr}  \,  \rho^r_S \, \phi^-(t) \, \phi^-(t) = \mathrm{Tr} \, \phi^+(t)  \rho^r_S\,  \phi^- (t)  \equiv \mathrm{Tr}\, \Psi^2(t) \, \rho^r_S \,,\label{twopt}\ee a similar analysis yields
\be \mathrm{Tr}\, R(t)\,\rho^r_S =0   \,,\label{exRs}\ee therefore the correlations of $\Psi$ will yield information on the environmental correlations.

 In terms of $\Psi,R$, the boundary conditions (\ref{bcfipm}) at the initial time $t=0$ become
\be \phi_i(\vx) = \phi^+(\vx,t=0)=   \Psi_i(\vx)+ \frac{R_i(\vx)}{2} ~~;~~ \phi'_i(\vx)  = \phi^-(\vx,t=0)=   \Psi_i(\vx)- \frac{R_i(\vx)}{2}  \,,\label{kelbc} \ee where
\be   \Psi_i(\vx)  \equiv \Psi(\vx,t=0)~;~ R_i(\vx) = R(\vx,t=0)\,. \label{kelinis}\ee This change of variable suggests to introduce the functional Wigner transform of the initial density matrix in field space as

\begin{equation}
{  W}(\Psi_i ; \Pi_i) = \int D R_i\, e^{i\int d^3x \Pi_i(\vec x)
R_i(\vec x)} \rho_S(\Psi_i+\frac{R_i}{2};\Psi_i-\frac{R_i}{2};0)\,. \label{wignerrho}\ee

Correlation functions of the system field $\phi$ are obtained by taking the trace of the reduced density matrix, namely taking   $\phi_f=\phi'_f$ leading to the boundary condition on the relative variable in the path integral representation
\be   R(\vec{x},t)= 0 \label{Rt} \,. \ee
Passing to spatial Fourier transforms
\be \Psi(\vx,t) = \frac{1}{\sqrt{V}}\,\sum_{\vk} \Psi_{\vk}(t) e^{i\vk\cdot\vx}~~;~~R(\vx,t) = \frac{1}{\sqrt{V}}\,\sum_{\vk} R_{\vk}(t) e^{i\vk\cdot\vx}\,,\label{FTs}\ee  and  integrating by
parts   accounting for the boundary conditions (\ref{kelbc},\ref{Rt}), the
non-equilibrium effective action (\ref{Leff}) with $S[\phi]=\phi$ in eqn. (\ref{finF}) becomes:

\begin{eqnarray}
iS_{eff}[\Psi,R] & = & \int_0^{t} dt_1 \sum_{\vec k} \left\{-i R_{-\vec k}(t_1)\left( \ddot{\Psi}_{\vec k}(t_1)+\Ok^2\,\Psi_{\vk}(t_1)  \right) \right\} \nonumber \\
                 & - & \int_0^{t} dt_1 \int_0^{t} dt_2 \left\{\frac{1}{2}\,R_{\vec k}(t_1)\,{\mathcal{N} }_k(t_1-t_2)\,R_{-\vec k}(t_2) + R_{-\vec k}(t_1)\,
i\Sigma^R_k(t_1-t_2)\, \Psi_{\vec k}(t_2) \right \} \nonumber \\
& -i & \int d^3x R_i(\vec x) \dot{\Psi}(\vec x,t=0)
\label{effkel}
\end{eqnarray}

\noindent where $\Ok^2=k^2+m^2_\phi$, and the last term arises after  integration by parts in time,
using the boundary conditions (\ref{kelbc}) and (\ref{Rt}).
The kernels in the above effective Lagrangian are given by (see
eqns. (\ref{ggreat}-\ref{antitimeordered}))
\begin{eqnarray}
\mathcal{N}_k(t_1-t_2) & = &
\frac{g^2}{2} \left[   C^>(k;t_1-t_2)+{ C}^<(k;t_1-t_2) \right] \label{kernelkappa} \\
i\Sigma^{R}_k(t_1-t_2) & = &  g^2 \left[{ C}^>(k;t_1-t_2)-{
C}^<(k;t_1-t_2) \right]\Theta(t_1-t_2) \equiv
i\Sigma_{k}(t_1-t_2)\Theta(t_1-t_2) \label{kernelsigma}
\end{eqnarray} where $C^{\lessgtr}(k;t-t')$ are the spatial Fourier transforms of the correlation functions in (\ref{ggreat}-\ref{antitimeordered}).

Up to a normalization constant, the term quadratic in the relative variable $R$ can be written in terms of a stochastic
noise as
\begin{eqnarray}
\exp\Big\{-\frac{1}{2} \int dt \int dt' R_{-\vec k}(t)\mathcal{N}_k(t-t')R_{\vec k}(t')\Big\} & = & \int {\cal D}\xi
\exp\Big\{-\frac{1}{2} \int dt \int dt' ~~ \xi_{\vec k}(t)
\mathcal{N}^{-1}_k(t-t')\xi_{-\vec k}(t')  \nonumber \\
 &+ & i \int dt ~~ \xi_{-\vec k}(t) R_{\vec k}(t)\Big\}
\label{noisefunc}
\end{eqnarray}

 Correlation functions of the center of mass Keldysh field $\Psi(\vx,t)$ are now obtained as

\begin{eqnarray}
&& \mathrm{Tr} \Big(\Psi  \cdots\Big)\,\rho^r_S    =    \int D \Psi_i \int D \Pi_i DR_i\int {\cal D} \Psi {\cal D}R {\cal D}\xi ~
{ W}(\Psi_i;\Pi_i\,
 e^{i \int d^3x R_i(\vec x) \left(\Pi_i (\vec x)-\dot{\Psi}(\vec x,t=0)\right)} {\cal P}[\xi]\, \Big(\Psi  \cdots\Big)
     \nonumber \\
& \times &    \exp\left\{-i \int_0^{t} dt_1 ~~ R_{-\vec k}(t_1) \left[
\ddot{\Psi}_{\vec k}(t_1)+\Ok^2\Psi_{\vec k}(t_1)+\int_0^{t_1} dt_2 ~~ \Sigma_k(t_1-t_2)\Psi_{\vec k}(t_2)-\xi_{\vec k}(t_1) \right] \right\} \eea where
\be{\cal P}[\xi]  =   \exp\left\{-\frac{1}{2} \int_0^{t} dt_1
\int_0^{t} dt_2 ~~ \xi_{\vec k}(t_1) \mathcal{N}^{-1}_k(t_1-t_2)
\xi_{-\vec k}(t_2) \right\} \,, \label{genefunc}
\ee and we have used the definition of $\Sigma^{R}_k(t-t')$
in terms of $\Sigma_k(t-t')$ given in equation (\ref{kernelsigma}).

The functional integral over $R_i$ can now be done, resulting in a functional delta function,
that fixes the boundary condition $\dot{\Psi}(\vec x,t=0) =
\Pi_i(\vec x)$.

Finally the path integral over the relative variable can be
performed, leading to a functional delta function and the final form
of correlations of $\Psi$ is given by
\begin{eqnarray}
\mathrm{Tr} \Big(\Psi  \cdots\Big) \rho^r_S & =  & \int D \Psi_i   D \Pi_i ~~{  W}(\Psi_i;\Pi_i)
 {\cal D} \Psi {\cal D}\xi ~~ {\cal P}[\xi]\,\Big(\Psi  \cdots\Big)\nonumber \\
 & & \delta\left[
\ddot{\Psi}_{\vec k}(t)+\Ok^2\Psi_{\vec k}(t)+\int_0^{t} dt'
~\Sigma_k(t-t')\Psi_{\vec k}(t')-\xi_{\vec k}(t) \right]
\label{deltaprob}
\end{eqnarray}
with the boundary conditions on the path integral on $\Psi$ given by
\begin{equation}
\Psi(\vec x,t=0) = \Psi_i(\vec x) \; \; ; \; \; \dot{\Psi}(\vec
x,t=0)= \Pi_i(\vec x) \label{bcfin}
\end{equation}

The meaning of the final equation (\ref{deltaprob}) for the correlation functions of $\Psi$  is the following:
\begin{itemize}
\item{ Obtain  the solution of the classical
{\em stochastic} Langevin equation of motion
\begin{eqnarray}
&& \ddot{\Psi}_{\vec k}(t)+\Ok^2\Psi_{\vec k}(t)+\int_0^t dt'
~ \Sigma_{k}(t-t')
\Psi_{\vec k}(t')=\xi_{\vec k}(t)\, \nonumber \\
&& \Psi_{\vec k}(t=0)= \Psi_{i,\vec k} ; ~~
\dot{\Psi}_{\vec k}(t=0)= \Pi_{i,\vec k}\,
\label{langevin}
\end{eqnarray}
for arbitrary noise term $\xi$.}

\item{ Average the products of
$\Psi$ over the stochastic noise with the Gaussian probability
distribution ${\cal P}[\xi]$ given by (\ref{genefunc})}.

\item{ Finally
average over the initial conditions $\Psi_i(\vec x);
\Pi_i(\vec x)$ weighted by the Wigner function ${
W}(\Psi_i,\Pi_i)$, which plays the role of an initial (semiclassical)
 phase space distribution function.}

 \end{itemize}

 In summary, the time evolution of the Keldysh field $\Psi(\vx,t)$ is determined by the Langevin equation (\ref{langevin}), and equal time correlation functions of $\Psi$ involve  two different averages:

\begin{itemize}
\item{ The average over the  stochastic noise term, which up to
this order is Gaussian. We denote the average of a functional
$\mathcal{A}[\xi]$  over the noise with the probability distribution
function $P[\xi]$ given by eqn. (\ref{genefunc}) as

\begin{equation}\label{stocha}
\langle \langle \mathcal{A}[\xi] \rangle \rangle \equiv \frac{\int
\mathcal{D}\xi P[\xi] \mathcal{A}[\xi]}{\int \mathcal{D}\xi P[\xi]}.
\end{equation}

Since the noise probability distribution function is Gaussian the
only necessary correlation functions for the noise are given by

\begin{equation}
\langle \langle \xi_{\vec{k}}(t)\rangle \rangle =0 \; , \; \langle
\langle \xi_{\vec{k}}(t)\xi_{-\vec{k}'}(t')\rangle \rangle =
 \mathcal{N}_k(t-t')\,\delta_{\vk,\vk'}
\label{noisecorrel}
\end{equation}
\noindent and  the higher order correlation functions are obtained
from Wick's theorem. In general the  noise kernel
$\mathcal{N}_k(t-t')\neq \delta(t-t')$ corresponding to
\emph{colored} noise.

The two-point correlation functions of bath operators in the space-time translational invariant initial density matrix $\rho_E(0)$ (\ref{rhochi})   admits a non-perturbative Lehmann representation in terms of a spectral density, which is obtained in appendix (\ref{app:fd}), with the results
\bea i\Sigma_{k}(t-t')  & = &  g^2 ~ \int \frac{dk_0}{(2\pi)}\,\rho(k_0,k) \, e^{-ik_0(t-t')}  \label{isig} \\
\mathcal{N}_k(t-t') & = & \frac{g^2}{2}~  \int \frac{dk_0}{(2\pi)}\,\rho(k_0,k) \,\mathrm{coth\big[ \frac{\beta k_0}{2}\big]} e^{-ik_0(t-t')}  \label{noiset}\eea therefore the self energy $i\Sigma$ and noise $\mathcal{N}$ kernels obey the generalized fluctuation dissipation relation\cite{kms} derived in appendix (\ref{app:fd}).

In the literature it is usually \emph{assumed} that the noise term $\mathcal{N}_k(t-t')$ has very short range temporal correlations, namely
\be \mathcal{N}_k(t-t') \propto \delta(t-t') \label{delnoise}\ee which implies  that
\be \rho(k_0,k) \,\mathrm{coth\big[ \frac{\beta k_0}{2}\big]} \simeq \mathrm{constant}\,.\label{const} \ee In the classical limit $\mathrm{coth[k_0/2T]}\simeq 2T/k_0$ this implies an Ohmic noise spectral density\cite{breuer}, namely \be \rho(k_0,k) \propto k_0 \,.\label{ohm} \ee However, this spectral density is in general not compatible with that arising in a relativistic quantum  field theory as it will be discussed in detail in the next sections, therefore, in general an  ohmic spectral density is not expected to  reliably describe the temporal correlations of a realistic bath described by a quantum field theory.}

\item{The average over the initial conditions with the Wigner
distribution function ${  W}(\Psi_i,\Pi_i)$ which we denote as

\begin{equation}
\overline{\mathcal{A}[\Psi_i,\Pi_i]} \equiv  \frac{\int D \Psi_i
\int D\Pi_i ~~{  W}(\Psi_i;\Pi_i) \mathcal{A}[\Psi_i,\Pi_i]}{\int
D \Psi_i \int D\Pi_i ~~{  W}(\Psi_i;\Pi_i) } \label{iniaverage}
\end{equation}

}

\end{itemize}

Therefore correlation functions of the Keldysh variable $\Psi$  in the reduced density matrix of the system are given by

\begin{equation}
\mathcal{C}[\Psi] \equiv \mathrm{Tr}\Big(\Psi \cdots\Big)\,\rho^r_S \equiv   \overline{\langle \langle
\Big(\Psi \cdots\Big) \rangle \rangle}=  \overline{ \frac{\int
\mathcal{D}\xi P[\xi]~  \Big(\Psi\cdots\Big)}{\int \mathcal{D}\xi P[\xi]}}\,. \label{totave}
\end{equation} where $\Psi$ is the  solution of the Langevin equation (\ref{langevin}) and depends explicitly on the noise $\xi$.

 Calling
the solution of (\ref{langevin}) $\Psi_{\vec
k}(t;\xi;\Psi_i;\Pi_i)$,  the expectation value and equal time two point correlation function of $\Psi$  are given by
\be  \langle \phi_{\vk}(t)  \rangle =  \overline{\langle \langle \Psi_{-\vec k}(t)  \rangle \rangle}    ~~;~~\langle \phi_{\vk}(t)\phi_{-\vk}(t) \rangle =  \overline{\langle \langle \Psi_{-\vec k}(t) \Psi_{\vec k}(t) \rangle \rangle}  \,.  \label{eqtimfi}\ee where the averages on the left hand side of this equation are in the reduced density matrix of the system, and $\Psi$ on the right hand side is the solution of the Langevin equation (\ref{langevin}), which is an explicit function of $\Psi_i,\Pi_i,\xi$.

This result is  remarkably similar to the Martin-Siggia-Rose\cite{msr}   formulation for the path integral representation of
stochastic \emph{classical} systems.

It follows from this analysis that the expectation value of the spatial Fourier transform of the scalar field in the reduced density matrix obeys the \emph{retarded} equation of motion

\be  \frac{d^2}{dt^2} {\langle \phi_{\vk}(t)  \rangle} +\Ok^2\langle \phi_{\vk}(t)  \rangle+\int_0^t
~ \Sigma_{k}(t-t')
\langle \phi_{\vk}(t')  \rangle \, dt'\, = 0 \,, \label{exvalred}   \ee a result that will establish a link with the framework of the quantum master equation in the next section.

The solution of the Langevin equation (\ref{langevin}) is obtained by a Laplace transform, defining the Laplace transforms

\begin{eqnarray}
\widetilde{\Psi}_{\vec k}(s)  \equiv \int^{\infty}_0 dt
e^{-st}\Psi_{\vec k}(t)\label{laplapsi}\\
\widetilde{\xi}_{\vec k}(s)  \equiv \int^{\infty}_0 dt
e^{-st}\xi_{\vec k}(t)\label{laplaxi}
\end{eqnarray}

\begin{equation}
\widetilde{\Sigma}(k,s)\equiv \int^{\infty}_0 dt e^{-st}\Sigma_k(t)=
-\frac{g^2}{2\pi} \int^{\infty}_{-\infty}
\frac{\rho(k_0,k)}{k_0-is}~dk_0\,.
\label{laplasig}
\end{equation}

\noindent in terms of which we find

\begin{equation}\label{solutionlap}
\widetilde{\Psi}_{\vec k}(s)=\frac{\Pi_{i,\vec k}+s \, \Psi_{i,\vec
k}+\widetilde{\xi}_{\vec
k}(s)}{s^2+\Ok^2+\widetilde{\Sigma}(k,s)}
\end{equation}
\noindent where we have used the initial conditions (\ref{bcfin}).
  Introducing the function ${G}_k(t)$ that obeys the
following equation of motion and initial conditions

\begin{equation}
\ddot{ G}_{k}(t)+\Ok^2\,{G}_{ k}(t)+\int_0^t dt' ~
\Sigma_{k}(t-t') G_{ k}(t')=0~~;~~  {G}_k(t=0)= 0; ~~ \dot{{G}}_{
k}(t=0)=1 \label{functionf}
\end{equation}

\noindent whose Laplace transform is given by

\begin{equation}\label{laplaf}
\widetilde{G}_k(s) =
\frac{1}{s^2+\Ok^2+\widetilde{\Sigma}(k,s)}
\end{equation}
we find that the solution of the Langevin
equation (\ref{langevin}) in real time is given by

\begin{equation}
\Psi_k(t) = \mathcal{H}_{\vk}(t)+ \int^t_0
G_k(t-t')~\xi_{\vec k}(t') dt' \,,\label{inhosolution}
\end{equation}  where
\be \mathcal{H}_{\vk}(t) = \Psi_{i,\vec k}~ \dot{G}_k(t) +
 \Pi_{i,\vec k}~ G_k(t) \,,\label{Psiini} \ee is the solution of the homogenoues Langevin equation (\ref{langevin}), it depends only on initial conditions but not on the noise and
the last term in (\ref{inhosolution}) highlights the stochastic nature of the ``in-in'' effective field theory and depends solely on the noise and not
on the initial conditions. As a consequence of this separation, and the Gaussian correlations of the noise (\ref{noisecorrel}), it follows that
\be
 \overline{\langle \langle \Psi_{\vk}(t) \rangle \rangle}  =    \overline{\mathcal{H}_{\vk}(t)} \,,\label{avesolPsi}\ee
\be  \overline{\langle \langle \Psi_{\vk}(t)\,\Psi_{-\vk}(t) \rangle \rangle}    =    \overline{\mathcal{H}_{\vk}(t)\mathcal{H}_{-\vk}(t)}+ \frac{g^2}{2}~  \int \frac{dk_0}{(2\pi)}\,\rho(k_0,k) \,\mathrm{coth\big[ \frac{\beta k_0}{2}\big]}\Big[\int^t_0 G_k(\tau)\,e^{-ik_0\tau}\,d\tau\,\int^t_0 G_k(\tau) \,e^{ik_0\tau}\,d\tau \Big]
\,. \label{avesolPsisq} \ee

 $G_k(t)$ is obtained  by the inverse Laplace
transform of (\ref{laplaf})

\begin{equation}\label{bromw}
G_k(t) = \int_{C}\frac{ds}{2\pi i} \frac{e^{st}
}{s^2+\Ok^2+\widetilde{\Sigma}(k,s)}
\end{equation}
\noindent where $C$ stands for the Bromwich contour, parallel to the
imaginary axis in the complex $s$ plane to the right of all the
singularities of $\widetilde{G}_k(s)$ and along the semicircle at
infinity for $\mathrm{Re}\,s < 0$. The singularities of
$\widetilde{G}_k(s)$ in the physical sheet are isolated single
particle poles  and multiparticle cuts along the imaginary axis. Decaying states correspond to complex poles with a \emph{negative} real part. Thus the contour   runs parallel to the imaginary
axis with a small positive real part, namely
$s=i\omega+\epsilon\,;\,-\infty \leq \omega \leq \infty$ with $\epsilon \rightarrow 0^+$.
 Therefore
\be G_k(t) = -\int \frac{d\omega}{2\pi } \frac{e^{i\omega\,t}
}{\Big[(\omega-i\epsilon)^2-\Ok^2- {\Sigma}_R(k,\omega) -i\Sigma_I(k,\omega)\Big]} \label{goft2} \ee which is recognized as the Fourier transform of the \emph{retarded} propagator,
where
\bea \Sigma_R(k,\omega) & = &  \frac{g^2}{2\pi}\, \int dk_0 \mathcal{P} \Big[\frac{\rho(k_0,k)}{\omega-k_0} \Big] \label{sigmaR}\\
\Sigma_I(k,\omega) & = & \frac{g^2}{2} \, \rho(\omega,k) \label{sigmaI}\,, \eea where $\mathcal{P}[\cdots]$ is the principal part and  $\Sigma_R(k,\omega), \Sigma_I(k,\omega)$ are even and odd functions of $\omega$ respectively as a consequence of the property (\ref{oddros}) which was used to write (\ref{sigmaR},\ref{sigmaI}).

In the non-interacting theory the integrand in (\ref{goft2}) features poles at $\omega^\pm = \pm \Ok$, including self-energy corrections the positions of the complex poles are determined by
\be \Omega^2_k -\Ok^2-\Sigma_R(k,\Omega_k)-i\Sigma_I(k,\Omega_k)=0\,. \label{poles}\ee Consistently with a perturbative expansion in $g$, this equation is solved by writing
\be \Omega^\pm_k = \pm \omega_k + \delta_\pm   ~~;~~ \frac{|\delta_{\pm}|}{|\omega_k|} \propto g^2 \ll 1\,,\label{delomes}\ee and to
  leading order in $g^2$ we find
  \be \delta_{\pm} = \pm \frac{\Sigma_R(k,\Ok)}{2\,\Ok} + i \frac{\Gamma_k}{2} \,,\label{delomsol}\ee with the decay width
  \be  \Gamma_k  =  \frac{\Sigma_I(k,\Ok)}{{\Ok}} =  \frac{ g^2}{2\Ok}\,\rho(\Ok,k) \,,   \label{width}\ee in agreement with the S-matrix (FGR) decay rate.
  The \emph{real} part of the poles are given by the renormalized frequencies
\be     \pm \Okr = \pm\Ok \pm \frac{\Sigma_R(k,\Ok)}{2\,\Ok}\,. \label{polerenor}\ee Near these poles and for small coupling, the propagator may be approximated   by writing $\omega = \pm\Omega_k + (\omega\mp \Omega_k)$ and keeping the linear term in $(\omega\mp\Omega_k)$ leading to the Breit-Wigner form of  the propagator in (\ref{goft2})
\be \frac{Z(k)}{\pm 2\,\Omega^R_k} \frac{1}{\Big[\omega\mp\Omega^{R}_k-i\frac{\Gamma_k}{2}\Big]}\,, \label{BWprop}\ee where to $\mathcal{O}(g^2)$  the wave function renormalizaton is given by
\be Z(k)   =    1+ \frac{\Sigma'_R(k,\Ok)}{2\,\Ok} = 1- \frac{g^2}{4\pi\Ok}\, \int dk_0 \mathcal{P} \Big[\frac{\rho(k_0,k)}{(\Ok-k_0)^2} \Big] \label{wavefuncren} \ee  In (\ref{wavefuncren}) $\Sigma'_R = d\Sigma_R/d\omega$. Because $\Sigma_R,\Sigma_I$ are even and odd in $\omega$ respectively, it follows that $Z,\Gamma$ are the same for both poles.  At long time the Fourier transform in (\ref{goft2}) is dominated by the complex poles of the Breit-Wigner propagator near the real axis, yielding
\be G_k(t)=  {Z}(k) \,e^{-\frac{\Gamma_k}{2}\,t}\,\frac{ \sin[\Okr\, t]}{\Okr}\,. \label{solutionG}\ee

 Introducing this solution into (\ref{inhosolution}) it is   straightforward   to obtain the \emph{long time} averages of  $\Psi_{\vk}(t)$ and  $\Psi_{\vk}(t)\Psi_{-\vk}(t)$ with the initial Wigner distribution function and the stochastic noise as in eqn. (\ref{totave}). We find
  \be \overline{\langle \langle \Psi_{\vk}(t)  \rangle \rangle } =  \overline{\mathcal{H}_{\vk}(t)} =  Z e^{-\frac{\Gamma_k}{2} t} \Big[ \overline{\Psi_{i,\vk}}~\cos[\Okr \, t] + \frac{\overline{\Pi_{i,\vk}}}{\Okr}~\sin[\Okr \,t] \Big] \,. \label{avePsi} \ee

We can now obtain the long time limit of the equal time correlation function (\ref{avesolPsisq}): the first term on the right hand side that depends on the initial conditions falls off $\propto e^{-\Gamma_k t}$, with the result (\ref{solutionG}) the time integrals in the second term can be done straightforwardly yielding four terms, the $k_0$ integrals are performed by contour integrations. In the narrow width approximation $\Gamma_k/\Okr \ll1$, two of the terms feature poles at $\pm \Omega^R_k$ with residues $\propto1/\Gamma_k$, two other poles with residues $\propto 1/(\pm 2\Omega^R_k + \Gamma_k)$ and poles from the thermal function at the Matsubara frequencies $k_0 = i 2\pi\,m/\beta\,;m = \pm 1,\pm 2\cdots$ (the pole at $m=0$ yields vanishing residue because $\rho(0,k) =0$), depending on the analytic structure of $\rho(k_0,k)$ there may also be cut- discontinuities in the complex $k_0$ plane. Of all of these contributions, the poles at $\pm \Omega^R_k$ dominate since $1/\Gamma_k \propto 1/g^2$, all the others yield terms of $\mathcal{O}(g^2)$, therefore, using the result (\ref{width}) we obtain
 \be \overline{\langle \langle \Psi _{\vk}(t)\,\Psi_{-\vk}(t)  \rangle\rangle} = e^{-\Gamma_k t}\,\,\Big(\mathrm{initial~conditions}\Big)+ \frac{1}{2\Omega_k}\,\Big[ 1 + 2 n(\Omega_k)\Big]\, \Big(1-e^{-\Gamma_k t}\Big)+ \mathcal{O}(g^2)\,, \label{asquareave} \ee For $\Gamma_k \, t \gg1$ this result implies thermalization of the scalar field with the thermal bath.

Although this result is physically in agreement with the expectations from Brownian motion, it inputs very limited information on the spectral density of the bath, since $\Gamma_k$ only depends on $\rho(k_0,k)$ evaluated at $k_0= \Omega^R_k$, as we will see in detail below, $\rho(\omega_k,k)$ vanishes in threshold and infrared cases, yielding $\Gamma_k =0$, which neglects important transient dynamics because even in these cases $\rho(\omega,k) \neq 0$ for $\omega \neq \omega_k$ . This limitation is a consequence of focusing solely on the long time dynamics.

The Breit-Wigner approximation yielding the result (\ref{solutionG}) neglects the   contribution from the continuum along the branch cuts, describing the propagator in terms of simple narrow  resonances with complex poles near the real axis, it is only reliable in the long time limit. The contribution from the continuum determines early transient dynamics and   also gives rise to power law long time tails dominated by the behavior of the spectral density near the thresholds\cite{fonda,misra} which are neglected in the Breit-Wigner approximation.

The solution of equation (\ref{functionf}) for $g=0$ ($\Sigma_k =0$) is $G_k(t) = \sin[\omega_k t]/\omega_k$, and including self-energy corrections under the perturbative assumption $\Sigma(\omega,k)/\omega^2_k \propto g^2 \ll 1$ yields (\ref{solutionG}) valid in the long time limit, thus the radiative corrections amount to a renormalization of frequency and a decay width both $\propto g^2$. This observation motivates us to seek a solution of (\ref{functionf}) in a multiple time scale expansion, proposing
\be G_k(t) = A_k(t)\,e^{-i\omega_k t} \,+\, B_k(t)\,e^{i\omega_k t}\,,\label{multim}\ee where $A_k(t);B_k(t)$ are slowly varying functions of time, namely
\be \dot{A}_k(t)~;~\dot{B}_k(t) \propto g^2\,,\label{slovar}\ee
inserting this \emph{ansatze} in equation (\ref{functionf}) yields
\bea
&& e^{-i\omega_k t}\,\Bigg[\ddot{A}_k(t)-2i\omega_k\,\dot{A}_k(t) + \int^t_0 \Big[\Sigma_k(t-t')\,e^{i\omega_k(t-t')}  \Big]\,A_k(t')\,dt'  \Bigg] + \nonumber \\ && e^{i\omega_k t}\,\Bigg[\ddot{B}_k(t)+2i\omega_k\,\dot{B}_k(t) + \int^t_0 \Big[\Sigma_k(t-t')\,e^{-i\omega_k(t-t')}  \Big]\,B_k(t')\,dt'   \Bigg] \,=0 \,, \label{GABeq}
\eea  and request that each bracket vanishes separately.

Consistent with a perturbative expansion and the condition (\ref{delomes}),  we will require self-consistently that
\be  \ddot{A}_k(t) \ll \omega_k\,\dot{A}_k(t)~;~ \ddot{B}_k(t) \ll \omega_k\,\dot{B}_k(t)\,,\label{ddotAB}\ee therefore, the second derivative terms in (\ref{GABeq}) can be neglected. Let us first consider the resulting equation for $A_k(t)$,
 \be \dot{A}_k(t) = \frac{1}{2i\omega_k}\,   \int^t_0 \Big[\Sigma_k(t-t')\,e^{i\omega_k(t-t')}  \Big]\,A_k(t')\,dt'   \,,\label{dotA}\ee which is consistent with (\ref{slovar}) since $\Sigma \propto g^2$, hence $A_k(t)$ is slow. Although (\ref{dotA}) can again be solved by Laplace transform, we rewrite the non-local time integral as
\be \int^t_0 \Sigma_k(t-t') \, e^{i\omega_k\,(t-t')}\,{A_k}(t')\,dt' \equiv \int^t_0 \frac{d}{dt'}\Big[\int^{t'}_0 \Sigma_k(t-t') \, e^{i\omega_k\,(t-t')}\,dt''\Big] \,{ {A_k}}(t')\,dt'\,,\label{trick}\ee integrating by parts on the right hand side and neglecting the term $\simeq \dot{ {A}}_k\,\Sigma_k \propto g^4$ to leading order in $g^2$  equation (\ref{dotA}) becomes local, namely
\be \dot{A}_k(t) = \Big[\frac{1}{2i\omega_k}\, \int^t_0 \Sigma_k(t-t') \, e^{i\omega_k\,(t-t')}\, \,dt'\Big] \,{ {A}_k}(t)\,. \label{dotAloc} \ee Using the representation (\ref{isig}), the time integral can be done yielding
\be \dot{A}_k(t) = \Big[ -i \Delta(k,t) - \frac{\Gamma(k,t)}{2} \Big] A_k(t)\,,\label{dotAfin} \ee where we introduced
\be \Delta(k,t) = \frac{g^2}{2\,\omega_k} \,\int \frac{dk_0}{2\pi} \,\rho(k_0,k)\,\frac{\Big[1-\cos[(\omega_k-k_0)t] \Big]}{(\omega_k-k_0)}\,, \label{deltat}\ee
\be \Gamma(k,t) =   \frac{g^2}{\omega_k} \,\int \frac{dk_0}{2\pi} \,\rho(k_0,q)\,\frac{ \sin[(\omega_k-k_0)t] }{(\omega_k-k_0)}\,   \,. \label{gamat}\ee A similar analysis applies to $B_k(t)$ with $\omega_k \rightarrow -\omega_k$, which results in $\Delta_k(t) \rightarrow -\Delta_k(t) ~;~ \Gamma_k(t) \rightarrow \Gamma_k(t)$ as a consequence of $\rho(k_0,k) = - \rho(-k_0,k)$. With the initial conditions (\ref{functionf}), we finally find
\be G_k(t) = \frac{1}{\omega_k}\,{e^{-\frac{1}{2}\int^t_0 \Gamma(k,t') dt'}} \,\sin\Big[\omega_k\,t +\int^t_0 \Delta(k,t')\,dt'\Big] \,.\label{Goftfina}\ee

Now, we can self-consistently assess the assumption   (\ref{ddotAB}),
\be \ddot{{A}}_k = \Big[ -i \dot{\Delta}(k,t)-\frac{1}{2}\,\dot{\Gamma}(k,t)   \Big] \,{ {A}_k} + \mathcal{O}(g^4) \,.\label{ddA}\ee  The derivatives $\dot{\Delta},\dot{\Gamma}$ do not feature the resonant denominators and  average out on short time scales vanishing in the long time limit, whereas both $\Delta(k,t);\Gamma(k,t)$ approach a constant in the long time limit (see next section), hence $\dot{\Delta}_k/\omega_k \ll \Delta_k~;~ \dot{\Gamma}_k/\omega_k \ll \Gamma_k$ consistently with weak coupling and (\ref{delomes}). These aspects   will be discussed in more detail in the next section within the context of the quantum master equation.

Combining this result with equations (\ref{eqtimfi},\ref{Psiini},\ref{avesolPsi}) we obtain

\be  \langle \phi_{\vk}(t)  \rangle =  \overline{\langle \langle \Psi_{-\vec k}(t)  \rangle \rangle} = {e^{-\frac{1}{2}\int^t_0 \Gamma(k,t') dt'}} \,\Big[ {A}_k(0)\, e^{-i\omega_k t}\,e^{-i\int^t_0 \Delta(k,t')dt'}+  {B}_k(0)\, e^{i\omega_k t}\,e^{i\int^t_0 \Delta(k,t')dt'}\Big]\,,\label{exfifin}\ee where the constant coefficients $ {A}_k(0); {B}_k(0)$ depend on the initial conditions.

In particular, note that taking the infinite time limit of $\Gamma(k,t)$ in equation (\ref{gamat}), and  using
\be \frac{ \sin[(\omega_k-k_0)t] }{(\omega_k-k_0)}~~{}_{\overrightarrow{t\rightarrow \infty}}~~\pi\,\delta(k_0-\omega_k)\,, \label{sinc}\ee in the sense of a distribution as in Fermi's golden rule,  yields
\be \Gamma(k,t) ~~ ~~{}_{\overrightarrow{t\rightarrow \infty}}~~\frac{g^2}{2\omega_k}\,\rho(\omega_k,k) \,,\label{gif}\ee which is precisely the S-matrix (FGR) result for the decay rate (\ref{width}).

However, the functions $\Delta(k,t);\Gamma(k,t)$ describe early time transient dynamics, depend on the full spectral density, and their exponentials in (\ref{exfifin}) depend on the history, consequently  they describe non-Markovian dynamics. They are a bridge to  and   play a fundamental role in the (QME) approach  in the next section where   their properties and asymptotic behavior   will be analyzed in detail.

\subsection{The Quantum Master Equation.}\label{subsec:QME}

A complementary framework for quantum open systems is the quantum master equation. In this section we derive the (QME) linking its derivation to the ``in-in'' formulation of the previous section and focusing  explicitly on obtaining the transient dynamics in the cases in which the environment features critical behavior.

In the Schroedinger picture, the time evolution   of an initial density matrix is determined by equation (\ref{rhooft}) with the total Hamiltonian (\ref{hami}), however, it is more convenient to pass to the interaction picture,
\be \rho_I(t) = U(t)\,\rho(0)\,U^{-1}(t)~~;~~ U(t) = e^{i(H_S+H_E)t}\,e^{-iHt}\,,\label{rhoip} \ee where the system and environment Hamiltonians are assumed to commute $[H_S,H_E]=0$. The time evolution operator in the interaction picture $U(t)$ obeys
\be \frac{d}{dt} U(t) = -iH_I(t) \,U(t)~~;~~ U(0) =1 \,,  \label{dUdt}\ee and
\be H_I(t)= e^{iH_S t}\, e^{iH_E t}\,H_{SE}\,e^{-iH_E t}\,e^{-iH_S t} \,,\label{Hinip}\ee is the interaction Hamiltonian in the interaction picture. The Heisenberg  system (S) and environment (E) operators evolve in time with the total Hamiltonian (\ref{hami}), namely
\be \mathcal{O}^{S} (t) =  e^{iH t}\,\mathcal{O}^{S}(0)\, e^{-iH t}~~;~~\mathcal{O}^{E} (t) =  e^{iH  t}\,\mathcal{O}^{E}(0)\, e^{-iHt}\,,\label{OSEheis}\ee while in the interaction picture

\be \mathcal{O}^{S}_{I}(t) =  e^{iH_S t}\,\mathcal{O}^{S}(0)\, e^{-iH_S t}~~;~~\mathcal{O}^{E}_{I}(t) =  e^{iH_E t}\,\mathcal{O}^{E}_{I}(0)\, e^{-iH_E t}\,,\label{OSEin}\ee therefore Heisenberg and interaction picture operators are related by
\be \mathcal{O}^{S} (t) = U^{-1}(t)\, \mathcal{O}^{S}_I (t)  \,U(t)~~;~~ \mathcal{O}^{E} (t) = U^{-1}(t)\, \mathcal{O}^{E}_I (t)  \,U(t)\,,\label{Osheisip}\ee therefore
\be \langle \mathcal{O}^{S,E}(t) \rangle \equiv \mathrm{Tr} \mathcal{O}^{S,E}(t)\,\rho(0) = \mathrm{Tr} \mathcal{O}^{S,E}_I(t)\,\rho_I(t) \,.\label{equiv}\ee

 The interaction picture density matrix $\rho_I(t)$ obeys the time evolution equation
\be \dot{\rho}_I(t) = -i [H_I(t),\rho_I(t)] \,,\label{rhodot}\ee with formal solution
\be \rho_I(t) = \rho_I(0)-i\,\int^t_0 [H_I(t'),\rho_I(t')] \,dt' \,. \label{solu}\ee This solution is inserted back into (\ref{rhodot}) leading to the iterative equation
\be \dot{\rho}_I(t) = -i[H_I(t),\rho_I(0)]- \int^t_0[H_I(t),[H_I(t'),\rho_I(t')]]\,dt' \,.\label{itereq}\ee This equation cannot be solved exactly and several approximations are usually invoked\cite{breuer,linme}.

The interaction picture \emph{reduced density matrix} is defined as

\be \rho^r_S(t) = \underset{\{E\}}{\mathrm{Tr}} \, \rho_I(t)\,. \label{rhored1}\ee

\textbf{i): Factorization:}  the total interaction picture density matrix is assumed to factorize at all times as
\be \rho_I(t) = \rho^r_{S}(t)\otimes \rho_E(0)  \,,\label{fact}\ee where $\rho_E(0)$ is given by (\ref{rhochi}) and $\rho^r_S(t)$ is the \emph{reduced density matrix for the system} given by (\ref{rhored1}). As a consequence of this factorization, for any interaction picture operator of the system's degrees of freedom $\mathcal{O}^S(t)$ it follows that
\be \langle \mathcal{O}^S \rangle(t) = \mathrm{Tr} \, \mathcal{O}^S(t)\,\rho^r_S(t)\,.\label{OSval}\ee

Although factorization is generally an \emph{assumption}\cite{breuer,linme}, the ``in-in'' framework analyzed in the previous section provides a justification, as can be understood clearly and directly from the general form
of the influence functional (\ref{Fln}) as obtained by expectation values in the initial environmental density matrix $\rho_E(0)$. We will discuss this important aspect in section (\ref{sec:discussion}). Since the initial density matrix is in factorized form and the composite operator $\mathcal{O}[\chi]$ in the interaction Hamiltonian $H_{SE}$ (\ref{hami}) has been normal ordered (see equation (\ref{NO})) it follows that the first term on the right hand side of (\ref{rhodot}) vanishes.

The \emph{reduced} density matrix of the system is obtained by tracing over the environmental fields (\ref{rhored1}), therefore, carrying out this trace in the evolution equation (\ref{itereq})  with the factorized form (\ref{fact}), the equation of motion for the system fields becomes
\bea \dot{\rho}^r_{S}(t) & =  & -g^2\int^t_0 dt' \int d^3x \int d^3 x'\Bigg\{C^>(\vx-\vx',t-t')\,\Big[ S_I(\vx,t) \, S_I(\vx^{\,'},t')\, {\rho}^r_{S}(t')-  S_I(\vx^{\,'},t') \,  {\rho}^r_{S}(t')\,S_I(\vx,t)\Big] \nonumber \\ & + & C^<(\vx-\vx',t-t')\,\Big[ {\rho}^r_{S}(t') \,S_I(\vx^{\,'},t')\, S_I(\vx,t)- S_I(\vx,t) \, {\rho}^r_{S}(t') \, S_I(\vx^{\,'},t')  \Big] \Bigg\}\,,\label{qmeS}\eea where $C^\lessgtr(\vx-\vx',t-t')$ are precisely the (connected) environmental correlation functions (\ref{ggreat},\ref{lesser}). Note the similarity with the   influence functional (\ref{Funravel}) and its time derivative (\ref{dtF}) with the equivalence (\ref{equivalence}). However an important aspect that is revealed by the derivation of the quantum master equation (\ref{qmeS}) is that the reduced density matrix $\rho^r_S$ is evaluated at $t'$, namely the resulting quantum master equation is time non-local.

The quantum master equation (\ref{qmeS}) is \emph{exact} up to second order in the system-environment interaction, the environmental correlation functions $C^{\lessgtr}$ include \emph{all orders} in the self interactions of the environmental fields, and the composite operators $S$ of the system are arbitrary operators that may include self-interactions of the system field. The main objective is to obtain the time evolution of expectation values or correlation functions of system variables in the reduced density matrix solution of (\ref{qmeS}).

With the factorization (\ref{fact}), for any interaction picture operator $\mathcal{O}_{IS}(\vx,t)$ that acts solely on the system degrees of freedom,   its expectation value in full density matrix $\rho_I(t)$   is given by
\be \langle \mathcal{O}_{IS}(\vx,t) \rangle \equiv \mathrm{Tr}\,\mathcal{O}_{IS}(\vx,t)\,\rho_{I}(t) = \mathrm{Tr}_S\,\mathcal{O}_{IS}(\vx,t)\,\rho^r_S(t) \,,\label{aveO}\ee consequently
\be \frac{d}{dt} \langle \mathcal{O}_{IS}(\vx,t) \rangle = \langle \dot{\mathcal{O}}_{IS}(\vx,t) \rangle + \mathrm{Tr}_S\,\mathcal{O}_{IS}(\vx,t)\,\dot{\rho}^r_{S}(t) \,.\label{timeder}  \ee

 \subsubsection{Free scalar fields coupled to the environment: quantum master equation}\label{subsub:freeqme}

 We now focus on the simple case of the system field $\phi$ as a free scalar field, with $S[\phi]\equiv \phi$  as in equation (\ref{free}) with two objectives: i) to compare to the results obtained with the ``in-in'' framework in the previous section, thereby establishing a direct link between the `in-in'' and the quantum master equation, ii) to discuss decoherence solely as a consequence of the interaction of the system with the environment. In this case the quantum master equation becomes
 \bea \dot{\rho}^r_{S}(t) & =  & -g^2\int^t_0 dt' \int d^3x \int d^3 x'\Bigg\{C^>(\vx-\vx',t-t')\,\Big[ \phi_I(\vx,t) \, \phi_I(\vx^{\,'},t')\, {\rho}^r_{S}(t')-  \phi_I(\vx^{\,'},t') \,  {\rho}^r_{S}(t')\, \phi_I(\vx,t)\Big] \nonumber \\ & + & C^<(\vx-\vx',t-t')\,\Big[ {\rho}^r_{S}(t') \,\phi_I(\vx^{\,'},t')\, \phi_I(\vx,t)- \phi_I(\vx,t) \, {\rho}^r_{S}(t') \, \phi_I(\vx^{\,'},t')  \Big] \Bigg\}\,,\label{qmephi}\eea where $\phi_I(\vx,t)$ is the field in free field interaction picture.

  The Hamiltonian equations of motion for the field $\phi$    in the Heisenberg picture with the full Hamiltonian (\ref{hami}) (where system bath interactions are assumed to not depend on time derivative of field operators), is
 \be \dot \phi(\vec{y},t) = \pi(\vec{y},t) \,, \label{heiseom}\ee and because in the Heisenberg picture the density matrix does not depend on time, it follows that
 \be   \frac{\partial}{\partial t} \langle \phi(\vec{y},t) \rangle = \mathrm{Tr}\, \pi_I(t) \rho_I(t) = \mathrm{Tr}_S\, \pi_{I} (t) \rho^r_S(t) \,,  \ee  where we used the identity (\ref{equiv}), and the factorization (\ref{fact}) since $\pi$ is a system operator, so that the trace over the environmental variables yields the reduced density matrix for the system. .
 Taking another time derivative and using
 \be \frac{\partial^2}{\partial t^2} \langle \phi(\vec{y},t) \rangle = \mathrm{Tr}\, \big[ \dot{\pi}_I(t) \rho^r_S(t) + \pi_I(t) \dot{\rho}^r_{S}(t)\big] \,.\label{eom2}\ee
 In  the interaction picture of the free field theory of the scalar field $\phi$, the  equation of motion
 \be \dot{\pi}_I(\vy,t) = \nabla^2\phi_I(\vy,t)\big)  - m^2 \phi_I(\vy,t) \,,\label{ffeom}\ee it remains to obtain the second term in the bracket in (\ref{eom2}), in which the time derivative of the reduced density matrix is given by the quantum master equation (\ref{qmephi}).    Using the cyclic property of the trace, along with the canonical commutation relation $[\pi_I(\vy,t),\phi_I(\vx,t)] = -i\,\delta^{(3)}(\vy-\vx)$, and the identity (\ref{equiv})  we obtain
 \be \frac{\partial^2}{\partial t^2}{\langle \phi(\vec{y},t) \rangle} -\nabla^2\langle \phi(\vec{y},t) \rangle +m^2_\phi\,\langle \phi(\vec{y},t) \rangle + \int^t_0 dt' \int d^3 x\,\Sigma(\vy-\vx,t-t')\,\langle \phi(\vec{y},t') \rangle = 0   \,,\label{reteom}  \ee where
 \be \Sigma(\vy-\vx,t-t')= -i g^2 \Big[ C^>(\vy-\vx,t-t') - C^<(\vy-\vx,t-t')\Big] \,. \label{selfen}\ee  This is precisely the same equation of motion as for the expectation value of the center of mass Keldysh field variable $\Psi$ in the ``in-in'' formulation (\ref{eqtimfi}, \ref{exvalred}) after a spatial Fourier transform.

 We note that the retardation in the equation of motion for the expectation value is a consequence of the fact that $\rho^r_S$ enters at time $t'$ in (\ref{qmephi}).

 \textbf{(Partial) Markov approximation:} The second approximation is the Markov approximation where $\rho^r_S(t') \rightarrow \rho^r_S(t)$ in the time integral in (\ref{qmephi}) thereby rendering the quantum master equation local in time. This approximation is justified in weak coupling, up to $\mathcal{O}(g^2)$ as we now show explicitly.
 Let us consider the first term in (\ref{qmephi}) as an example: introducing
 \be   \mathcal{H}(t') \equiv \int^{t'}_0 \phi_I({\vx}',t'') \, C^>(\vx-{\vx}',t-t'')dt''\,,\label{Hdefi}\ee the first term can be written as
 \be -g^2 \, \phi_I(\vx,t) \int^t_0 \frac{d\mathcal{H}(t')}{dt'} \, {\rho}^r_{S}(t') \,dt' = -g^2 \,\phi_I(\vx,t)  \mathcal{H}(t) {\rho}^r_{S}(t) + g^2 \,  \phi_I(\vx,t) \int^t_0  \mathcal{H}(t') \,\frac{d{\rho}^r_{S}(t')}{dt'} dt' \label{incha2}\ee where we have integrated by parts. Since $\dot{\rho}^r_{S} \propto g^2$, the second term in (\ref{incha2}) is $\propto g^4$ and can be safely neglected up to $\mathcal{O}(g^2)$. A similar analysis can be applied to all the other terms in (\ref{qmephi}) with the result
  \bea \dot{\rho}^r_{S}(t) & =  & -g^2\int^t_0 dt' \int d^3x \int d^3 x'\Bigg\{C^>(\vx-\vx',t-t')\,\Big[ \phi_I(\vx,t) \, \phi_I(\vx^{\,'},t')\, {\rho}^r_{S}(t)-  \phi_I(\vx^{\,'},t') \,  {\rho}^r_{S}(t)\, \phi_I(\vx,t)\Big] \nonumber \\ & + & C^<(\vx-\vx',t-t')\,\Big[ {\rho}^r_{S}(t) \,\phi_I(\vx^{\,'},t')\, \phi_I(\vx,t)- \phi_I(\vx,t) \, {\rho}^r_{S}(t) \, \phi_I(\vx^{\,'},t')  \Big] \Bigg\}\,.\label{qmephi2}\eea Note the similarity of this analysis with the multi-time scale argument leading up to the time-local evolution equation (\ref{dotAloc}). We refer to this approximation as a \emph{partial} Markov approximation because usually, in the full Markov approximation the upper limit of the time integral is taken to infinity\cite{breuer}, instead, we keep the finite time interval   to study transient dynamics that reveals properties of the spectral density of the environment.

  {\textbf{c:) Rotating wave approximation: }}
    In interaction picture of free fields the scalar field $\phi_I(\vx,t)$ is expanded as
  \be  \phi_I(\vx,t) =  \frac{1}{\sqrt{V }}\,\sum_{\vk}   \phi_I(\vk;t)\, e^{i\vk\cdot\vec{x}} ~~;~~ \phi_I(\vk;t) =  \frac{1}{\sqrt{2 \omega_k}}\,\Big[a(\vk)\,e^{-i\omega_kt}\,  + a^\dagger(-\vk)\,e^{i\omega_kt} \Big]~~;~~ \omega_k = \sqrt{k^2+m^2_\phi}\,.\label{phiip}\ee
   In writing the products $\phi_I(\vx,t)~\phi_I(\vx^{\,'},t')$ of interaction picture field operators (\ref{phiip}) in (\ref{qmephi2}) there are two types of terms  with very different time evolution. Terms of the form
\be a^\dagger(\vk)~a(\vk)~e^{  i{ \omega_k(t-t')}}\,, \label{ada} \ee   are ``slow'', and terms of the form
\be a^\dagger(\vk)~a^\dagger(-\vk)~ e^{2i\omega_k t'}~e^{i{\omega_k(t-t')}}~~;~~ a(\vk)~a(-\vk)~ e^{-2i\omega_k t'}~e^{-i{\omega_k(t-t')}}\,, \label{nonrot}\ee are fast, the extra rapidly varying phases $e^{ \pm 2i\omega_k t'}$ lead to rapid dephasing on time scales $\simeq 1/\omega_k$ and do not yield resonant (nearly energy conserving) contributions (see below). Neglecting these terms is tantamount to neglecting non-resonant terms that average out over   time scales  $\simeq 1/\omega_k$.   These terms only give perturbatively small transient non-secular contributions and are discussed in section (\ref{sec:discussion}). Keeping only the slow terms  and neglecting the   terms (\ref{nonrot})  defines the ``rotating wave approximation'' ubiquitous in quantum optics\cite{breuer,zoeller,carmichael}.

Implementing the  Markov  approximation ${\rho}^r_{S}(t')\rightarrow {\rho}^r_{S}(t)$, and the    rotating wave  approximation keeping only terms of the form $a^\dagger a, a a^\dagger$,  using the spectral representation of the correlators $C^\lessgtr$ given by (\ref{Ggfd},\ref{Glfd}),along with the relations (\ref{relas})   and carrying out the spatial and temporal integrals we finally obtain the  \emph{Lindblad} form\cite{breuer,zoeller,lin,gori,pearle,carmichael,linme} of the quantum master equation (QME)
\bea \dot{{\rho}}^r_{S}(t)  & = &  \sum_{\vq} \Bigg\{   -i\,\Delta(\vq,t)~\Big[a^\dagger(\vq)~a(\vq), {\rho}^r_{S}(t) \Big] \nonumber \\
& - & \frac{1}{2}\,\Gamma^>(q,t) \Big[a^\dagger(\vq)~a(\vq)~ {\rho}^r_{S}(t) + {\rho}^r_{S}(t)~a^\dagger(\vq)~a(\vq) - 2 \, a(\vq)~{\rho}^r_{S}(t)~a^\dagger(\vq) \Big]\nonumber \\
& - & \frac{1}{2}\,\Gamma^<(q,t) \Big[a(\vq)\,a^\dagger(\vq)~ {\rho}^r_{S}(t) + {\rho}^r_{S}(t)~a(\vq)\,a^\dagger(\vq) - 2 \, a^\dagger(\vq)~{\rho}^r_{S}(t)~a(\vq) \Big] \Bigg\} \,,
\label{Linfin} \eea  where
\be \Delta(q,t) = \frac{g^2}{2\,\omega_q} \,\int \frac{dq_0}{2\pi} \,\rho(q_0,q)\,\frac{\Big[1-\cos[(\omega_q-q_0)t] \Big]}{(\omega_q-q_0)}\,, \label{Roftim}\ee
\be \Gamma^>(q,t) = \frac{g^2}{ \omega_q}  \,\int \frac{dq_0}{2\pi} \, \rho(q_0,q)\,\big[1+n(q_0)\big]\frac{ \sin[(\omega_q-q_0)t] }{ (\omega_q-q_0)} \,, \label{gamgre}\ee
\be \Gamma^<(q,t) = \frac{g^2}{\omega_q}  \,\int   \, \rho(q_0,q)\, n(q_0) \frac{ \sin[(\omega_q-q_0)t] }{(\omega_q-q_0)}\, \frac{dq_0}{2\pi} \,, \label{gamles}\ee  where $n(q_0) = [e^{\beta q_0}-1]^{-1}$.

The second and third lines in (\ref{Linfin}) are called the \emph{dissipator}\cite{breuer}, these are non-Hamiltonian, purely dissipative terms, however it follows from the (QME) (\ref{Linfin}) that the trace of the reduced density matrix is conserved. It is argued in refs. \cite{lin,gori,pearle,carmichael}   that the equation (\ref{Linfin}) is the most general linear evolution equation that preserves unit trace and  Hermiticity of the density matrix.

In the usual Born-Markov approximation the coefficient functions in the Lindblad (QME), $\Delta(\vq,t),\Gamma^{\lessgtr}(\vq,t)$ are evaluated at $t =\infty$\cite{breuer,linme,nonmarkov,vega}, instead, we keep the full time dependence of these functions, in particular $\Gamma^{\lessgtr}(\vq,t)$ to study transient phenomena as a probe of the properties of the spectral density of the environment.

Since one of our objectives is to study the time evolution of the expectation value of the scalar field $\phi(\vx,t)$, we consider an initial pure state   density matrix for the system $\rho_S(0)$,
\be \rho_S(0) = \ket{\alpha}\bra{\alpha} \,,\label{rhosalfa}\ee where $\ket{\alpha}$ is a coherent state of the free scalar field, namely an eigenstate of the annihilation operator $a(\vk)$ in the free field expansion (\ref{phiip}),
\be a({\vk}) \ket{{\alpha}} = \alpha({\vk}) \ket{\bf{\alpha}} \,,  \label{eigenb}\ee with
 \be \ket{{\alpha}} = \Pi_{\vk}\, e^{- \frac{1}{2}|\alpha({\vk})|^2}\,e^{-\alpha({\vk}) a^\dagger({\vk})}\,\ket{0}\,, \label{cs} \ee yielding the following expectation values at the initial time
 \bea \mathrm{Tr} \,a(\vk) \rho_S(0)  & = &  \alpha(\vk)~~;~~  \mathrm{Tr}\, a^\dagger(\vk) \rho_S(0) = \alpha^*(\vk) \label{ai}\\
  \mathrm{Tr}\, a(\vk)\,a(-\vk) \rho_S(0)  & = &  \alpha(\vk)\,\alpha(-\vk)\,,\label{a2i}\\ \mathrm{Tr}\, a^\dagger(\vk) \, a^\dagger(-\vk)\rho_S(0) & = & \alpha^*(\vk) \, \alpha^*(-\vk) \,,\label{a2icc} \\ \mathrm{Tr}\, a^\dagger(\vk)\,a(\vk) \rho_S(0) & = &  N(0)= |\alpha(\vk)|^2 \,,\label{Nzero}\eea yielding

 \bea \mathrm{Tr} \phi_I(\vx,0)\rho_S(0) & = &   \sum_{\vk} \frac{1}{\sqrt{2V \omega_k}}\,\Big[\alpha(\vk)\, \,e^{i\vk\cdot \vx} + \alpha^*(\vk)\, \,e^{-i\vk\cdot \vx}  \Big]  \,,\label{fiexp}\\
 \mathrm{Tr} \pi_I(\vx,0)\rho_S(0) & = & -i \sum_{\vk} \frac{\omega_k}{\sqrt{2V \omega_k}}\,\Big[\alpha(\vk)\, \,e^{i\vk\cdot \vx}-\alpha^*(\vk)\, \,e^{-i\vk\cdot \vx}  \Big]   \,,\label{piexp} \eea for each value of $\vk$ the $\alpha(\vk)$ are determined by the initial conditions on the expectation values of the field and its canonical momentum.

 In interaction picture, the operators $a(\vk),a^\dagger(\vk)$ are time independent, therefore, using the relation (\ref{timeder}) along with the canonical commutation relation of these operators we find the following evolution equations

\bea \frac{d}{dt} \langle a({\vq})\rangle(t)  & = &  \Big[-i\,\Delta(q,t)-\frac{1}{2}\,\Gamma(q,t) \Big]\, \langle a({\vq}) \rangle(t) \nonumber \\\frac{d}{dt}\langle a^\dagger({\vq})\rangle(t)   & = &  \Big[i\,\Delta(q,t)-\frac{1}{2}\,\Gamma(q,t) \Big]\, \langle a^\dagger({-\vq})\rangle(t) \,, \label{aveaad}\eea
\bea \frac{d}{dt} \langle a(\vq)~ a(-\vq) \rangle (t) & = &  \Big[-2i\,\Delta(q,t)  - \Gamma(q,t)\Big] \langle a(\vq)~ a(-\vq) \rangle (t) \nonumber \\
 \frac{d}{dt} \langle a^\dagger(\vq)~ a^\dagger(-\vq) \rangle(t)  & = &  \Big[2i\,\Delta(q,t)  - \Gamma(q,t)\Big] \langle a^\dagger(\vq)~ a^\dagger(-\vq) \rangle(t)\,,  \label{bilins}\eea

  where $\langle (\cdots)\rangle \equiv \mathrm{Tr}(\cdots)\,{\rho}^r_{S}(t)$,  and
\be \Gamma(q,t) = \Gamma^>(q,t)-\Gamma^<(q,t)= \frac{g^2}{\omega_q} \,\int   \,\rho(q_0,q)\,\frac{ \sin[(\omega_q-q_0)t] }{(\omega_q-q_0)}\, \frac{dq_0}{2\pi} \,   \label{gamadif}\ee
Note that $\Delta(q,t)~;~\Gamma(q,t) $ are the same functions that enter in the multi-time scale analysis of the evolution of the expectation value in the ``in-in''-influence functional framework (\ref{deltat},\ref{gamat}).

 The solutions of these equations with the initial conditions (\ref{ai}-\ref{a2icc}) are given by
 \bea  \langle a({\vq})\rangle(t) & = &  \Big[e^{-i \int^t_0 \Delta(q,t')\,dt'}\,e^{-\frac{1}{2}\int^t_0 \Gamma(q,t')\,dt'}  \Big] \,\alpha(\vq)\,, \label{aoftsol}\\
 \langle a_{\vq}~ a_{-\vq} \rangle (t) & = &   \Big[e^{-2i \int^t_0 \Delta(q,t')\,dt'}\,e^{- \int^t_0 \Gamma(q,t')\,dt'}  \Big] \alpha(\vq)\alpha(-\vq) \,, \label{aaoftsol}
 \eea and their hermitian conjugates.

 The time derivative of the expectation value of the number operator
\be \dot{N}(q,t) =  {\text{Tr}}\, \dot{\rho}^r_{S}(t)\, a^\dagger({\vq}) \, a({\vq})\,\label{numop}\ee is obtained again by using the canonical commutation relations of the creation and annihilation operators, it is found to obey the quantum kinetic equation

\be \frac{dN(q,t)}{dt} =   {\text{Tr}}\Big\{ a^\dagger({\vq}) \, a({\vq}) ~\dot{{\rho}}^r_{S}(t)\Big\} = -\Gamma(q,t) N(q,t) + \Gamma^<(q,t) \,, \label{qke}\ee which  can be written  as
 \be \frac{dN(q,t)}{dt} =  \Gamma^<(q,t)\,\Big(1+N(q,t)\Big) - \Gamma^>(q,t)\, N(q,t)  \,.\label{qkebe}  \ee
 In this form the quantum kinetic equation   has a simple   interpretation as the Boltzmann equation for the distribution function $N(q,t)$ of the form $dN/dt = (\mathrm{gain})- (\mathrm{loss})$  with a loss  term $\Gamma^>(q,t)\, N(q,t)$ and a gain term $\Gamma^<(q,t)\,\Big(1+N(q,t)\Big)$.

 The time dependent distribution function    solution of the quantum kinetic equation (\ref{qke})   is given by
   \be  N(q,t) = e^{-\int^t_0 \Gamma(q,t')\,dt'}\,\Bigg[|\alpha(\vq)|^2 + \int^t_0 \Gamma^<(q,t')\,e^{\int^{t'}_0 \Gamma(q,t'')\,dt''}\,dt' \Bigg]\,,\label{disfun} \ee whereas the first term inside the bracket depends on the initial condition, the second does not and only depends on the forward $\Gamma^<$ and backward $\Gamma^>$ rates.

   Armed with these results we now obtain the expectation values of the spatial Fourier transforms
   \be \langle \phi(\vq,t) \rangle = \mathrm{Tr}\phi_I(\vq,t)\rho^r_S(t) = \frac{1}{\sqrt{2\omega_q}}\,e^{-\frac{1}{2}\int^t_0 \Gamma(q,t')\,dt'}\, \Big[e^{-i\omega_q\,t}\,e^{-i \int^t_0 \Delta(q,t')\,dt'}  \,\alpha(\vq) + e^{i\omega_q\,t}\,e^{i \int^t_0 \Delta(q,t')\,dt'}  \,\alpha^*(\vq)    \Big] \,,\label{onept}\ee
   note the similarity with the result (\ref{exfifin}) obtained from the ``in-in'' formulation with multitime analysis.
   The equal time two-point average is obtained following the same steps and using the results (\ref{aaoftsol},\ref{disfun}), it is given by

    \bea & & \langle \phi(\vq,t)\,\phi(-\vq,t) \rangle = \mathrm{Tr}\phi_I(\vq,t)\phi_I(-\vq,t)\rho^r_S(t)  = \nonumber \\ & &  \frac{1}{ {2\omega_k}}\, e^{-\int^t_0 \Gamma(q,t')\,dt'}\, \Big[e^{-2i\omega_kt}\,e^{-2i \int^t_0 \Delta(q,t')\,dt'}  \,\alpha(\vq)\alpha(-\vq) + e^{2i\omega_kt}\,e^{2i \int^t_0 \Delta(q,t')\,dt'}  \,\alpha^*(\vq) \alpha^*(-\vq) + |\alpha(\vq)|^2  \Big]\nonumber\\ & + & \frac{1}{ {2\omega_k}}\,\Big[1 + 2 \,\int^t_0 \Gamma^<(q,t')\,e^{-\int^{t}_{t'} \Gamma(q,t'')\,dt''}\,dt' \Big]   \,.\label{twopoint}\eea  These results depend, crucially, on the functions $\Delta(q,t);\Gamma(q,t)$ whose time integrals highlight the \emph{history} of the dynamical evolution, namely non-Markovian dynamics.

    The first bracket in (\ref{twopoint}) depends on the initial conditions, whereas the second does not, therefore, comparing to the result (\ref{avesolPsisq}) from the ``in-in'' formulation, we identify the second bracket with the \emph{noise} contribution from the influence functional.

    \textbf{Markov approximation:} In the Markov approximation the functions $\Delta(q,t)~;~\Gamma(q,t)$ are evaluated in the $t\rightarrow \infty$ limit
    \be \Delta(q,t)~~{}_{\overrightarrow{t\rightarrow \infty}} ~~ \frac{g^2}{2\,\omega_q} \,\int   \,\mathcal{P}\Bigg[ \,\frac{\rho(q_0,q)}{(\omega_q-q_0)}\Bigg]\, \frac{dq_0}{2\pi} \label{Delasy}\ee
     \bea &&  \Gamma(q,t)~~{}_{\overrightarrow{t\rightarrow \infty}} ~~ \Gamma(q,\infty) = \frac{g^2}{2\omega_q} \, \rho(\omega_q,q)  \label{gamafgr}\\
&&  \Gamma^>(q,t)~~{}_{\overrightarrow{t\rightarrow \infty}} ~~ \Gamma^>(q,\infty)=\frac{g^2}{2\omega_q} \, \rho(\omega_q,q)\,\big[1+n(\omega_q)\big] \label{gamagrsy}\\
&&  \Gamma^<(q,t)~~{}_{\overrightarrow{t\rightarrow \infty}} ~~ \Gamma^<(q,\infty)=\frac{g^2}{2\omega_q} \, \rho(\omega_q,q)\, n(\omega_q)  \label{gamalesy}\,. \eea
which follow from the identities
\bea && \frac{1-\cos[(\omega_q-q_0)t]}{(\omega_q-q_0)}~~{}_{\overrightarrow{t\rightarrow \infty}} ~~\mathcal{P} \Bigg[\frac{1}{(\omega_q-q_0)} \Bigg] \nonumber \\
&&  \frac{ \sin[(\omega_q-q_0)t] }{(\omega_q-q_0)}~~{}_{\overrightarrow{t\rightarrow \infty}} ~~\pi\,\delta(\omega_q-q_0)\,.  \label{inftimi}\eea The result (\ref{gamafgr}) is the S-matrix (FGR) decay rate, it is worth emphasizing that the asymptotic long time limits (\ref{Delasy},\ref{gamafgr}) are precisely the energy renormalization and decay width obtained in the Breit-Wigner approximation in the ``in-in'' formulation (\ref{delomsol}),\ref{width}, \ref{sigmaR}), establishing a direct link between the (QME) and the ``in-in'' formulation, furthermore, in this Markov long-time limit
\be \frac{\Gamma^>(q,\infty)}{\Gamma^<(q,\infty)} = \frac{\big[1+n(\omega_q)\big]}{ n(\omega_q)}= e^{\beta \omega_q}\,,\label{detbal}\ee which is the \emph{detailed} balance condition. As a consequence of this long time limit, replacing $\Gamma(q,t) \rightarrow \Gamma(q,\infty)~;~\Delta(q,t) \rightarrow \Delta(q,\infty) $ in the time integrals, the distribution function (\ref{disfun}) becomes
\be N(q,t) = e^{-\Gamma(q,\infty)t}\,N(q,0) +  \frac{\Gamma^<(q,\infty)}{\Gamma(q,\infty)}\,\Big(1- e^{-\Gamma(q,\infty)t}\Big) = n(\omega_q)+ \Big(N(q,0)- n(\omega_q)\Big) \, e^{-\Gamma(q,\infty)t}\,,\label{noftfgr} \ee where we used the long time limits (Fermi's Golden rule) (\ref{gamafgr},\ref{gamalesy}). This is the statement that the system thermalizes with the environment. A similar analysis applies to  the equal time two-point average (\ref{twopoint}),  we find
\be \langle \phi(\vq,t)\,\phi(-\vq,t) \rangle = e^{-\Gamma(q,\infty) t}\,\Big(\mathrm{Initial~Conditions}\Big)+ \frac{1}{2\omega_k}\,\Big[1+ 2 n(\omega_k)\Big] \,\Big(1-e^{-\Gamma(q,\infty) t}\Big)\,, \label{twoptfgr}\ee where the last term is a consequence of the asymptotic limit
 \be N(q,\infty) = n(\omega_k)\,,\label{Nfty}\ee obtained after replacing $\Gamma(q,t)\rightarrow \Gamma(q,\infty)$ in the time integrals in (\ref{disfun}), implying thermalization. The   result (\ref{twoptfgr}) is in agreement with the ``in-in'' result in the Breit-Wigner approximation (\ref{asquareave}), confirming that the Breit-Wigner approximation is reliable only in the long time limit and is equivalent to the Markov approximation in the (QME).

 We highlight that, in agreement with the textbook Brownian motion example, while the expectation value of the  system scalar field decays as a consequence of \emph{dissipation} the number of $\phi$ particles and the equal time two point function approach a non-vanishing asymptotic limit, a consequence of the \emph{noise}. Therefore, the solutions above reflect both the dissipative and noise contributions from the environment. Just as with the simple example of a Brownian oscillator immersed in the environment, the systems averages and
 correlations manifest the interplay between the dissipation and noise in the dynamics of the degrees of freedom of the system, which in fact are probes of the environmental correlations.

The solutions (\ref{aoftsol},\ref{aaoftsol}), their complex conjugates and (\ref{disfun}) all involve the time integrals of $\Delta(q,t);\Gamma(q,t)$, obviously taking the infinite time limit of their integrands is not consistent. The time integrals yield
 \be \int^t_0 \Delta(q,t') dt'  =  t\,\frac{g^2}{2\,\omega_q} \int^\infty_{-\infty}   \frac{\rho(q_0,\vq)}{(\omega_q-q_0)} \,\Bigg[ 1-\frac{\sin(\omega_q-q_0)\,t}{(\omega_q-q_0)\,t} \Bigg]\,\frac{dq_0}{(2\pi)} ~~{}_{\overrightarrow{t\rightarrow \infty}} ~~ t\,\delta \omega_q  \label{realpartofE}  \ee where
  \be \delta \omega_q = \frac{g^2}{2\,\omega_q}\, \int^\infty_{-\infty} \mathcal{P}\Bigg[ \frac{\rho(q_0,q)}{(\omega_q-q_0)} \Bigg]\,\frac{dq_0}{2\pi}\,,\label{renfreq}\ee is a renormalization of the single particle energy  $\omega_q$ and  $\mathcal{P}$ stands for the principal part, and
  \bea  \int^t_0 \Gamma(q,t')\,dt'  & = &   \frac{g^2}{ \omega_q}\, \int_{-\infty}^{\infty} \frac{dq_0}{2\pi} \,  {\rho(q_0,q)}\,\Bigg[\frac{ 1-\cos\big[(q_0-\omega_q)t\big]}{( q_0-\omega_q)^2} \Bigg]\nonumber \\ & {}_{\overrightarrow{t\rightarrow \infty}}& \Gamma(q,\infty) \, t +  \frac{g^2}{\omega_q}\,  \int_{-\infty}^{\infty} \frac{dq_0}{2\pi} \,\mathcal{P}  \frac{\rho(q_0,q)}{( \omega_q-q_0)^2}\,, \label{gamasy}\eea where $\Gamma(q,\infty)$  is given by (\ref{gamafgr}).  This long time limit is understood as follows: at long time, the region of $q_0$ in the integral around $\omega_q$ of width $\Delta q_0 \simeq 2\pi/ t$ yields a contribution to the integral that grows $\propto \omega_q\,t$, this is familiar from Fermi's golden rule. The oscillating cosine terms integrated in the remaining domain of $q_0$  namely   $|q_0-\omega_q| > 2\pi/t$ average out by Riemann-Lebesgue, leaving solely the contributions that are not oscillatory (the ``1'' inside the bracket in equation (\ref{gamasy})) in the long time   limit after ``excising'' the region $|q_0-\omega_q| \lesssim 2\pi/t$, yielding the principal value prescription, hence equation (\ref{gamasy}) follows from \emph{assuming} that $\rho(\omega_q,q) \neq 0$ and finite, otherwise the limit must be studied more carefully by analyzing the behavior of $\rho(q_0,q)$ as $q_0 \rightarrow \omega_q$. In fact, in the next sections we study important cases where this assumption \emph{fails} and the full time evolution of the decoherence rate reveals unexpected behavior that is \emph{not} captured by Fermi's golden rule.

      The results (\ref{realpartofE},\ref{gamasy}) imply that in the long time limit  the solution  (\ref{aoftsol}) and its complex conjugate become
  \be \langle a_{\vq} \rangle(t) ~ {}_{\overrightarrow{t\rightarrow \infty}}~ {Z}(q)\,  e^{-i\delta \omega_q\, t}\,e^{-\frac{1}{2}\,\Gamma(q,\infty)t } \,  \alpha({\vq})  ~~;~~ \langle a^\dagger({\vq}) \rangle(t)  ~ {}_{\overrightarrow{t\rightarrow \infty}}~ {Z}(q)\,  e^{ i\delta \omega_q\, t}\,e^{-\frac{1}{2}\,\Gamma(q,\infty)t } \,\alpha^*(\vq)\,,\label{solubs}  \ee yielding the long time behavior of the expectation value (\ref{onept})
   \be \langle \phi(\vq,t) \rangle  ~ {}_{\overrightarrow{t\rightarrow \infty}}~  \frac{Z(q)}{\sqrt{2\omega_q}}\,e^{-\frac{1}{2}\,\Gamma(q,\infty)t}\, \Big[e^{-i\omega_q t}\,e^{-i\delta \omega_q \, t}  \,\alpha(\vq) + e^{i\omega_q t}\,e^{i \delta \omega_q\, t}  \,\alpha^*(\vq)    \Big] \,,\label{asyexv}\ee
   where    to leading order in $g^2$ ,
  \be Z(q) = 1- \frac{g^2}{ 2\,\omega_q}\,  \int_{-\infty}^{\infty} \frac{dq_0}{2\pi} \,\mathcal{P} \Bigg[ \frac{\rho(q_0,q)}{( \omega_q-q_0)^2}\Bigg] \label{wfun} \ee is recognized with the wave function renormalization (\ref{wavefuncren}) in complete agreement with the ``in-in'' result (\ref{avePsi}).

\vspace{1mm}

\textbf{Decoherence:}
The initial coherent state density matrix for the system (\ref{rhosalfa}) is off-diagonal in the occupation number basis, namely the basis of energy eigenstates of the system, we refer to the off-diagonal density matrix elements in this basis as \emph{coherences} and the diagonal ones as \emph{populations}.

The solutions (\ref{solubs},\ref{asyexv},\ref{aaoftsol},\ref{twoptfgr}) indicate that the reduced density matrix is undergoing \emph{decoherence} and is becoming diagonal in this basis, the diagonal matrix elements, namely the population approaches a thermal fixed point if the bath is at non-zero temperature, while the off-diagonal elements, the coherences  decay. As these solutions show, decoherence is driven by the time integral of $\Gamma(q,t)$, therefore, we \emph{define} the \emph{decoherence function}
\be \Phi(q,t) = \int^t_0 \Gamma(q,t')\,dt' =  \frac{g^2}{ \omega_q}\, \int_{-\infty}^{\infty}     {\rho(q_0,q)}\,\Bigg[\frac{ 1-\cos\big[(q_0-\omega_q)t\big]}{( q_0-\omega_q)^2} \Bigg]\,\frac{dq_0}{2\pi} \,, \label{decofun}\ee This function  will be the focus of study for several different  environments below, with typical dependence $e^{-\Phi(q,t)}$ of the physical observables such as the amplitude of the coherent field oscillations and the dynamics of thermalization and population build-up. We also refer to $\Gamma(q,t) = \dot{\Phi}(q,t)$ as the \emph{decoherence rate} since in several cases studied below it cannot be directly associated with a decay rate, yet it determines the rate at which the coherences decay.

Decoherence, defined with respect to the energy basis of the system, indicates that this basis is privileged, if the initial state is a pure state diagonal in this basis with a definite populations for each $\vk$, in other words of the form $\ket{n_{\vk}}\bra{n_{\vk}}$, the expectation values $\langle a_{\vk} \rangle, \langle a_{\vk} a_{-\vk}\rangle$ etc. will vanish at all times.    Therefore, we infer from this analysis that the basis of energy eigenstates of the system is a \emph{pointer basis}\cite{zurek}.

\vspace{1mm}

\textbf{Particle production:} The time dependent particle number (\ref{disfun}) features two contributions, the first term is determined by the initial conditions and vanishes if the initial state is the vacuum of the system, for which $\alpha(\vq)=0 \, \forall \vq$. The second term describes the \emph{production} of particles of the system as a consequence of the coupling to the environment, in particular, this term is a manifestation of the environmental \emph{noise} correlation function. As discussed above, approximating the rates by their  S-matrix  expressions  (\ref{gamafgr}-\ref{gamafgr}) and considering the environment to be in thermal equilibrium, this second term describes the thermalization of the system.

The distribution function of \emph{produced particles} as a function of time is given by
\be N_P(q,t) =   \int^t_0 \Gamma^<(q,t')\,e^{-\big[\Phi(q,t)-\Phi(q,t')\big]}\,dt'   \,,\label{parprod}\ee which again shows   non-Markovian dynamics and the importance of the decoherence function $\Phi(q,t)$.

 The main questions that we study for different \emph{slow} environments are the following: \textbf{i:)} how long does it take for the Fermi's golden rule asymptotic results (\ref{gamafgr})  to set in for the rates and (\ref{gamasy}) for the decoherence function, \textbf{ii:)} if the spectral density $\rho(q_0,q)$ \emph{vanishes} on the mass shell of the system's particle, namely $q_0=\omega_q$, in which case Fermi's golden rule yields a vanishing rate, how is dissipation manifest?, namely what is the behavior of the decoherence function $\Phi(q,t)$ as a function of time?, \textbf{iii:)} if the environment is in its ground state, rather than in thermal equilibrium, in which case $n(\omega_q)=0$, what is the   time evolution  of the distribution function of produced particles (\ref{parprod})?.

\vspace{1mm}

\section{Anomalous environments}\label{sec:anomalous}

\subsection{Unparticles}\label{subsec:unparticles}
Unparticles refers to an effective field theory\cite{unparticle} that describes a conformal sector as an  extension beyond the Standard Model that includes fields associated with a non-trivial, strongly coupled  infrared (IR) fixed point of the renormalization group\cite{banksZaks} that interact with standard model degrees of freedom via a very massive mediator. The (IR) fixed point describes a Lorentz invariant critical theory, in terms of interpolating fields that feature anomalous scaling dimensions $\du$, characterized by the scaling form of two-point correlation (Wightman) function (for scalar unparticle operators)\cite{unparticle}
\be \langle 0 |\mathcal{O}(x)\mathcal{O}(0)     |0\rangle \propto \frac{1}{(x^2)^{\du}}\,. \label{unop}\ee

Environmental unparticle fields have been recently studied in ref.\cite{bare} analyzing the temporal correlations of the degrees of freedom of such environment, in particular the dissipation and noise kernels at zero spatial momentum $\vq$. As stated above, our objective is to study how these environmental correlations manifest as dissipative aspects  and environment-induced decoherence  of the reduced density matrix of  a scalar field that acts as a probe of this ``anomalous environment''.

Following the original proposal\cite{unparticle} we consider a Lorentz invariant scalar unparticle field, described by an interpolating operator $\mathcal{O}(x)$ that couples locally to the system field $\phi(x)$  as in (\ref{lagra}), with $S[\phi]=\phi(x)$ (see eqn. \ref{free}). We consider the unparticle degrees of freedom to be in their ground state since in this case there is no other scale such as temperature,  and correlation functions of these fields are Lorentz invariant and  transform covariantly under scale transformations with anomalous scaling dimensions. The correlation function (\ref{unop}) and the results of reference\cite{unparticle} translate directly onto the Lehmann representation of the correlation functions (\ref{siggreat}), yielding the spectral density
\be \rho(q_0,q) =
\mathcal{A}_{\du}\,\mathrm{sign}(q_0)\,\Theta(q^2_0-q^2)\,\Big[q^2_0-q^2\Big]^{\du-2}\,\label{rhounp} \ee with\cite{unparticle}
\be \mathcal{A}_{\du} = \frac{16\pi^{5/2}}{(2\pi)^{2 d_{\mathcal U}}}\,
\frac{\Gamma\!\left(d_{\mathcal U}+\tfrac12\right)}
{\Gamma(d_{\mathcal U}-1)\,\Gamma(2 d_{\mathcal U})}\,\label{Adu} \ee the single particle pole corresponds to the limit $\du \rightarrow 1^+$\cite{unparticle}. For $\du-1 >0$ the unparticle describes not a single particle (or quasiparticle) pole but a \emph{continuum} so that the scalar probe field $\phi(x)$ is actually coupled to a continuum,   this is the case of relevance in our study.

The connection with \emph{static} critical phenomena\cite{ma,gold} becomes clear in Euclidean space-time, with the momentum space propagator at the critical point
 $G(p) \propto  p^{-2+\eta}$\cite{ma,gold}  and upon analytic continuation to Minkowski space-time with $\rho \propto \mathrm{Im}[G(-p^2-i\epsilon)]$ one can identify the anomalous critical exponent $\eta =  {2}(\du-1)$, in agreement with the free field limit $\du \rightarrow 1^+$ of the spectral density. This similarity with static critical phenomena intuitively suggests that if the anomalous scaling dimension $\du-1$  is large the unparticle field is associated with a strongly coupled infrared   ``Wilson-Fisher'' (rather Banks-Zaks)\cite{banksZaks} fixed point.

 Since the spectral density is an odd  function of $q_0$, we write the decoherence rate (\ref{gamadif}), the gain (forward) rate (\ref{gamles})  and  decoherence function (\ref{decofun}) as
 \be \Gamma(q,t) =  \frac{g^2\,\mathcal{A}_{\du}}{2\pi\,\omega_q}\,\int^{\infty}_{q}\,\Big[ {q^2_0-q^2}\Big]^{ (\du-2)}\Big[S_-(q_0,q,t)-S_+(q_0,q,t)  \Big] \,dq_0 \equiv \Gamma_-(q,t)-\Gamma_+(q,t) \,,\label{gamaun}\ee

 \be \Gamma^<(q,t) =  \frac{g^2\,\mathcal{A}_{\du}}{2\pi\,\omega_q}\,\int^{\infty}_{q}\,\Big[ {q^2_0-q^2}\Big]^{ (\du-2)}\Big[n(q_0)\,S_-(q_0,q,t)+ (1+n(q_0))\,S_+(q_0,q,t)  \Big] \,dq_0   \,,\label{gamagainun}\ee

 \be \Phi(q,t)= \frac{g^2\,\mathcal{A}_{\du}}{2\pi\,\omega_q}\,\int^{\infty}_{q}\,\Big[ {q^2_0-q^2}\Big]^{ (\du-2)}\Big[C_-(q_0,q,t)-C_+(q_0,q,t)  \Big] \,dq_0 \equiv \Phi_-(q,t)-\Phi_+(q,t)  \,,\label{fiun}\ee in obvious notation, where
 \be S_{\mp}(q_0,q,t) = \frac{ \sin\big[(q_0\mp\omega_q)t\big]}{( q_0\mp\omega_q)}~~;~~ C_{\mp}(q_0,q,t) = \frac{ 1-\cos\big[(q_0\mp\omega_q)t\big]}{( q_0\mp\omega_q)^2} ~~;~~ \omega_q = \sqrt{q^2+m^2_\phi} \,.\label{Cis}\ee

  As a consequence of the resonant denominators, at large time the functions $S_-(q_0,q,t);C_-(q_0,q,t)$ are  strongly peaked at $q_0 = \omega_q$ which is inside the integration domain, with $S_-(q_0,q,t) \propto t: C_-(q_0,q,t) \propto t^2$ as $q_0 \rightarrow \omega_q$, resulting in a secular growth in time of the decoherence function in agreement with Fermi's golden rule, whereas the functions $S_+,C_+$  do not feature resonant denominators within the domain of integration  these contributions remain always bound in time.

\vspace{1mm}

 \textbf{Non-relativistic:}\,$ {\vec{q}}=0$: The temporal correlations of the environmental noise (\ref{noiset}) for the case $\vq=0$ have been studied in ref.\cite{bare}. Instead here we study the effects on the dynamics of the probe field $\phi$, in particular the decoherence rate $\Gamma(0,t)$ and decoherence function $\Phi(0,t)$. This case is of special importance since it affords a closed analytic expression for $\Gamma(0,t)~;~\Phi(0,t)$ as a consequence of the scaling property of the spectral density. Let us consider the two contributions for these expressions separately, and define $q_0 t \equiv  u~;~ m_\phi t \equiv \tau~;~ G= g^2\mathcal{A}_{\du}/2\pi\, m_\phi $ yielding
\be \Gamma_\mp (0,t) = G\,t^{-\sigma} \,\int^\infty_0 u^{\sigma}\,\frac{\sin(u\mp\tau)}{(u\mp\tau)}\,du ~~;~~ \sigma\equiv 2(\du-2)\label{gampsca}\ee  we find
\bea \Gamma_-(0,t) &  = &   -G\,m^{\sigma}_\phi\,\Gamma[\sigma+1]\, \mathrm{Im}\Big\{ e^{i\pi\,\sigma}\, \gamma[-\sigma;i\tau]   \Big\}\,,\label{gamin}\\
  \Gamma_+(0,t)  & = &    G\,m^{\sigma}_\phi\,\Gamma[\sigma+1]\, \mathrm{Im}\Big\{ \gamma[-\sigma;i\tau]   \Big\}\,,\label{gamplu}\eea where $\gamma[a,b]$ is the  lower incomplete gamma function, yielding
\be \Gamma(0,t) =  -2\,G\,m^{\sigma}_\phi\,\Gamma[\sigma+1]\,\cos\Big[\frac{\pi}{2} \sigma \Big] \,\mathrm{Im}\Big\{ e^{i\frac{\pi}{2} \sigma}\, \gamma[-\sigma;i\tau]   \Big\}\,,\label{gamunzero}\ee where convergence of the integrals requires that $-1<\sigma<1$.

In the limit $\tau \rightarrow \infty$ we find
\be \Gamma(0,t)~~{}_{\overrightarrow{\tau \rightarrow \infty}}~~  \pi\,G\,m^{\sigma}_\phi \Big[1 - \frac{2}{\pi}\,\Gamma(\sigma+1) \cos\Big( \frac{\pi}{2}\,\sigma\Big)\, \frac{\cos(\tau)}{\tau^{\sigma+1}}+\cdots\Big]\,.\label{asygamun}  \ee

The first term on the right hand side of (\ref{asygamun}) is the Fermi's golden rule result (\ref{gamafgr}), the ``anomalous exponent'' $\du-1$ determines the approach to the asymptotic long time Fermi's golden rule limit.

Figure (\ref{fig:Ioftau}) displays $I(\sigma,\tau)= \Gamma(0,t)/ \pi\,G\,m^{\sigma}_\phi$ vs. $\tau$ for $\sigma = -0.5,0,0.5$, it clearly shows that for smaller $\sigma$ the approach to the asymptotic Fermi's golden rule result is slower.

 \begin{figure}[H]
  \centering
  \includegraphics[width=4in]{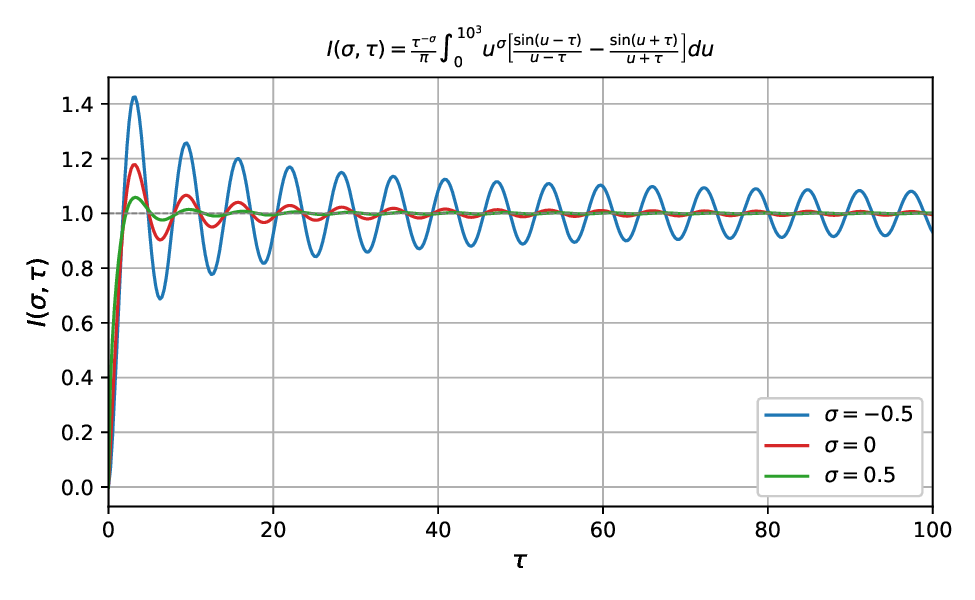}
  \caption{$I(\sigma,\tau)= \Gamma(0,t)/ \pi\,G\,m^{\sigma}_\phi$ vs. $\tau$ for $\sigma = -0.5,0,0.5$, corresponding to large ``anomalous dimensions'' $2(\du-1) = \frac{3}{2}, 2, \frac{5}{2}$ respectively.}
  \label{fig:Ioftau}
\end{figure}
With the same change of variables,
\be \Phi(0,t) =  G\,t^{1-\sigma} \,\int^\infty_0 u^{\sigma}\,\Bigg[ \frac{1-\cos(u-\tau)}{(u-\tau)^2} - \frac{1-\cos(u+\tau)}{(u+\tau)^2}  \Bigg]\,du \,,\label{finu}\ee with analytic expression
\be
\Phi(0,t)= G\,t^{1-\sigma}\Bigg[2\,\Gamma(\sigma+1)\cos\!\left(\frac{\pi\sigma}{2}\right)\left\{\tau^{\sigma-1}\,\mathrm{Re}\Big[(\sigma+i\tau)\,e^{i\pi\sigma/2}\,\gamma(-\sigma,i\tau)\Big]+
\frac{\cos\tau}{\tau}\right\}\Bigg] \ee its  asymptotic limit for large $\tau$ is given by
\be \Phi(0,t)~{}_{\overrightarrow{\tau \rightarrow \infty}}~  \pi\,G\,t\,m^\sigma_\phi\,\Big[ 1-  \cot\Big( \frac{\pi\,\sigma}{2}\Big) \,\frac{\sigma}{m_\phi t} +\cdots \Big]\,,\label{asyfizero}  \ee which is \emph{exactly} the result (\ref{gamasy}) as can be checked for the second time independent term in (\ref{gamasy}) from the integral representation of the standard beta function.\footnote{ $\int^\infty_0 \frac{u^\sigma}{u+x}\,du = - \frac{\pi x^\sigma}{\sin(\pi\sigma)}$.}

Figure (\ref{fig:Hoftau}) displays $H(\sigma,\tau)=\Phi(0,t)/ \pi\,G\,m^{\sigma}_\phi\,t$ vs. $\tau$ for $\sigma = -0.5,0,0.5$, showing the asymptotic limit to the usual secular term linear in time from Fermi's golden rule, clearly showing the transient behavior.

 \begin{figure}[H]
  \centering
  \includegraphics[width=4in]{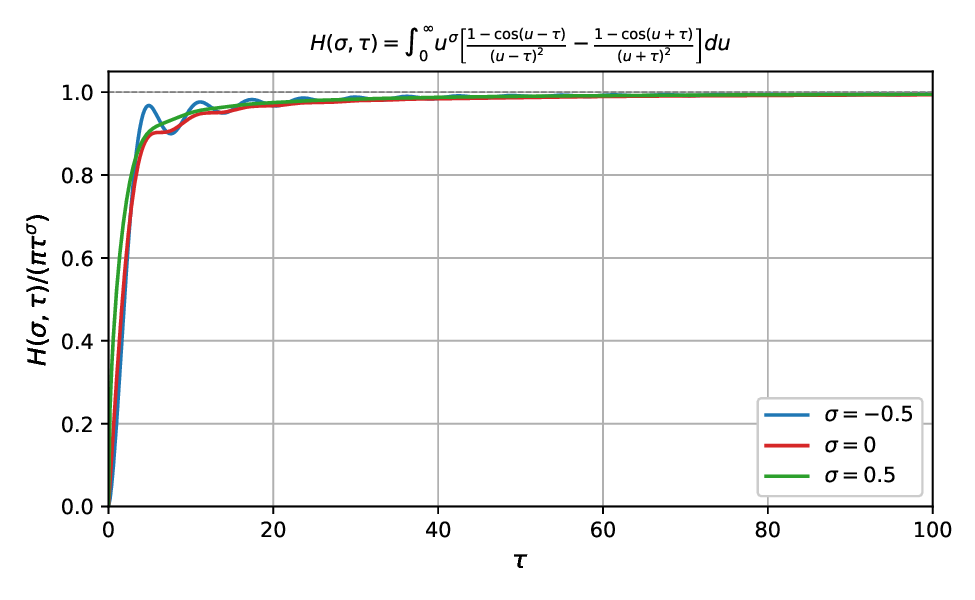}
  \caption{$H(\sigma,\tau)= \Phi(0,t)/ \pi\,G\,m^{\sigma}_\phi\,t$ vs. $\tau=m_\phi t$ for $\sigma = -0.5,0,0.5$, corresponding to large ``anomalous dimensions'' $2(\du-1) = \frac{3}{2}, 2, \frac{5}{2}$ respectively. The Fermi's Golden rule result sets in at $\tau \simeq 50$.}
  \label{fig:Hoftau}
\end{figure}

In this case it is clear from the exact solution that whereas the asymptotic long time limit is in agreement with the S-matrix  rate  and a linearly growing decoherence function as predicted by Fermi's golden rule, the transient behavior features an ``anomalous''  time dependence that reveals the anomalous exponent $\sigma= 2(\du-2)$.

\vspace{1mm}

\textbf{Ultrarelativistic limit:}

We now study the relativistic and ultrarelativistic limits. For these cases, it is convenient to do a different change of variables: define $q_0 = \omega_q x~;~ \tau = \omega_q\,t~;~ v\equiv q/\omega_q ~;~\alpha\equiv (\du-2)$ and keeping the same definition for $G=g^2\,\mathcal{A}_{\du}/2\pi\,m_\phi$, the rate (\ref{gamaun}) and coherence function (\ref{fiun}) are given by

\be \Gamma(q,t)= G\,m_\phi\,\omega^{2\alpha}_q\,\int^{\infty}_{v}\,\Big[ {x^2-v^2}\Big]^{ \alpha}\Bigg[\frac{\sin\big[(x-1)\tau\big]}{(x-1)}-\frac{\sin\big[(x+1)\tau\big]}{(x+1)}  \Bigg] \,dx  \,, \label{gamaunur}\ee

\be \Phi(q,t)= G\,m_\phi\,\omega^{(2\alpha-1)}_q\,\int^{\infty}_{v}\,\Big[ {x^2-v^2}\Big]^{ \alpha}\Bigg[\frac{1-\cos\big[(x-1)\tau\big]}{(x-1)^2}-\frac{1-\cos\big[(x+1)\tau\big]}{(x+1)^2}  \Bigg] \,dx  \,, \label{fiunur}\ee where $\gamma(q)= m/\omega_q$ is the Lorentz factor. We note that as $ v\rightarrow 1^-$ the resonant denominators at $x=1$ approach the threshold, namely the lower limit in the integrals and the spectral density evaluated
on the ``mass shell'' at $x=1$, either vanishes for $\alpha >0$ or diverges for $\alpha <0$ in this limit, therefore the usual Fermi's golden rule result would naively vanish or diverge, respectively in this limit. The resonant term with $\sin[(x-1)\tau]/(x-1)$ dominates the long time behavior of the rate, whereas the other, non-resonant term averages out. The resonant term peaks at $x=1$ with height $\tau$ and width $\simeq 2\pi/\tau$, therefore, in the ultrarelativistic limit as $v\rightarrow 1^-$ the decoherence rate will be sensitive to the behavior of the spectral density at threshold up to a very long time $\tau \propto 2\pi/(1-v)$.

Figure (\ref{fig:Jvoftau}) shows $J(\alpha,\tau,v)= \Gamma(q,t)/(G\,\gamma(q)\,\omega^\alpha_q) $ vs. $\tau$ for $\alpha = -0.5,0,0.5~,~ v= 0.95,0.98,1$

 \begin{figure}[H]
  \centering
  \includegraphics[width=3in]{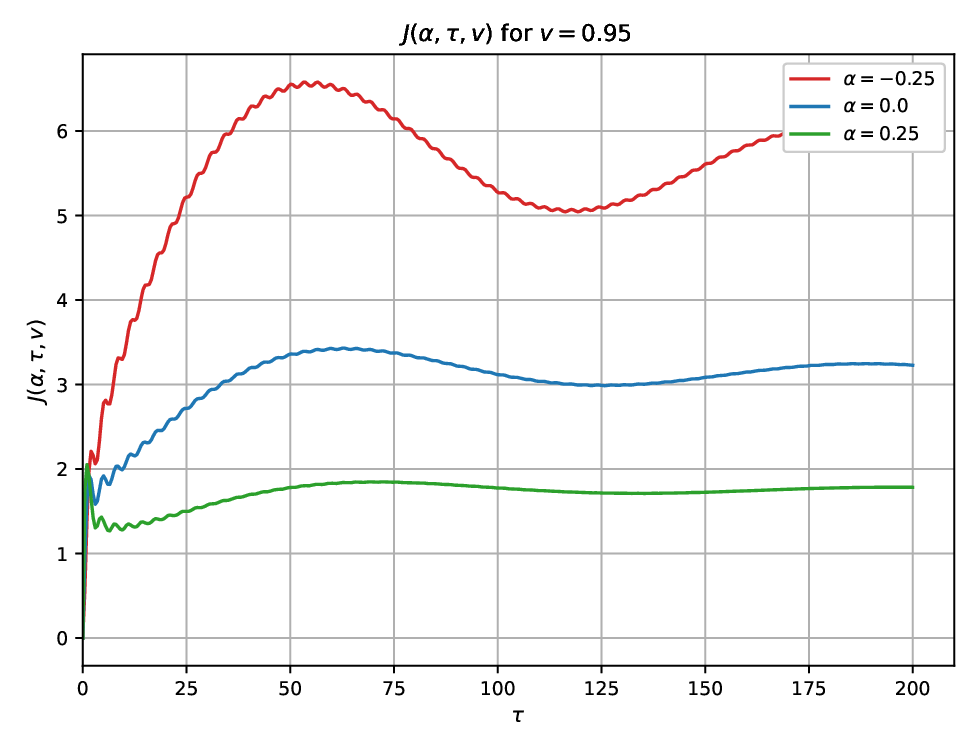}
   \includegraphics[width=3in]{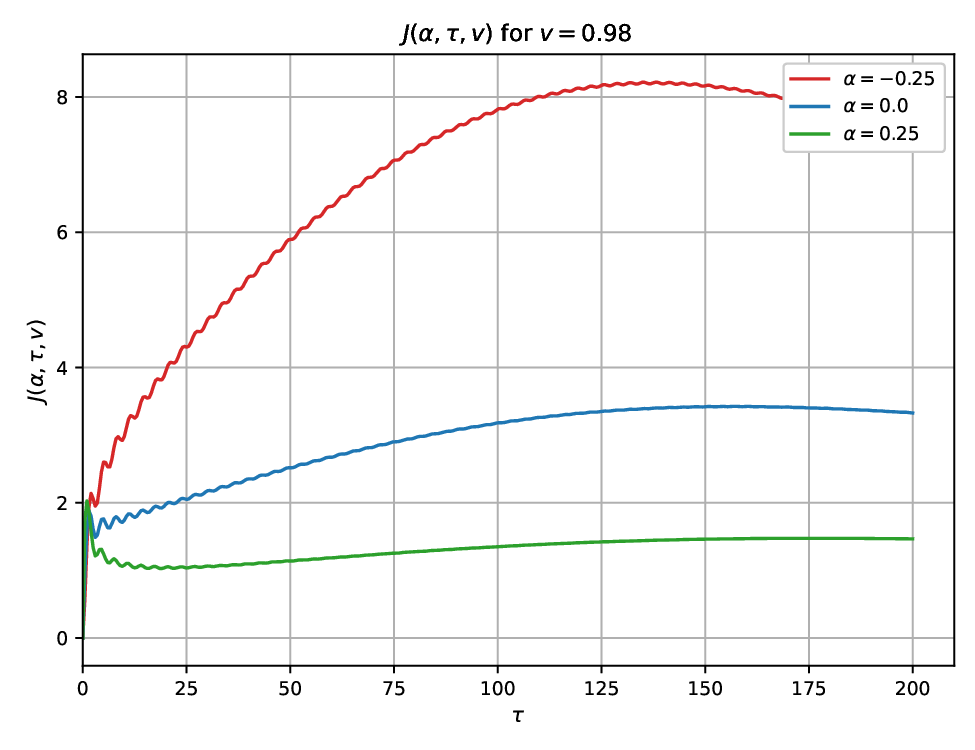}
   \includegraphics[width=3in]{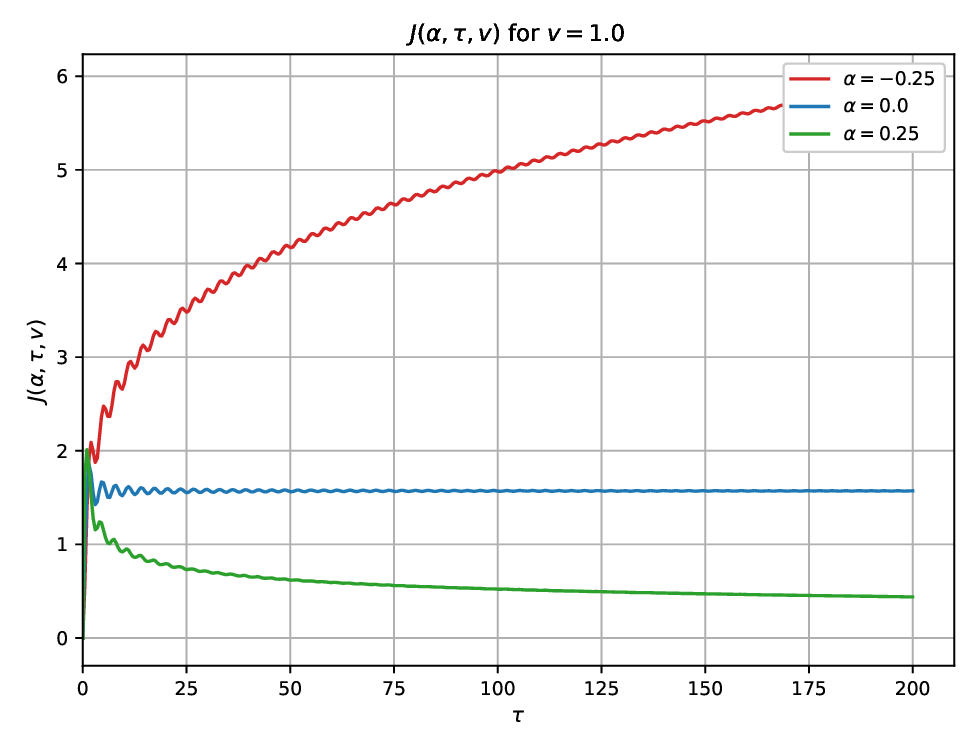}
  \caption{$J(\alpha,\tau,v)= \Gamma(q,t)/(G\,\gamma(q)\,\omega^\alpha_q) $ vs. $\tau=\omega_q t$ for $\alpha = -0.5,0,0.5~,~ v= 0.95,0.98,1$ }
  \label{fig:Jvoftau}
\end{figure}

In particular, the case $v=1$ provides a simple yet illuminating example where an analytic expression for the decoherence rate is available. Let us focus on the resonant contribution as the non-resonant part does not feature secular growth, is always oscillatory and vanishes at long time by the Riemann-Lebesgue lemma. Define $s = (x-1)\tau$, in terms of which the resonant term at long time becomes\footnote{The result for the integral is simply the Mellin transform of the sine function, valid for $-1< \alpha <1$.}
\be J_{res}(\alpha,\tau, 1)~{}_{\overrightarrow{\tau \rightarrow \infty}}~ \simeq 2^\alpha\, \tau^{-\alpha}\,   \underbrace{\int^\infty_0 s^{\alpha-1}\,\sin(s)\,ds}_{\substack = \,\Gamma(\alpha)\, \sin \big( \frac{\pi\,\alpha}{2}\big)}    \,,\label{asyJv1}\ee this result agrees with the numerical one for $v=1$ to percent accuracy. The figures for   $v<1$ reflect the fact that at early and intermediate time, the behavior at threshold is important featuring characteristics similar to the case $v=1 $, however, at very long time $\tau \gg 1/(1-v)$ the Fermi's golden rule limit
is attained $J_{FGR}(\alpha,\infty,v<1) = \pi (1-v^2)^{\alpha}$.

  \begin{figure}[H]
  \centering
  \includegraphics[width=3in]{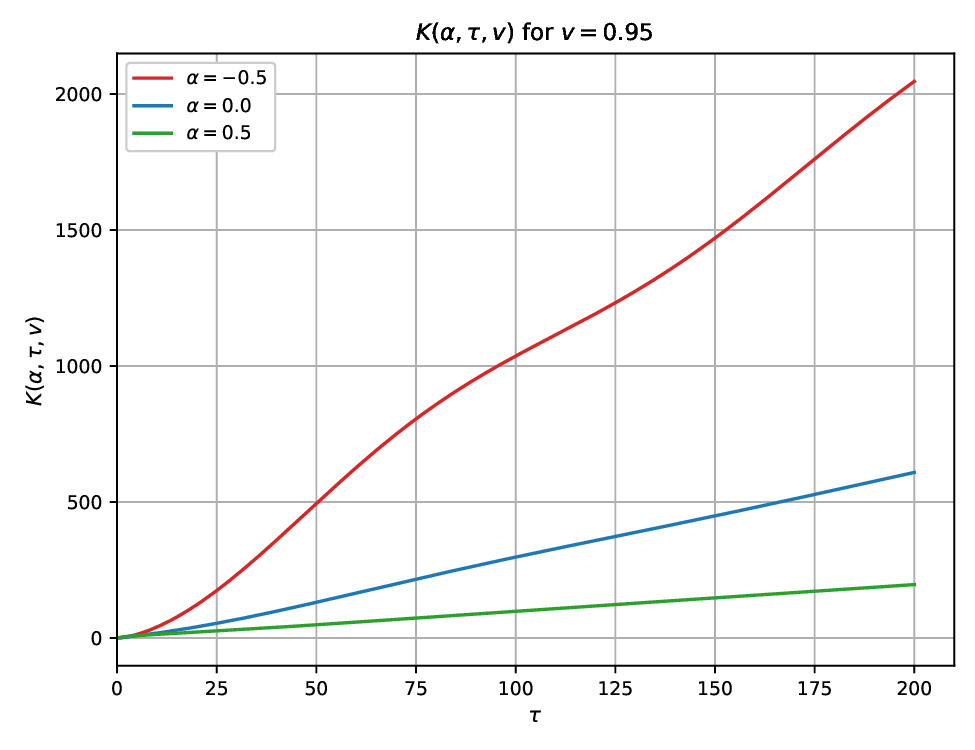}
   \includegraphics[width=3in]{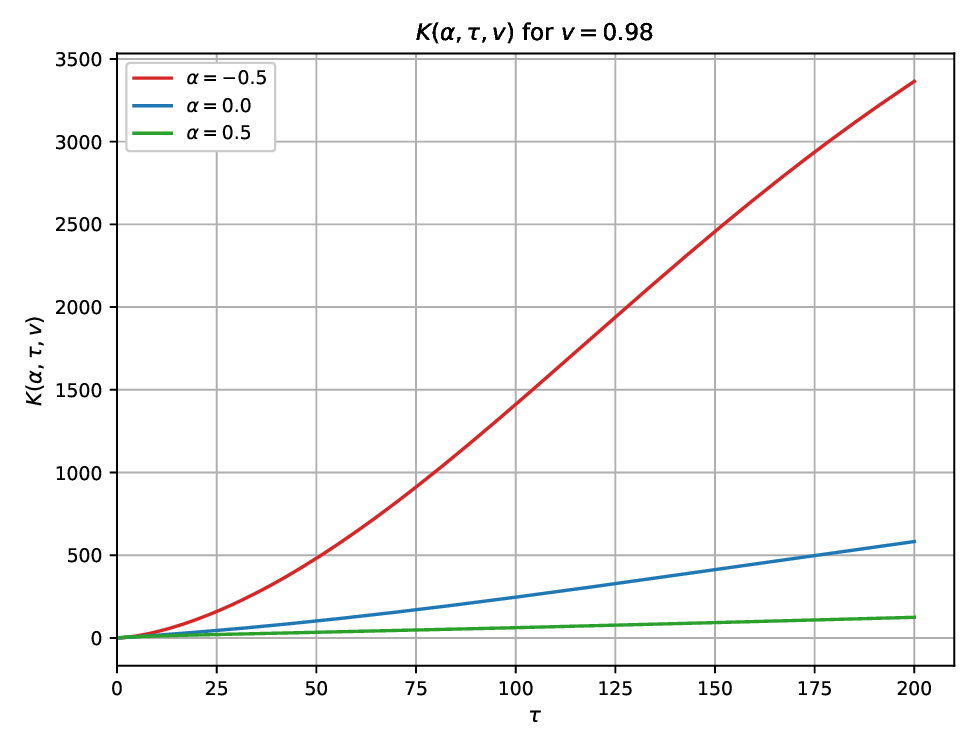}
   \includegraphics[width=3in]{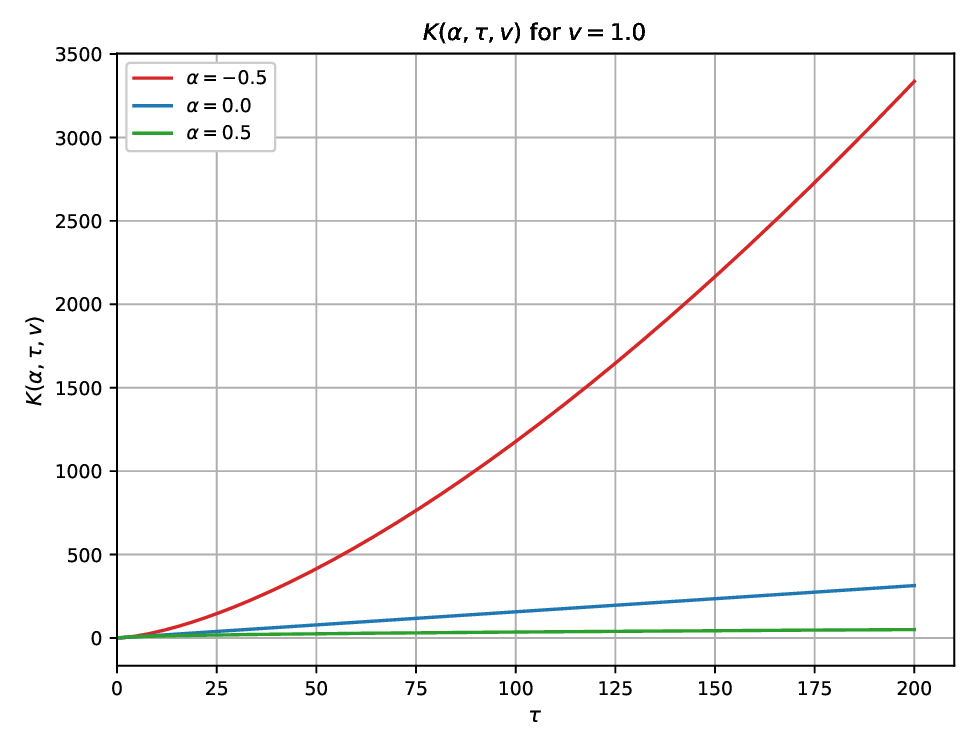}
  \caption{$K(\alpha,\tau,v)= \Phi(q,t)/(G\,\gamma(q)\,\omega^{\alpha-1}_q) $ vs. $\tau$ for $\alpha = -0.5,0,0.5~,~ v= 0.95,0.98,1$ }
  \label{fig:Kvoftau}
\end{figure}

Figures (\ref{fig:Kvoftau}) show $ K(\alpha,\tau,v)= \Phi(q,t)/(G\,\gamma(q)\,\omega^{\alpha-1}_q) $ vs. $\tau=\omega_q t$ for $\alpha = -0.5,0,0.5~,~ v= 0.95,0.98,1$.
An analysis similar to that for the resonant contribution to the function $J(\alpha,\tau,v=1)$, yields the asymptotic long time behavior for the resonant contribution to $K(\alpha,\tau, v=1)$
\be K_{res}(\alpha,\tau,1)~{}_{\overrightarrow{\tau \rightarrow \infty}}~ \simeq  2^\alpha\, \frac{\Gamma(\alpha)}{ (1-\alpha)}\, \sin \big( \frac{\pi\,\alpha}{2}\big) \,\tau^{1-\alpha} ~~;~~ \alpha < 1\,, \label{kres}\ee which obviously coincides with the time integral of the rate (\ref{asyJv1}). Note that as $\alpha \rightarrow 0$ the rate $\rightarrow \pi/2$ rather than $\pi$, this is a consequence of the threshold coinciding with the resonance, yielding ``$1/2$'' the delta function in the long time limit. Using the standard integral representation of the Beta function, we obtain the non-resonant contribution in the asymptotic long time limit where the oscillatory terms $\cos[(x+1)\tau]$ average out by the Riemann-Lebesgue lemma yielding

\be K_{nonres}(\alpha,\infty,1) = \int^\infty_1 \frac{(x^2-1)^\alpha}{(x+1)^2} dx = 2^{2\alpha-1}\,\frac{\Gamma(1+\alpha)\,\Gamma({1}-2\alpha)}{\Gamma(2-\alpha)}~~;~~ -1< \alpha < \frac{1}{2}\,. \label{nonresK}\ee

The case $\alpha=1$ requires a separate treatment because the long time  and $\alpha \rightarrow 1$ limits do not commute as evidenced by the result (\ref{kres}), furthermore,  the integral for $K(1,\tau,1)$ is logarithmically divergent. We introduce a (dimensionless) upper frequency cutoff $\tilde{\Lambda} \gg 1$ in the integral in (\ref{fiunur}) with the result
\be K(1,\tau,1) =  2\ln\Big(\frac{\tilde{\Lambda}^2}{2}\Big) \;+\; 2\gamma_E + 2\ln\!\left( {\tau}\right) + 2\,\mathrm{Ci}(2\tau) - \frac{\sin(2\tau)}{\tau} \;+\; \frac{2\sin\tau}{\tau}\cos(\tilde{\Lambda}\tau) \;+\; O\!\left(\tilde{\Lambda}^{-1}\right)\,,\label{Kalpha1} \ee where $\gamma_E$ is the Euler-Mascheroni constant and $Ci$ is the cosine integral function.

Since we are focusing on massive scalar probes, obviously, setting $v=1$  is unphysical, however, the analytic solution of the long time dynamics provides a useful guide to interpret the numerical results displayed in the figures (\ref{fig:Jvoftau},\ref{fig:Kvoftau}).

The physical reason for the ``anomalous'' time dependence in these cases is that in the ultrarelativistic limit the mass shell of the probe approaches the light-cone from below and because the critical bath features Lorentz invariance with massless excitations, the spectral density vanishes continuously on the light cone for $\du > 2$, thus the vanishing of the spectral density conspires with the ``pole'' of the resonant term yielding the long time transient.

Above and beyond the technical aspects, the main lessons from this study are that there are important non-Markovian effects that lead to transient dynamics over long time scales, and that in these cases it takes a long time for the Fermi's golden rule asymptotic limit of the rate to set in.   In the explicit cases above, taking the long time in the rate $\Gamma(q,t)$ yields, misleadingly,  either a vanishing rate and coherence function for $\alpha >1$ or a divergent rate for $\alpha<1$, obviously incorrect.

Therefore, taking the infinite time limit  and invoking  uncritically S-matrix (Fermi's golden rule) rates, thereby neglecting  non Markovian processes in environments with ``slow'' degrees of freedom, glosses over  very important   long transient dynamics, with potentially relevant consequences in an expanding cosmology where the expansion of the Universe competes with the dynamical time evolution.

\textbf{Particle production:} We now consider the production of scalar probe particles as a consequence of the ``unparticle vacuum noise'', focusing on the distribution function of produced particles (\ref{parprod}) which using that the spectral density is an odd function of $q_0$, becomes

\be N_P(q,t)= \frac{g^2\,e^{-\Phi(q,t)}}{2\pi\, \omega_q}  \,\int^t_0 \Bigg\{ \,\int^{\infty}_0    \, \rho(q_0,q)\, \Bigg[ n(q_0) \frac{ \sin[(\omega_q-q_0)t'] }{(\omega_q-q_0)} + (1+n(q_0)) \frac{ \sin[(\omega_q+q_0)t'] }{(\omega_q+q_0)} \Bigg]\,  {dq_0}\Bigg\}\, e^{\Phi(q,t') }\,dt' \,\label{parprod2}  \ee

  Consistent with the main assumptions\footnote{Temperature introduces a scale, therefore the arguments in\cite{unparticle} based on conformal invariance may not hold for a thermal bath.}   in ref.\cite{unparticle} we consider   the unparticle sector to be in its ground state, which implies $n(q_0) =0$ in (\ref{parprod2}), therefore only the ``1'' in the second term inside the bracket contributes. The resulting distribution function $N_P(t)$ is in general non-vanishing, therefore describing particle production. As argued in the previous section particle production is a distinct consequence of the ``noise'' contribution from the environment, therefore this production mechanism is a consequence of the unparticle vacuum noise.

  The time evolution of the distribution function (\ref{parprod2}) must be studied numerically as no closed analytic expression is available, however, before we do so, we can gain intuition into the time evolution by analyzing the quantum kinetic equation  (\ref{qke}) with initial condition $N(q,t=0) =0$ with $\Gamma(q,t)$ and $\Gamma^<(q,t)$ given by equations (\ref{gamaunur}) and (\ref{gamagainun}) for $n(q_0)=0$  respectively.
   The asymptotic long time fixed point of the kinetic equation with this initial condition $\dot{N}(q,\infty) =0$ is the thermal fixed point

\be N(q,\infty) = \frac{\Gamma^{<}(q,\infty)}{\Gamma(q,\infty)} = n(\omega_q)~{}_{\overrightarrow{T\rightarrow 0}}~0 \,,\label{asyfp}\ee where we used the asymptotic results (\ref{gamafgr},\ref{gamalesy}). Therefore, taking the infinite time (Markov) limit \emph{before} solving the evolution equation, and setting the rate $\Gamma^< =0$ yields a vanishing population at all time for the initial condition $N(q,0)=0$.

 However, keeping the time dependence of the rate in the evolution equation, for $N(q,0)=0$ at early time the population grows $\propto t$ with a slope proportional to the  integral $\int^{\infty}_0 \rho(q_0,q) dq_0$, therefore it must reach a maximum and decrease towards zero asymptotically. Figure (\ref{fig:Noftau}) is a representative of the numerical solution for  generic $\alpha,v$ clearly confirming the expected behavior. In fact for a large range of parameters we find qualitatively the same characteristic behavior for zero temperature, namely the population rises rapidly, reaches a maximum  and becomes negligible on a time scale $\omega_q t \gtrsim 30-40$, but is clearly non-vanishing within a time window, whereas taking the Markov limit from the beginning would yield a vanishing population at all time. This transient behaviour is a distinct consequence of non-Markovian dynamics solely determined by the ``vacuum noise'' of the environment, and may have potentially interesting implications in cosmology.

    \begin{figure}[H]
  \centering
  \includegraphics[width=3in]{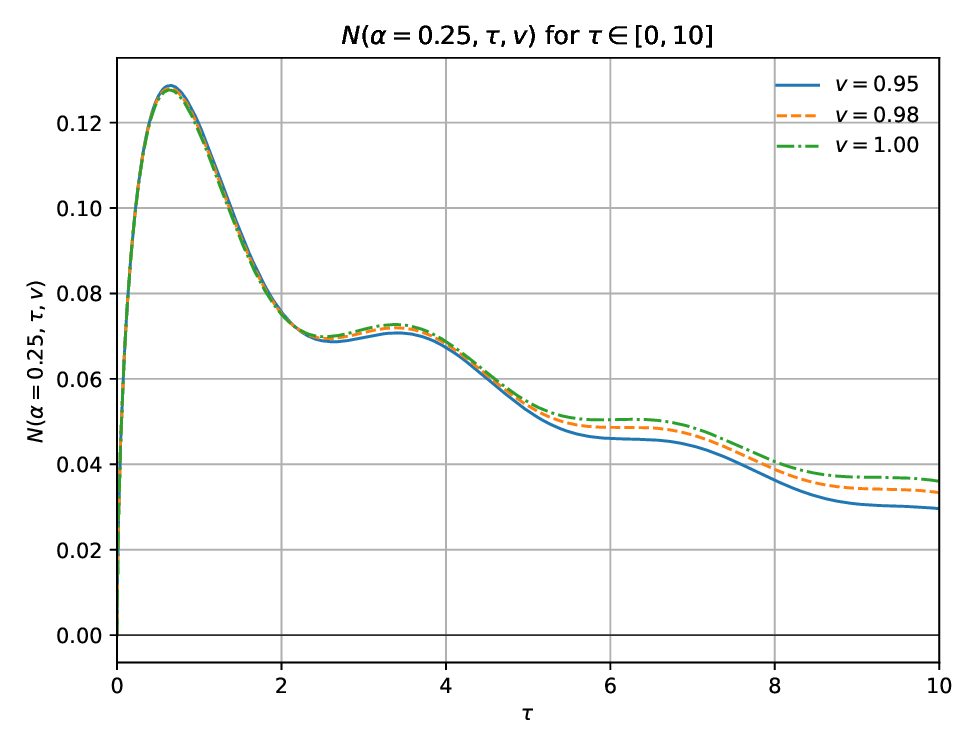}
     \caption{$N(\alpha,\tau,v) $ vs. $\tau=\omega_q t$ for $\alpha = 0.5~,~ v= 0.95,0.98,1$ for $ = G\,\gamma(q)\,\omega^\alpha_q= 0.1$.}
  \label{fig:Noftau}
\end{figure}

\subsection{Threshold and Infrared}\label{subsec:threshold}

The results (\ref{kres},\ref{Kalpha1}) for the decoherence function $\Phi(q,t) \propto t^{\alpha-1}~;~\propto \ln(\omega_q\,t)$ for $\alpha = (\du-2) < 1; 1$ respectively in the ultrarelativistic limit $q/\omega_q \rightarrow 1^-$,  is a consequence of the ``pole'' (resonance) at $\omega_q$ coinciding with the threshold, and the spectral density vanishing as a power law  at threshold, resulting in a ``pinching'' of the resonant part by the vanishing of the spectral density at threshold. This ``anomalous'' time dependence of the decoherence function is in striking contrast with the expectation from Fermi's golden rule, which would yield a vanishing decoherence rate in this limit because the spectral density vanishes. This is an important example of long time transient dynamics characterized by anomalous time dependence of both the decoherence function and the rate as a consequence of the mass shell of the probe being close to the threshold at which the spectral density of the bath vanishes. A similar situation has been previously recognized in ref.\cite{boyir} where renormalization group arguments suggested a notion of critical behavior near threshold or associated with infrared divergences from emission and absorption of massless quanta by a massive particle.

In this section we establish a direct link between the unparticle and threshold and infrared cases by considering  the case of two bosonic environmental fields $\chi_1,\chi_2$, and in order to compare to the unparticle case, we consider a zero temperature environment. The total Lagrangian density   is given by (\ref{lagra}) with $\mathcal{L}_S[\phi]$ and $S[\phi]$ given by (\ref{free}) and
\bea \mathcal{L}_E[\chi_1,\chi_2]  & = & \frac{1}{2} \Big[ \partial_\mu \chi_1 \partial^\mu \chi_1 - m^2_1 \chi^2_1+ \partial_\mu \chi_2 \partial^\mu \chi_2 - m^2_2 \chi^2_2\,\Big] \,\label{lagen} \\ \mathcal{O}[\chi_1,\chi_2] & = & \chi_1(x)\,\chi_2(x) \,,\label{enoper}\eea without loss of generality, we consider that $m_1 \geq m_2$. The correlation function of this composite operator is obtained in appendix (\ref{app:twofields}) and  corresponds to the one-loop diagram shown in figure (\ref{fig:loop}).

    \begin{figure}[H]
  \centering
  \includegraphics[width=2in]{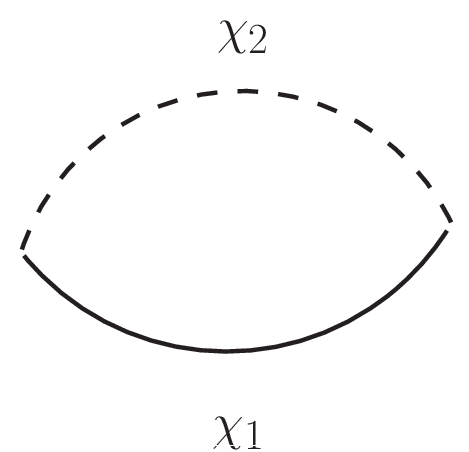}
     \caption{ The correlation function $  \langle 0|\mathcal{O}(\vx,t)\mathcal{O}(\vx',t')|0\rangle \equiv \langle 0|\chi_1(\vx,t)\chi_1(\vx',t')|0\rangle \,\langle 0|\chi_2(\vx,t)\chi_2(\vx',t')|0\rangle $.}
  \label{fig:loop}
\end{figure}

Threshold and infrared divergences emerge in this theory when the mass $m_\phi$ of the scalar probe field $\phi(x)$ fulfills the relation $m_\phi = m_1+m_2$ and we will study this case in detail since it reveals a time evolution with striking similarities to the case of unparticles and feature aspects of critical phenomena as anticipated in ref.\cite{boyir}.

The spectral density for this case is obtained in appendix (\ref{app:twofields}), it is given by
\be \rho(q_0,q) = \frac{\mathrm{sign}(q_0)}{8\pi [q^2_0 - q^2]}\, \Bigg\{ \Big[q^2_0-q^2 -(m_1+m_2)^2 \Big]\,\Big[q^2_0-q^2 -(m_1-m_2)^2 \Big]   \Bigg\}^{\frac{1}{2}}\,\Theta\Big(q^2_0-q^2 -(m_1+m_2)^2 \Big)\,.\label{rho2f}\ee We focus on two cases with $m_\phi = m_1+m_2 \neq 0 $:

\vspace{1mm}

\textbf{I): Threshold singularity:} corresponding to $m_1=m_2$ for which

\be \rho_T(q_0,q) = \frac{\mathrm{sign}(q_0)}{8\pi  }\,  \frac{\Big[q^2_0-\omega^2_q \Big]^{\frac{1}{2}}}{\big[q^2_0 - q^2\big]^{\frac{1}{2}}}\,\Theta\Big(q^2_0-\omega^2_q \Big)~~,~~\omega_q = \sqrt{q^2+m^2_\phi}\,.\label{rhoT}\ee Since the probe is massive, the denominator never vanishes, the similarity with the unparticle case is clear since  the spectral density vanishes at threshold with a power $\alpha = 1/2$, therefore we expect that the long-time dynamics of the decoherence function and rate will be similar to the unparticle case for an ultrarelativistic particle with $\alpha = \du-2 = 1/2$.

\vspace{1mm}

\textbf{II): Infrared divergence:} corresponding to $m_2=0 $ for which

\be \rho_{Ir}(q_0,q) = \frac{\mathrm{sign}(q_0)}{8\pi  }\,  \frac{\Big[q^2_0-\omega^2_q \Big] }{\big[q^2_0 - q^2\big]}\,\Theta\Big(q^2_0-\omega^2_q \Big)~~,~~\omega_q = \sqrt{q^2+m^2_\phi}\,.\label{rhoIr}\ee This case   again  features a spectral density vanishing at threshold with a power law $\alpha =1$, and the long time behavior should be similar to the unparticle case for an ultrarelativistic probe with $\alpha = \du-2 = 1$.

In both cases the phase space closes on the mass shell of the $\phi$ particle, the ``infrared'' case corresponds to the massive emitted particle of the same mass as the $\phi$ along with the emission of a massless particle, yielding infrared divergences akin to the case of fermions coupled to massless gauge fields. In both cases taking the infinite time limit of the rate, namely  the S-matrix limit or equivalently invoking Fermi's golden rule, yield vanishing rates, whereas as discussed below in detail for both cases, non-Markov dynamics leads to long time phenomena very similar to the critical unparticle environment  with an ultrarelativistic probe analyzed in the previous section.

We focus on the decoherence function $\Phi(q,t)$ since the decoherence rate can be simply obtained from its time derivative.  We define $q_0 = \omega_q x~;~ \tau = \omega_q t~;~ v = q/\omega_q $,  both cases can be combined by using this change of variables to write
\be \Phi(\alpha,\tau,v) = \frac{g^2}{16 \,\pi^2\,\omega^2_q} \, I(\alpha,\tau, v) \,,\label{fhith} \ee with
\be I(\alpha,\tau, v)= \int^{\infty}_{1}\,\Bigg[\frac{x^2-1 }{ x^2-v^2 }\Bigg]^{ \alpha}\Bigg[\frac{1-\cos[\big(x-1)\tau]}{(x-1)^2}-\frac{1-\cos[\big(x+1)\tau]}{(x+1)^2}  \Bigg] \,dx  \,,\label{phitir}\ee where $\alpha=\frac{1}{2}$ corresponds to the
case of threshold singularity and $\alpha=1$ to the case of infrared divergence.

To extract the asymptotic long $\tau$ limit it proves convenient to define $z =x-1$, to separate the resonant term  and write the integral in (\ref{phitir}) as
\be I(\alpha,\tau,v) = I_{res}(\alpha,\tau,v) + I_{nr}(\alpha,\tau,v) \,,\label{Isplit}\ee  with
\be   I_{res}(\alpha,\tau,v) =  \int^{1}_{0}\,\Bigg[\frac{z+2 }{ (z+1)^2-v^2 }\Bigg]^{ \alpha}z^{\alpha-2} \Big(1-\cos\big(z\tau\big)\Big)\,dz  \,,\label{Ires} \ee
\be  I_{nr}(\alpha,\tau,v) =  \int^{\infty}_{1}\,\Bigg[\frac{z+2 }{ (z+1)^2-v^2 }\Bigg]^{ \alpha}z^{\alpha-2} \Big(1-\cos\big(z\tau\big)\Big)\,dz - \int^{\infty}_0 \Bigg[\frac{(z+2)\,z }{ (z+1)^2-v^2 }\Bigg]^{ \alpha}\,  \frac{1-\cos\big((z+2)\tau\big) }{(z+2)^2} \,dz  \,.\label{Inores}\ee In the resonant contribution, a further change of variables to $w= z\tau$ yields
\be I_{res}(\alpha,\tau,v) = \tau^{1-\alpha} \, \int^{\tau}_0 \Bigg[\frac{\frac{w}{\tau}+2 }{ (\frac{w}{\tau}+1)^2-v^2 }\Bigg]^{ \alpha}w^{\alpha-2} \Big(1-\cos\big(w\big)\Big)\,dw \,,\label{Iresw}\ee For both cases $\alpha = 1/2, 1$  we can neglect the $w/\tau$ terms inside the bracket in the long time limit $\tau \rightarrow \infty$, since these give corrections $\mathcal{O}(1/\tau^{3/2}),\mathcal{O}(1/\tau)$ respectively,    and for the case $\alpha = 1/2$ the upper limit can be safely taken to infinity,  yielding for $\alpha = 1/2$
\be I_{res}(\frac{1}{2},\tau,v) ~{}_{\overrightarrow{\tau \rightarrow \infty}}~    \Big[\frac{\sqrt{4\pi}\, \omega_q }{ m_\phi }\Big] \,\sqrt{\tau}\,.\label{Iresonehalf}\ee  For $\alpha =1$ the upper limit must be kept, yielding
\bea I_{res}(1,\tau,v) & ~{}_{\overrightarrow{\tau \rightarrow \infty}}~ &  \Bigg[\frac{ 2 }{ 1-v^2 }\Bigg] \, \underbrace{\int^{\tau}_0   \frac{\Big(1-\cos\big(w\big)\Big)}{w}\,dw }_{\substack \gamma_E + \ln(\tau) + Ci(\tau) } \nonumber \\ & \simeq &  \Bigg[\frac{ 2\,\omega^2_q }{ m^2_\phi }\Bigg] \,\Big[ \ln(\tau) + \gamma_E + \mathcal{O}(1/\tau) \Big]\,,\label{Iresoneasy} \eea where we used that $1/(1-v^2) = \omega^2_q/m^2_\phi$. In the long time limit, the oscillatory contributions in the non-resonant terms average out yielding a finite constant
\be I_{nr}(\alpha,\tau,v) ~{}_{\overrightarrow{\tau \rightarrow \infty}}~ \mathcal{Z} = \int^{\infty}_{1}\,\Bigg[\frac{(z+2)\,z }{ (z+1)^2-v^2 }\Bigg]^{ \alpha}  \frac{ 1}{z^2}\,dz- \int^{\infty}_0 \Bigg[\frac{(z+2)\,z }{ (z+1)^2-v^2 }\Bigg]^{ \alpha}\,  \frac{1}{(z+2)^2} \,dz \,.\label{Inrtir}\ee The decoherence rate follows from the time derivative of the decoherence function, in both cases it vanishes in the asymptotic long time limit in agreement with Fermi's golden rule, but obviously the ``transient'' behavior is important over very long time scales. Figure (\ref{fig:Itir}) displays the integral in (\ref{phitir}) for $\alpha = 1/2,1$ and $v=0.95$  as a representative case, the numerical results confirm the asymptotic limits in both cases to high accuracy.

    \begin{figure}[H]
  \centering
  \includegraphics[width=4in]{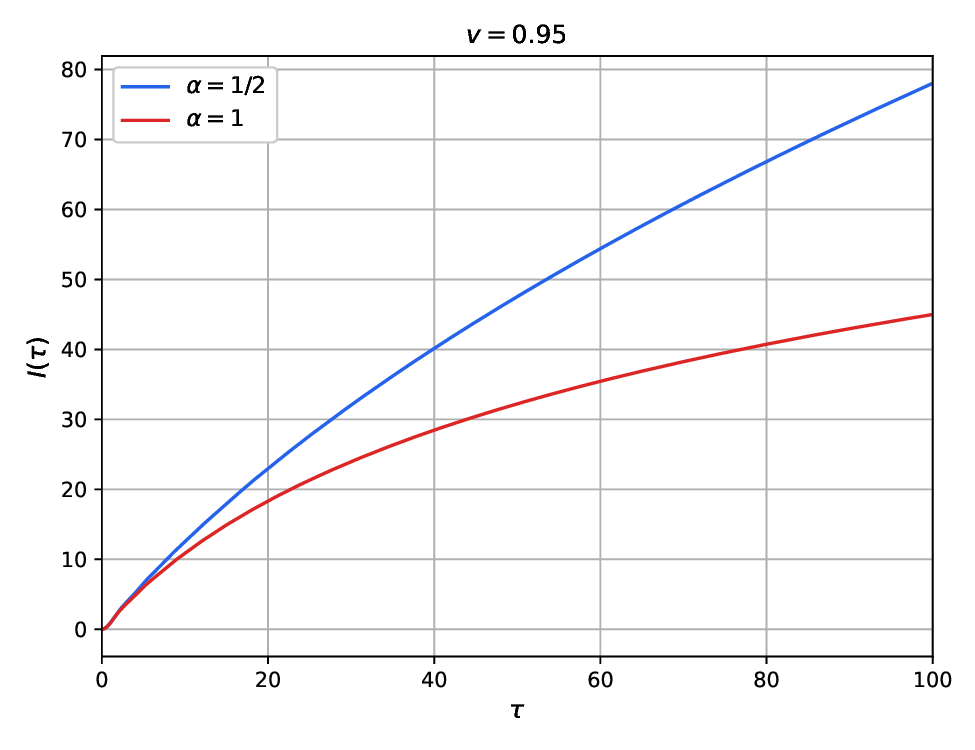}
     \caption{ The integrals $I(\alpha,\tau,v)= \Phi(q,t)/(g^2/16\,\pi^2 \omega^2_q)$ in (\ref{phitir})  for $\alpha = 1/2,1$ and $v =0.95$ vs $\tau = \omega_q t$.}
  \label{fig:Itir}
\end{figure}

  Since the long time behavior is dominated by the resonant term at $x\simeq 1$, it is clear that in the cases of threshold and infrared divergences the decoherence function features the \emph{same} long time asymptotic behavior as that for the unparticle case in the ultrarelativistic regime, in particular the limiting unparticle cases  described by the asymptotic limits (\ref{kres}) yielding $\tau^{1/2}$ for $\alpha=1/2$ or (\ref{Kalpha1}) yielding $\simeq \ln(\tau)$ for $\alpha=1$,  confirming the observation in\cite{boyir} that the long time dynamics determined by the threshold behavior of the spectral density are a manifestation of ``critical phenomena'' in common with unparticle fields. Certainly the similarity does not apply to non-resonant terms which in general are dominated by the large frequency components.

  \vspace{1mm}

  \textbf{Away from thresholds:}  Although the analysis above focused on the case when the mass shell of the probe particle coincides with threshold, because such cases yield critical-like near-threshold behavior and anomalous time evolution of the decoherence function, it is important to understand the case when  $\omega_q$ does not coincide with the threshold.  Let us define $\Omega^2_q = q^2+(m_1+m_2)^2$ and the ratio $R^2_q = \Omega^2_q/\omega^2_q$, and consider again both cases: $m_1=m_2$ and $m_2=0$, for which the integral (\ref{phitir}) now becomes
  \be I(\alpha,\tau, v)= \int^{\infty}_{R_q}\,\Bigg[\frac{x^2-R^2_q }{ x^2-v^2 }\Bigg]^{ \alpha}\Bigg[\frac{1-\cos[\big(x-1)\tau]}{(x-1)^2}-\frac{1-\cos[\big(x+1)\tau]}{(x+1)^2}  \Bigg] \,dx  \,.\label{phitirR}\ee
   For $R_q \neq 1$  the probe mass shell is away from threshold and the decoherence rate $\Gamma(q,t)$  approaches the asymptotic S-matrix limits:
       \be  \Gamma(q,\infty) =  \frac{g^2}{16\,\pi\,\omega_q}\,\Bigg[\frac{1-R^2_q}{1-v^2}\Bigg]^\alpha \neq 0 ~\mathrm{for}~  R_q< 1\,,\label{aboth}\ee
       since the ``pole'' at $x=1$ is above threshold and
        \be \Gamma(q,\infty) =0 ~~;~~ \Phi(q,\infty) =   \frac{g^2}{2\,\pi\,\omega_q}\,  \int_{-\infty}^{\infty}   \,\mathcal{P}  \frac{\rho(q_0,q)}{( \omega_q-q_0)^2} \,dq_0 ~~~\mathrm{for}~  R_q>1\,,\label{belth}\ee  because in this case the ``pole'' at $x=1$ is below threshold and the probe particle is stable. However,   the resonant term in the bracket features a width in the $x$ variable $\Delta x \simeq 2\pi/\tau$, therefore for $R_q \neq 1$ these terms probe the threshold region of the spectral density over a time scale $\Delta \tau \simeq 2\pi/| R_q -1|$ during which threshold effects affect the dynamics and the coherence function does not grow linearly (for $R_q < 1$) or vanishes (for $R_q > 1$), but nearly follows the cases in which the ``pole'' merges with the threshold, $R_q \simeq 1$. This is clearly exhibited in fig. (\ref{fig:away}) in both cases, where the curvature of the coherence function as a function of time, and the   delay in establishing the linear secular growth is a direct consequence of threshold effects.

     \begin{figure}[ht]
  \centering
  \includegraphics[width=6in]{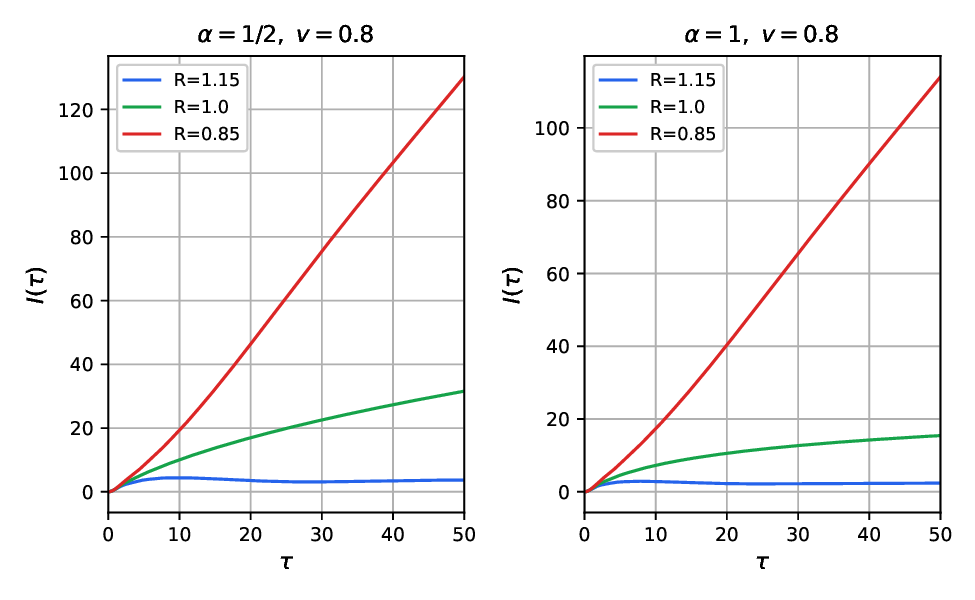}
     \caption{ The integral $I(\alpha,\tau,v)= \Phi(q,t)/(g^2/16\pi^2 \omega^2_q)$ in (\ref{phitirR})  for $\alpha = 1/2,1$ and $v =0.8$ and $R_q=0.85,1.0,1.15$ vs $\tau = \omega_q t$.}
  \label{fig:away}
\end{figure}

The similarity with the unparticle case studied in the previous section is noteworthy: in the threshold and infrared cases the resonance is brought to the edge of
the spectral support by a nonzero kinematic scale: the mass shell, with the
condition $m_\phi=m_1+m_2$ pinning the resonance at threshold. The
ultrarelativistic limit of the unparticle case  (\ref{gamaunur},\ref{fiunur})   provides the corresponding
kinematic limit through $v \rightarrow 1^-$, as the mass shell approaches the threshold
according to $\omega_q-q\simeq m_\phi^2/2q$.

 These cases all   share a common mechanism:
the resonance at $x=q_0/\omega_q=1$ approaches the lower edge of the spectral
support, and the resulting long-time behavior is controlled by the spectral
density near threshold which entail strong departures from the S-matrix result of linear secular growth determined by the (FGR) rate (\ref{gamafgr}).

  Both the unparticle case\cite{unparticle,bare} and the threshold and infrared cases\cite{boyir} are manifestations  of critical behavior, where
  the spectral density vanishes at  threshold yielding  anomalous scaling in time of the coherence
function, reflecting non-Markov evolution as a consequence of ``slow'' degrees of freedom of the bath. This behavior is  in striking contrast with the   Markov approximation ubiquitous in the literature and  the S-matrix expectations of linear secular growth in time of the decoherence function, which is ultimately a manifestation of (FGR).

It is then natural to ask how these results extend to the case when the bath is characterized by degrees of freedom that feature slow dynamics, in contrast to the usual assumptions underpinning the Markov approximation.   In particular we refer to the critical case when
the characteristic frequency of the soft modes  \emph{vanishes} and their \emph{relaxation time diverges} for long-wavelength fluctuations, namely  a bath that is at
a critical point of a second order phase transition, whose long wavelength dynamics  features critical slowing
down and there emerges a new dynamical critical exponent that characterizes the slow relaxation. We turn to this case next.

\subsection{Critical slowing down}\label{subsec:slowdown}

Dynamical critical phenomena extends the concepts of \emph{static} (equilibrium) critical phenomena at second order phase transitions to the realm of dynamical non-equilibrium evolution of the fluctuations and correlations of the order parameter\cite{hohenberg,ma,tauber}. At the critical point of a second order phase transition the correlation length diverges and there are no intrinsic scales for long-wavelength, low frequency fluctuations,  leading to scaling laws of the correlation functions.

In the case of \emph{dynamical} critical phenomena a new critical exponent emerges, $z$,  that describes the scaling behavior as a function of \emph{frequencies} in complement to
the scaling behavior as a function of momenta. This is encoded in new dynamical scaling form of the correlation function of the order parameter in frequency $\omega$ and momentum $k$ space (assuming rotational invariance) at the critical point,
\be G(\omega,k) = \frac{1}{k^{2-\eta}\,\mu^\eta}\,\mathcal{G}\Big[ \frac{\omega}{k^z}\Big]\,, \label{gscale}\ee where $\eta$ is the usual static scaling exponent, $\mu$ is a renormalization scale,  $z$ is the new dynamical critical exponent, and $\mathcal{G}$ is a dimensionless function of its argument.

Furthermore, at the critical point the relaxation time scale of fluctuations $\tau_k$   diverges in the long wavelength limit
\be \tau_k \propto k^{-z} \,,  \label{relatau}\ee this is the hallmark of \emph{critical slowing down} confirmed in experiments of critical fluids or systems approaching the critical temperature\cite{ma,hohenberg,tauber}. Just as the exponent $\eta$ along with the other usual static critical exponents define universality classes, the new dynamical critical exponent $z$ also defines universality classes in critical dynamics\cite{hohenberg}.

For the case of unparticles\cite{unparticle} where the two point correlation function   is a scaling function of the Lorentz invariant $\omega^2-k^2$, it follows that the dynamical exponent  $z=1$ which is the \emph{only} possible dynamical critical exponent consistent with Lorentz invariant and consistent with a zero temperature theory as envisaged in\cite{unparticle}. In contrast,  a finite temperature quantum field theory at the critical point where the fully renormalized temperature dependent   mass vanishes,  does not necessarily feature Lorentz invariant correlation functions because  the thermal bath defines a particular frame (its rest frame).

We consider the system-bath coupling $S[\phi]\,\mathcal{O}(x)$ as in (\ref{lagra}) with $S[\phi] = \phi$  and $\mathcal{O}(x)$ the   order parameter associated with a scalar field at the critical point of a second order phase transition, whose correlation functions features the scaling form (\ref{gscale}). Since the spectral density is completely determined by the correlation functions (see appendix (\ref{app:fd})),  dynamical scaling  (\ref{gscale}) implies a scaling form of  $\rho(q_0,q)$  at the critical point, namely

\be
\rho(q_0,q) = q^{-2+\eta}\,\mu^{-\eta}
\mathcal{F}\!\Big(\frac{q_0}{q^z}\Big)\,,
\label{rhodyn}
\ee where $\mathcal{F}$ is an odd function of its argument.

Consequently, the S-matrix rate (FGR)  is given by the asymptotic limit (\ref{gamafgr})
\be \Gamma(q,\infty) = \frac{g^2\,q^{-2+\eta}\,\mu^{-\eta}}{2\omega_q}\,
\mathcal{F}\!\Big(\frac{\omega_q}{q^z}\Big)\,. \label{fgrdyn}  \ee

A main lesson learned from the cases studied in the previous sections is that even when the S-matrix rate is attained in the asymptotic long time limit, the time scale at which this asymptotic result sets in  during which non-Markovian dynamics is  most important, is determined by the properties of the spectral density between the  threshold and the position of the mass shell of the probe $\omega_q$.  Therefore the  main question that we study in this case of a bath at a critical point is how long it takes for the asymptotic rate (\ref{fgrdyn}) to set in. The answer depends on the particular form of the scaling function $\mathcal{F}$, which   in general is \emph{not} determined by the scaling arguments or the scaling solution of the renormalization group equation at the fixed point. Its explicit form along with the   values of the critical exponents     require a detailed evaluation.

Most of the studies of dynamical critical phenomena, and indeed the classification of dynamical universality classes has focused on phenomenological descriptions of the dynamics based on purely dissipative time dependent Landau-Ginsburg equations (typically with Gaussian white noise obeying the fluctuation dissipation relation)\cite{hohenberg,tauber}, while the study in relativistic field theories has received much less attention.

As an explicit example, we borrow the results of reference\cite{hotboy} for a scalar field theory at finite temperature $T$ at the critical point, defined by the vanishing of the fully renormalized mass. In this reference, the renormalization group is implemented to study dynamical critical phenomena for long wavelength excitations. For $q_0, q \ll T$ the effective low energy theory is described by the zero Matsubara frequency thereby entailing a dimensional reduction but also an effective coupling $\lambda T$ with $\lambda$ the self-interaction quartic coupling. This effective field theory is renormalizable in $D =5$  Euclidean space-time dimensions, where $\lambda T$ is dimensionless, and an $\epsilon$ expansion is implemented in $D=5-\epsilon$, to approach the critical point in a controlled expansion, just as in static critical phenomena\cite{ma,gold,tauber}. In the Matsubara formulation of finite temperature quantum field theory, the frequencies are discrete, whereas momenta are continuous, this different treatment of frequencies and momenta leads to the need for a new counterterm beyond wave function renormalization identified as a renormalization of the speed of light\cite{hotboy}.  A main result of this reference is that there is an infrared ``Wilson-Fisher'' fixed point with the effective coupling $\propto \epsilon$ at which there emerges scaling behavior  with anomalous scaling exponents featuring the usual static exponents along with a new dynamical exponent $z$. In this reference the scaling function is obtained explicitly from a renormalization group- improved two-loop self-energy, with the result for the Euclidean Green's function at the critical point
\be G_E(s,q) = \frac{1}{\Big[q^{2-\eta}\,\mu^{\eta}+ \big(\frac{s}{v_R(\mu)}\big)^{2-\eta_t} \,\mu^{\eta_t} \Big]} \,,\label{GEscal}\ee where $s$ is the Euclidean frequency variable and the critical exponents
\be \eta = \frac{\epsilon^2}{54}~~;~~ \eta_t = \frac{2\epsilon}{27} \,,\label{etaexps}\ee $v_R(\mu)$ is the renormalized   speed of light and  $\mu$ is a renormalization scale. This function exhibits the general dynamical scaling form predicted by the renormalization group at the fixed point\cite{hotboy} and is in agreement with the result at the ``Gaussian'' fixed point (free field theory) with $\eta =\eta_t =0$. The critical exponents of the actual four dimensional theory must be extrapolated to the physical dimensionality $\epsilon =1$ as is the case with the familiar $\epsilon$ expansion in critical phenomena\cite{ma,gold,hohenberg,tauber}. The Euclidean Green's function (\ref{GEscal}) can be written in a manifestly scaling form that exhibits dynamical scaling (\ref{gscale})  as
\be   G_E(s,q) = \frac{1}{q^{2-\eta}\,\mu^{\eta}}\,\frac{1}{\Big[1+ \Big(\frac{s}{v_R(\mu)\,\mu^{1-z}q^z}\Big)^{2-\eta_t}   \Big]}\,,\label{GEzz} \ee with the dynamical scaling exponent
\be z= \frac{2-\eta}{2-\eta_t}\simeq 1+ \frac{1}{2} (\eta_t-\eta)  > 1 \,,\label{zetaex}\ee furthermore, the combination
$v_R(\mu) \,\mu^{1-z}$ is renormalization group invariant at the critical point\cite{hotboy}.

The retarded Green's function in Minkowski space-time is obtained from the analytic continuation
\be G_M(q_0,q) = - \, G_E(s=-iq_0+0^+)=  {\Bigg[ \Big(\frac{q^2_0}{v^2_R(\mu)}\Big)^{1-\frac{\eta_t}{2}} \,\mu^{\eta_t}\big[1+ i\,\pi \frac{\eta_t}{2}\,\mathrm{sign}(q_0) \big]- \big(q^2\big)^{1-\frac{\eta}{2}}\,\mu^{\eta}   \Bigg]}^{-1}   \,,\label{Gmink}\ee where   a term $\propto \eta^2_t$ has been neglected in the real part of the analytic continuation. From this expression, ref.\cite{hotboy} obtains the dispersion relation $\mathcal{W}_q$, group velocity $\mathcal{V}_q $ and width (the inverse of the relaxation time $\tau_q$) $\Upsilon_q$  of the critical quasiparticles from the complex poles of this Green's function with the results
\bea \mathcal{W}_q & = & v_R(\mu) \,\mu^{1-z}\,q^z\,,\label{disprel}\\ \mathcal{V}_q & = &  z\,v_R(\mu) \,\Big[\frac{q}{\mu}\Big]^{z-1}   \,,\label{grouvel}\\
\Upsilon_q & = &  \frac{1}{\tau_q} = \frac{\pi}{4}\,\eta_t\,v_R(\mu) \,\mu^{1-z}\,q^z = \frac{\pi}{4}\,\eta_t\,\mathcal{W}_q \,.\label{critwidth}\eea
The relation (\ref{grouvel}) implies that $v_R(\mu) = \mathcal{V}_\mu/z$, therefore measuring the group velocity at a scale $\mu$ determines the group velocity at any other scale $q$, furthermore, since $z>1$ it follows that the group velocity \emph{increases} with the  momentum of the critical quasiparticle excitation.

Critical slowing down is embodied in these results with $\Upsilon_q \propto q^z$, $\mathcal{V}_q \propto q^{\frac{1}{2}(\eta_t-\eta)}$, therefore long wavelength fluctuations feature long relaxation times and wavepackets with long wavelength components move and   disperse very slowly. Note that the dispersion relation can also be written as $\mathcal{W}_q =   q\,\mathcal{V}_q/ z $ with the group velocity $\mathcal{V}_q \rightarrow 0$ in the long wavelength limit for gapless excitations, unequivocally a sign of collective behavior, in striking contrast to free field theory.

The spectral density is obtained from $G_M(q_0,q)$ as\footnote{As $\eta_t , \eta \rightarrow 0$, $\rho(q_0,q) \rightarrow 2\pi \delta(q^2_0-q^2)\,\mathrm{sign}(q_0)$ as in free field theory.}
\be \rho(q_0,q) = -2\, \mathrm{Im}\big[G_M(q_0,q)\big] = \frac{\pi\,\eta_t \big| \frac{q_0}{v_R(\mu)} \big|^{2-\eta_t}\,\mu^{\eta_t}\,\mathrm{sign}(q_0) }{  \Big( \Big|\frac{q_0}{v_R(\mu)}\Big|^{2- \eta_t } \,\mu^{\eta_t} - q^{2-{\eta}}\,\mu^{\eta}   \Big)^2+ \Big(\frac{\pi}{2}\eta_t\,\Big|\frac{q_0}{v_R(\mu)}\Big|^{2- \eta_t } \,\mu^{\eta_t} \Big)^2   }\,,\ee which can be written explicitly in dynamical scaling form as
\be \rho(q_0,q) = q^{-2+\eta}\,\mu^{-\eta} \, \frac{\pi\,\eta_t \, \xi[\frac{q_0}{q^z}]\,\mathrm{sign}(q_0) }{\Big(  \xi[\frac{q_0}{q^z}] - 1   \Big)^2+ \Big(\frac{\pi}{2}\eta_t\,\xi[\frac{q_0}{q^z}] \Big)^2 }~~;~~ \xi[\frac{q_0}{q^z}] = \Big[\frac{|q_0|}{q^z\,v_R(\mu)\,\mu^{1-z}}\Big]^{2-\eta_t}\equiv \Big[\frac{|q_0|}{\mathcal{W}_q}\Big]^{2-\eta_t}\,,\label{rhoscale} \ee from which we can extract the scaling function $\mathcal{F}$ of equation (\ref{rhodyn})
\be \mathcal{F}[\frac{q_0}{q^z}] =   \, \frac{\pi\,\eta_t \, \xi[\frac{q_0}{q^z}]\,\mathrm{sign}(q_0) }{\Big(  \xi[\frac{q_0}{q^z}] - 1   \Big)^2+ \Big(\frac{\pi}{2}\eta_t\,\xi[\frac{q_0}{q^z}] \Big)^2 }~~;~~ \xi[\frac{q_0}{q^z}] = \Big[\frac{|q_0|}{q^z\,v_R(\mu)\,\mu^{1-z}}\Big]^{2-\eta_t}\equiv \Big[\frac{|q_0|}{\mathcal{W}_q}\Big]^{2-\eta_t}\,,\label{Funscal}\ee featuring a sharp resonance at a frequency corresponding to the quasiparticle energy $\mathcal{W}_q$ with $\Upsilon_q/\mathcal{W}_q \propto \eta_t \ll 1$.

Armed with these results we can now write the explicit form of the decoherence rate $\Gamma(q,t)$ (\ref{gamadif}) and decoherence function $\Phi(q,t)$ (\ref{decofun}) for which it is convenient again to define $q_0 = \omega_q\, x~;~ \tau= \omega_q t$ as in the previous case of ultrarelativistic limit  (\ref{gamaunur},\ref{fiunur}), yielding
\be \Gamma(q,t) =  \frac{g^2}{2\pi\,\omega_q}\,q^{-2+\eta}\mu^{-\eta} \, \mathcal{I}_{\Gamma}(\tau,\xi) \,,\label{gamcs} \ee with
\be \mathcal{I}_{\Gamma}(\tau,\xi) = \int^{\infty}_0 \Bigg[  \frac{\pi\,\eta_t \, \xi[\frac{\omega_q}{q^z}]\,x^{2-\eta_t} }{\Big(  \xi[\frac{\omega_q}{q^z}]\,x^{2-\eta_t} - 1   \Big)^2+ \Big(\frac{\pi}{2}\eta_t\,\xi[\frac{\omega_q}{q^z}]\,x^{2-\eta_t} \Big)^2 }  \Bigg]\, \Bigg[\frac{\sin\big[(x-1)\tau\big]}{(x-1)}-\frac{\sin\big[(x+1)\tau\big]}{(x+1)}  \Bigg] \,dx \,,\label{Igamcs}\ee and
\be \Phi(q,t) =  \frac{g^2}{2\pi\,\omega^2_q}\,q^{-2+\eta}\mu^{-\eta} \, \mathcal{I}_{\Phi}(\tau,\xi) \,,\label{phics} \ee with
\be \mathcal{I}_\Phi(\tau,\xi) =  \int^{\infty}_0 \Bigg[  \frac{\pi\,\eta_t \, \xi[\frac{\omega_q}{q^z}]\,x^{2-\eta_t} }{\Big(  \xi[\frac{\omega_q}{q^z}]\,x^{2-\eta_t} - 1   \Big)^2+ \Big(\frac{\pi}{2}\eta_t\,\xi[\frac{\omega_q}{q^z}]\,x^{2-\eta_t} \Big)^2 }  \Bigg]\, \Bigg[\frac{1-\cos\big[(x-1)\tau\big]}{(x-1)^2}-\frac{1-\cos\big[(x+1)\tau\big]}{(x+1)^2}  \Bigg]  \,dx \,.\label{Ifics} \ee
Although the upper limit of these integrals must be of $\mathcal{O}(T)$ since the scaling behavior emerges for wave vectors much smaller than this scale which determines the upper limit of validity of the critical low energy effective field theory, the resulting integrals are rapidly converging.

These expressions reveal the noteworthy feature that    a probe
of frequency $\omega_q$ is resonant with the critical bath at
$q_*\sim\omega_q^{1/z}$ revealing the dynamical exponent as well as the nature of the critical quasiparticles. The S-matrix (FGR) result would yield
\be \mathcal{I}_{\Phi}(\tau,\xi) =   \mathcal{I}_{\Gamma}(\infty,\xi)\,\tau ~~;~~ \mathcal{I}_{\Gamma}(\infty,\xi)=      \frac{\pi^2 \,\eta_t \, \xi[\frac{\omega_q}{q^z}]  }{\Big(  \xi[\frac{\omega_q}{q^z}]  - 1   \Big)^2+ \Big(\frac{\pi}{2}\eta_t\,\xi[\frac{\omega_q}{q^z}]  \Big)^2 }        \,.\label{Ifgr} \ee The asymptotic decoherence rate $\mathcal{I}_{\Gamma}(\infty,\xi)$   is strongly enhanced when the probe mode with momentum $q_*$ is resonantly coupled to a bath mode $q_* \propto \omega^{1/z}_q$ that is critically slow.

Figure (\ref{fig:iffy}) shows $\mathcal{I}_{\Phi}(\xi,\tau)$  vs. the S-matrix results for resonant and off-resonant cases, clearly displaying the importance of non-Markov transient phenomena with decoherence below and above the S-matrix predictions respectively. The resonant case is noteworthy, the reason that the actual coherence function is \emph{below} the (FGR) result is due to the principal part in the constant term  in the asymptotic limit (\ref{gamasy}), since the spectral density weighs strongly the region $x \simeq 1$ at the narrow resonance, the principal value part of the integral becomes large and negative. This  can be   demonstrated analytically by extracting the principal part expanding around $x=1$ which is the position of the resonance for $\xi =1$,  and has been  numerically confirmed, whereas in the non-resonant cases the contribution near $x\simeq 1$ are compensated by the remaining integral yielding a positive offset. Note also the large amplitude of the asymptotic rate in the resonant case, as a consequence of the narrowness of the resonance. Since $z>1$   it could be  possible  for a light scalar probe to satisfy the resonance condition, in which case decoherence would occur very fast. This seems rather counterintuitive that a mode of the probe in resonance with a \emph{critically slow} mode of the bath could rapidly decohere, but the resonant enhancement of the decoherence rate is a consequence of the sharpness of the resonance, in turn, a manifestation of critical slowing down.

The dynamical scaling is manifest in the asymptotic rate which is a function of $\omega_q/\mathcal{W}_q \simeq \omega_q/q^z$, along with the $q^{-2+\eta}$ in the prefactor, furthermore, the transient behavior is sensitive to the dynamical exponent $z$ through the terms $x^{2-\eta_t}$ in the integrands. The coherence function $\Phi(q,t)$ is dimensionless as it must be, since the scalar couples linearly to the order parameter, which has mass dimensions, therefore the coupling $g$ has dimensions of $(\mathrm{mass})^2$.

    \begin{figure}[ht]
  \centering
  \includegraphics[width=4in]{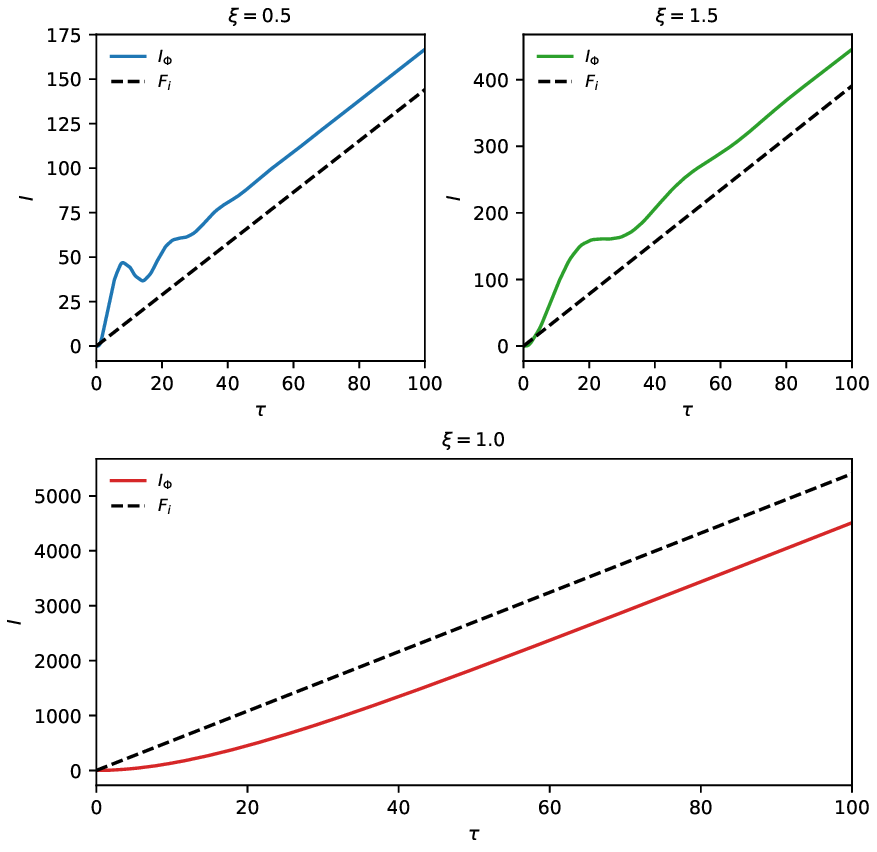}
     \caption{ The integrals $\mathcal{I}_{\Phi}(\xi,\tau)$ (solid lines)  compared to the linear result from S-matrix: $ \mathcal{I}_{\Gamma}(\infty,\xi)\,\tau $  (dashed lines)   for the off-resonant cases $\xi = 0.5,1.5$ (top panels) and the resonant case $\xi=1$ (bottom panel) vs.$\tau$ for $\epsilon = 1$.}
  \label{fig:iffy}
\end{figure}

The important corollary of this analysis is that as a consequence of the sharp resonance, a probe mode with wavevector $q$ that is resonant with a critically slowed down bath mode undergoes fast decoherence.

A noteworthy feature of the new dynamical exponent emerges when considering the decoherence rate of the probe particle at rest, namely $q \rightarrow 0$, yielding
\be \Gamma(q,\infty)~{}_{\overrightarrow{q\rightarrow 0}}~   \frac{\pi g^2\,\eta_t}{2 \, m^3_\phi} \, [v_R(\mu)]^{2-\eta_t}\,\Big[\frac{m_\phi}{\mu} \Big]^{\eta_t}\,,\label{gamqzero} \ee which does \emph{not} depend on $\eta$, as it must be the case, because the anomalous dimension $\eta$ is solely associated with (spatial) wavevector scaling, but depends on $\eta_t$ which is the new dynamical critical exponent in agreement with the interpretation of the rate as a \emph{dynamical} quantity. The decoherence rate  $\Gamma(0,\infty)$  features the correct dimensions of $(time)^{-1}$ since $g$ has dimensions of $(mass)^2$, but scales with \emph{anomalous dimension} $-3+\eta_t$ upon scaling the mass of the probe. There are several   aspects of this result that merit emphasizing:
\vspace{1mm}

\textbf{1:)} The limit $q\rightarrow 0$ of the decoherence rate $\Gamma(q,\infty)$ is highly non-trivial as it emerges from the limit $\xi\big[\frac{\omega_q}{\mathcal{W}_q}\big]\rightarrow \infty$ and as noted above \emph{cannot} depend on the anomalous dimension $\eta$ as it describes a particle at rest (zero momentum), instead it must depend solely on the
dynamical exponent $\eta_t$, as it reassuringly does.

\vspace{1mm}

\textbf{2:)} The  decoherence rate features anomalous mass dimension, scaling as $m^{-3+\eta_t}_\phi$ instead of the ``naive'' scaling dimension $(-3)$, a direct manifestation of dynamical scaling. The anomalous scaling in this case must be compared with the decoherence rate at zero momentum for the unparticle case (\ref{asygamun}) which also features anomalous mass dimension $\sigma = 2 (\du-2)$, however, unlike the unparticle case which describes Lorentz invariant critical correlations, here the dynamical anomalous exponent is different from the usual static exponent $\eta$. Nevertheless,    both are a manifestation of criticality and scaling of the correlation functions of the environmental fields.

\vspace{1mm}

\textbf{3:)} The long wavelength limit (\ref{gamqzero}) is determined by the \emph{tail} of the spectral density in the asymptotic limit (\ref{Ifgr}), namely the region $\xi \rightarrow \infty$ far away from the resonant (quasiparticle) peak. This is, indeed, the main reason for the scaling behavior with $m_\phi$ because in this region the spectral density is $\propto 1/\xi \propto q^{2-\eta}\,m^{-(2-\eta_t)}_\phi$ in the long-wavelength limit, which exactly cancels the $q^{-2+\eta}$ prefactor including the anomalous exponent $\eta$ rendering the long-wavelength decoherence rate independent of this \emph{static} critical exponent and solely dependent on the dynamical critical exponent $\eta_t$. Therefore, in a very clear manner, the dependence of the long wavelength decoherence rate solely on $\eta_t$ is a confirmation of dynamical scaling.

\vspace{1mm}

\section{Discussions}\label{sec:discussion}

\textbf{Factorization} An important approximation invoked in obtaining the (QME) is that even during time evolution the reduced density matrix and the bath density matrix factorize, this is explicit in eqn. (\ref{fact}), an approximation that is independent of both the Markov and rotating wave approximation. The Schwinger-Keldysh ``in-in'' formulation shows the validity of this approximation up to and including second order  because up to this order influence functional only contains the two-point correlations of the operators $\mathcal{O}$ in the environmental density matrix $\rho_E(0)$ as is explicit in equations (\ref{ggreat}-\ref{expec}).

\textbf{Beyond the rotating wave approximation:} In the derivation of the  quantum master equation (\ref{Linfin}) we  neglected terms of the form
\be a^\dagger(\vq)~a^\dagger(-\vq)~ e^{2i\omega_{q} t'}~e^{i{\omega_{q}(t-t')}}~~;~~ a(\vq)~a(-\vq)~ e^{-2i\omega_{q} t'}~e^{-i{\omega_{q}(t-t')}}\,. \nonumber \ee

  Their effect can be assessed  by including them into the (QME) after the (partial) Markov approximation. The time integral over $t'$ can be carried out following the steps leading to equation (\ref{Linfin}) yielding contributions of the form
$a_{\vq}\,a_{-\vq}\,e^{-2i\omega_{q} t} \rho^\lessgtr(q_0,q) {\rho}^r_{S}(t)$ etc. The contribution of these terms to the equations of motion for linear or bilinear forms of $aa^\dagger$ are straightforward to obtain. Although the equations of motion do not close in a simple manner as without these terms, they do not lead to  secular  growth in time because of the rapid oscillations  leading to rapid dephasing. These extra contributions result in non-resonant corrections   and yield perturbatively small subleading contributions    in weak coupling, as compared to those obtained from equation (\ref{Linfin}) which captures the secular growth in time because of the resonances and describes the leading behavior in the long time dynamics.   The long time analysis with the Breit-Wigner approximation and the multitime analysis of the equation of motion obtained without the rotating wave approximation in the Schwinger-Keldysh formulation  reproduce the results of the (QME) after such approximation, thereby confirming that at least to $\mathcal{O}(g^2)$ the counterrotating terms can be safely neglected. This is yet another confirmation of the validity of the approximations in the (QME) up to $\mathcal{O}(g^2)$ that is clarified by the comparison between the Schwinger-Keldysh and (QME) frameworks.

\vspace{1mm}

\textbf{Schwinger-Keldysh vs (QME) :}   The ``in-in'' and (QME) formulations are complementary, while both determine the time evolution of the reduced density matrix and allows to extract observables, each has advantages and disadvantages. In the Schwinger-Keldysh approach including the influence functional, one may introduce sources coupled to various operators and obtain multitime correlation functions from variational derivatives with respect to these sources. The (QME) most simply yields expectation values or equal time correlations, in order to obtain multi-time correlations within the (QME) one must resort to the quantum regression theorem\cite{breuer,zoeller} which involves attaching strings of the unitary time evolution operator. On the other hand, aspects of decoherence inferred from the time evolution of off-diagonal bilinear correlations such as those of equation (\ref{aaoftsol}) is easier to be extracted from the (QME).

\textbf{Higher orders in the influence functional:} The formal expression for the influence functional (\ref{Fln}) is the real time equivalent of the background field method and is valid to \emph{all orders} of the interactions of the bath degrees of freedom among themselves and also to all orders in the system-bath coupling. It can be expanded in a cumulant expansion rearranged in terms of powers of the system-bath coupling $g$ which involve higher point correlation functions of the bath operators in the initial environmental density matrix $\rho_E (0)$. However, there is no Lehmann representation for correlation functions beyond the two point correlator, therefore while formally exact the result (\ref{Fln}) does not seem to be very useful beyond the second order. When the higher order terms in the influence functional are included, taking expectation values of the system operators in the reduced density matrix will imply backreaction of the system onto the bath and generation of system-bath correlations. These are precisely the correlations discussed in the critical study in ref.\cite{linme}, therefore, again the ``in-in'' formulation of the influence functional clarifies the emergence of system-bath correlations beyond second order, in agreement with ref.\cite{linme}.

\textbf{Non-Markovian evolution from slow modes:} The cases studied above are by no means exhaustive of the possible scenarios where long-wavelength environmental degrees of freedom feature slow dynamics, for example Goldstone bosons or hydrodynamic modes corresponding to conserved quantities with diffusive relaxation\cite{hohenberg}. Once the spectral density associated with these degrees of freedom is obtained, the methods described in this study can be directly applied to understand the dynamics of fields coupled to these environments.

\textbf{Implications for cosmology and beyond:} The results of this study are likely to bear an impact in cosmology, for example in the ``misalignment'' dynamics of scalar or pseudoscalar fields coupled to other fields that act as a reservoir or environment, where transient dynamics competes with cosmological expansion and phenomenological descriptions in terms of damping rates obtained via  S-matrix are unjustified. Also most  studies of  freeze-out of hypothetical dark matter species implement   quantum kinetic equations with S-matrix rates and explicit energy conservation, glossing over the fact that in an expanding cosmology there is no global time-like Killing vector and particle energy is not conserved. Again, in these cases transient non-Markovian dynamics may prove to be very relevant. In condensed matter physics, deviations from (FGR) and transient phenomena has been experimentally confirmed with ultrafast experiments\cite{expdecay,navon}, perhaps similar experiments can be carried out with environments undergoing a second order phase transition and validate some of the results found in this study.

\textbf{Thermalization  and particle production:}  We have argued that the environmental noise leads to the production of probe particles with a distribution function given by (\ref{parprod}) which again depends on the decoherence function. For environments in thermal equilibrium particle production leads to the thermalization of the probe.  In the case of unparticles we considered the vacuum state since conformal invariance underpinning the unparticle concept may be broken in a heat bath resulting in correlations functions that are neither Lorentz invariant nor of scaling form. And in the case of threshold and infrared singularities we also considered that the environment is at zero temperature for the purpose of comparing with the unparticle case. However, we showed that even in this case, transient dynamics leads to a time window within which probe particles are produced and this may be of relevance also in cosmology, freeze out may be an example in which transient dynamics may yield an imprint on the distribution function. In the case of a critical bath, the effective field theory at the critical point does not feature the temperature since for long wavelengths and  small frequencies, temperature only enters in the effective coupling, which is at the ``Wilson-Fisher'' fixed point. It remains to understand particle production in this critical case which necessarily involves an exhaustive numerical analysis because of the rather complicated form of the spectral density, the coherence function and the memory explicit in the expression (\ref{parprod}). We postpone this analysis to a future study.

\section{Summary, conclusions and further questions}\label{sec:conclusions}

We have studied the non-equilibrium  and decoherence dynamics of a system field coupled to
``anomalous'' environments, characterized by critical behavior, slow degrees of freedom and long ranged temporal correlations,  within two complementary formulations of open
effective quantum field theory --- the Schwinger--Keldysh or ``in-in'' effective action with
the influence functional, and the quantum master equation in Lindblad form. Tracing (integrating) over the environmental degrees of freedom,  considered to be initially in
thermal equilibrium, yields  the time evolution of the reduced density matrix of the system.

By considering factorized interactions of the form $g S[\phi] \, \mathcal{O}[\chi]$ with $S[\phi],\mathcal{O}[\chi]$ being operators associated solely with the system (S) and environmental fields ($\mathcal{O}$)respectively,
and assuming factorization of the initial density matrix, we have obtained the   Schwinger-Keldysh effective action that evolves the reduced density matrix with
the \emph{exact} influence functional, to all orders in the system-environment coupling and to all orders in the interactions of environmental fields with themselves.
This influence functional can be obtained as a cumulant expansion in terms of the system-bath coupling $g$ and the correlation functions of the environmental operators $\mathcal{O}[\chi]$ in the initial (factorized) density matrix.   This exact formulation allows to clarify and justify ubiquitous approximations often invoked in the quantum master equation.

We then consider weak system-environment coupling $g$ and keep only up to order $(g^2)$ terms of the influence functional in the effective action. As in the case of quantum Brownian motion we consider a linear coupling of the system field $\phi$ to the environment, namely $S[\phi]\equiv \phi$ and consider $\phi$  as a free field as a \emph{probe} of the environment that allows us to focus on environment-induced decoherence and dissipative processes solely from the interaction of the system with the bath.  The `in-in'' effective action leads to a Langevin dynamics for the probe field with dissipative and noise kernels fulfilling generalized fluctuation-dissipation relations.

The quantum master equation is also obtained up to second order in the coupling $g$, which   becomes of Lindblad form but with \emph{time dependent rates} allowing the study of transient phenomena. Both formulations are indeed complementary and their comparison leads to a clarification and justification of the approximations involved,  making explicit where the Markov approximation and the use of S-matrix (Fermi's golden rule) rates break down, and leading  to a deeper understanding of  transient dynamics.

We focus on \emph{decoherence} defined in the basis of the energy eigenstates of the probe field $\phi$,  and find that the reduced density matrix becomes diagonal in this basis, off diagonal elements decay as $e^{-\Phi(t)}~;~ \Phi(t) \geq 0$, with $\Phi(t)$ a \emph{decoherence function} and $\Gamma(t) = \dot{\Phi}(t)$ a decoherence rate. If the bath is in equilibrium at finite temperature, the probe field thermalizes with the bath, a process that implies \emph{particle production},  it is a consequence of the noise term in the Langevin ``in-in'' formulation, which translates into a quantum kinetic equation for the population in the (QME). At zero temperature ``vacuum noise'' induces a transient production of particles.

Armed with these complementary formulations we study three cases of ``anomalous'' environments: \textbf{i:) unparticle)\cite{unparticle}}  a Lorentz invariant conformal sector beyond the Standard Model  defined by a strongly interacting infrared fixed point\cite{banksZaks}, with correlation functions with anomalous scaling dimensions, and a spectral density featuring a threshold. \textbf{ii:) threshold and infrared divergence} with $\phi$ coupled to a two particle continuum with a kinematic threshold right on or near to the  $\phi$ mass shell. An infrared enhancement emerges when one of the bath particles is massless. All these  cases share the feature that the spectral density vanishes with a power at threshold leading to ``anomalous'' decoherence functions of the form
$\Phi(t) \propto t^{-\mu}~;~\mu >0$ with the index $\mu$ being fractional and determined by the behavior of the spectral density at threshold, a manifestation of ``criticality'' in the long time dynamics, whereas the infrared case results in $\Phi(t) \propto \ln(t)$ therefore decoherence occurs via a \emph{power law decay} $\propto t^{-\Delta}$ of off-diagonal elements of the reduced density matrix. When the probe mass shell is away from threshold, transient dynamics is still sensitive to threshold effects during a time scale depending on the distance to threshold.   \textbf{iii:) critical environment} corresponds to a bath at the critical point of a second order phase transition. We study the case with $\phi$   coupled to the order parameter whose correlation functions feature both \emph{static} and \emph{dynamic} critical behavior. A  new dynamical critical exponent   emerges which is the frequency counterpart of the static critical exponent   associated with   scaling of momenta. Fluctuations in the critical bath are  characterized by ``critical slowing down'' with a typical relaxation time that diverges for long wavelengths\cite{ma,gold,tauber,hotboy}.  This case exhibits striking phenomena: a probe of frequency
$\omega_q$ is resonant with the critical bath at $q_\ast\sim\omega_q^{1/z}$, with $z$ a new dynamical exponent and
the sharpness of the critical resonance --- itself a manifestation of critical
slowing down --- makes a probe coupled to a critically \emph{slow} mode
to decohere very \emph{fast}, a counterintuitive consequence of the narrow spectral
weight of critical quasiparticles.
Early and intermediate time transient dynamics depend on the new dynamical critical exponent, moreover, the decoherence rate of a probe at rest acquires an anomalous scaling dimension with the mass of the probe  depending solely on this new
critical exponent.

  This dependence
on the dynamical exponent alone --- traced to the tail of the spectral density
far from the quasiparticle peak --- is a clean confirmation that the effect is
controlled by dynamical scaling, and provides a direct imprint of an anomalous
dimension on a physical, in-principle measurable decoherence rate.

Although \emph{ a prima facie} these examples may appear to be disconnected, the unifying theme is that an environment with scale-covariant correlations,  and/or
a spectral density that \emph{vanishes} at threshold with power laws,  features slow dynamics and  drives the non-equilibrium and decoherence
reduced dynamics of  the probe away from the Markovian (memoryless) approximation based on  S-matrix or Fermi's golden rule  rates.

While the examples studied here are by no means exhaustive, taken together, these results support the view of a quantum system as a
\emph{probe} of the correlations of its environment: the transient, non-Markovian
dynamics of decoherence, particle production and the approach to equilibrium
encode the scaling dimensions and slow modes of the bath in ways that the
Markovian, S-matrix/golden-rule description cannot capture. Several directions remain open.
In the weak system bath coupling approximation, the
influence functional and the master equation were obtained to second order in this coupling; extending the analysis beyond leading order, where backreaction on
the environment and corrections beyond the two-point (Gaussian) level enter, is
an important next step, as is a sharper characterization of the relation between
the ``in-in'' and master-equation descriptions in these non-Markovian settings.
The thermalization and build-up of populations, in other words, particle production in critical environments ---
whether the asymptotic distribution functions themselves display critical
exponents --- deserves a dedicated study. Finally, the breakdown of the Markov
approximation driven by slow environmental modes is likely to be relevant well
beyond the settings considered here, from scalar or pseudoscalar  misalignment dynamics and
particle production in the early Universe to the slow, long-wavelength modes of
hydrodynamics; we leave these to future work. Furthermore, fast time-resolved spectroscopy in condensed matter systems are beginning to probe
early time transient phenomena\cite{expdecay,navon,wilk} perhaps opening a window towards confirming anomalous decoherence in critical condensed matter
systems.

\acknowledgements
  GB is supported by PID2023-151418NB-I00 funded by MCIU/AEI/10.13039/501100011033/, and by the European ITN project HIDDeN (H2020-MSCA-ITN-2019/860881-HIDDeN).
 D.B.  gratefully acknowledges  support from the U.S. National Science Foundation through grant     NSF 2412374.

\appendix

\section{Lehmann representation and  fluctuation dissipation relation:}\label{app:fd}
Because the initial thermal density matrix for the environment $\rho_E(0)=e^{-\beta H_{E}}/\mathrm{Tr} e^{-\beta H_E} $ is translational invariant   we can write
\bea  C^>(\vx-\vx';t-t')  & = &  \langle \mathcal{O}(\vx,t)\mathcal{O}(\vx',t') \rangle = \int \frac{d^4k}{(2\pi)^4}~ \rho^>(k_0,\vk) e^{-ik_0(t-t')}\,e^{i\vk\cdot(\vx-\vx')} \label{Ggfd} \\
C^<(\vx-\vx';t-t')  & = &  \langle \mathcal{O}(\vx',t')\mathcal{O}(\vx,t) \rangle = \int \frac{d^4k}{(2\pi)^4}~ \rho^<(k_0,\vk) e^{-ik_0(t-t')}\,e^{i\vk\cdot(\vx-\vx')} \,. \label{Glfd} \eea
In interaction picture $\mathcal{O}(\vx,t) = e^{iH_{E}t}\,e^{-i\vec{P}_E\cdot\vx}\,\mathcal{O}(\vec{0},0)\,e^{i\vec{P}_E\cdot\vx}\,e^{-iH_{E}t}$ with $\vec{P}_E$ the momentum operator acting on environmental fields, therefore in the basis of the exact (common) eigenstates of $H_{E},\vec{P}_E$, namely $(H_{E},\vec{P}_E)\ket{m} = (E_m,\vec{P}_m)\ket{m}$
  it follows that the correlation functions of the spatial Fourier transformed operators in the thermal density matrix are given by
\begin{eqnarray}
\rho^>(k_0,\vk) & = &   {(2\pi)^4}~
\sum_{m,n}e^{-\beta E_n}
\langle n| {\cal O}_{\vk} (0) |m \rangle \langle m| {\cal O}_{-\vk} (0) |n \rangle \, \delta(k_0-(E_n-E_m))\,\delta^{(3)}(\vk-(P_n-P_m)) \nonumber \\ \label{siggreat} \\
\rho^<(k_0,\vk) & = &   {(2\pi)^4} ~
\sum_{m,n} e^{-\beta E_n}
 \langle n| {\cal O}_{-\vec k}(0) |m \rangle \langle m| {\cal O}_{\vec k}(0) |n
 \rangle \, \delta(k_0-(E_m-E_n))\,\delta^{(3)}(\vk-(P_m-P_n)) \,. \nonumber \\
 \label{sigless} \end{eqnarray}
  Upon relabelling
$m \leftrightarrow n$ in the sum in the definition (\ref{sigless})
we find the Kubo-Martin-Schwinger relation\cite{kms}

\begin{equation}
\rho^<(k_0,k)  = \rho^>(-k_0,k) = e^{-\beta k_0}
\rho^>(k_0,k) \label{KMS}
\end{equation}

\noindent where we have used   rotational invariance in
the second line above to assume that the spectral functions only
depend of the absolute value of the momentum. The spectral density is defined as
\be \rho(k_0,k) = \rho^>(k_0,k)-\rho^<(k_0,k) = \rho^>(k_0,k)\big[ 1-e^{-\beta k_0}\big] \label{specOs}\ee where
\be n(k_0) = \frac{1}{e^{\beta k_0}-1} \,,\label{BEdist}\ee
therefore
\be  \rho^>(k_0,k) = \rho(k_0,k)~\big[1+n(k_0)\big]~~;~~\rho^<(k_0,k) = \rho(k_0,k)~ n(k_0) \,. \label{relas}\ee Furthermore, from the first equality in (\ref{KMS}) it follows that
\be \rho(-k_0,k)= - \rho(k_0,k) \,. \label{oddros}\ee

In terms of the spectral densities we find
\be  \Big[C^>(\vx-\vx';t-t')-C^<(\vx-\vx';t-t')\Big]   = \int \frac{d^4k}{(2\pi)^4}\,\rho(k_0,k) \,  e^{-ik_0(t-t')}\,e^{i\vk\cdot(\vx-\vx')} \label{Godd} \ee
\be   \Big[C^>(\vx-\vx';t-t')+C^<(\vx-\vx';t-t')\Big]   \equiv       \int \frac{d^4k}{(2\pi)^4}\,\widetilde{\mathcal{K}}(k_0,k) \, e^{-ik_0(t-t')}\,e^{i\vk\cdot(\vx-\vx')} \label{Geven} \ee  where
\be \widetilde{\mathcal{K}}(k_0,k) =  \rho(k_0,k)\,\mathrm{coth\bigg[ \frac{  k_0}{2T}\bigg]}\,.\label{kernelK}\ee In the zero temperature limit
\be \mathrm{coth\bigg[ \frac{  k_0}{2T}\bigg]}~~{}_{ \overrightarrow{T \rightarrow 0 }}~~ \mathrm{sign}(k_0)\,. \label{zeroT} \ee

This is the general form of the fluctuation dissipation relation, with $\mathcal{K}$ proportional to the kernel for the noise correlation function (see eqn.(\ref{kernelkappa})). Note that $\rho(k_0,k)$ is \emph{odd}   whereas $\widetilde{\mathcal{K}}(k_0,k)$ is \emph{even} in $k_0$.

\section{Two environmental fields}\label{app:twofields}
Using Wick's theorem, the correlation function of the composite operator $\mathcal{O} = \chi_1\,\chi_2$ is given by
\be \langle 0|\mathcal{O}(\vx,t)\mathcal{O}(\vx',t')|0\rangle \equiv \langle 0|\chi_1(\vx,t)\chi_1(\vx',t')|0\rangle \,\langle 0|\chi_2(\vx,t)\chi_2(\vx',t')|0\rangle 0|\chi_2(\vx,t)\chi_2(\vx',t')|0\rangle\,\label{OOcorr}\ee with
\be \langle 0|\chi_1(\vx,t)\chi_1(\vx',t')|0\rangle = \int \frac{d^3 p_1}{(2\pi)^3\,2E_1} e^{-i E_1 (t-t')}\,e^{i\vec{p}_1\cdot(\vx-\vx')} \,\label{corre1}\ee and a similar expression for the two point function of $\chi_2$. With the definition of $\rho^>(k_0)$ given by equation (\ref{Ggfd}) we find
\be \rho^>(k_0,\vk) = \int \int \frac{d^3 p_1}{(2\pi)^3\,2E_1}\,\frac{d^3 p_2}{(2\pi)^3\,2E_2}\,\delta^{(4)}(K-p_1-p_2) \,,\label{2body}\ee this is precisely the invariant two-body phase space
\begin{equation}
\rho^>(k_0,\vk) = \frac{\Theta(k_0)}{8\pi}\,\frac{\sqrt{\lambda(s,m_1^2,m_2^2)}}{s}\,\Theta\Big(s -(m_1+m_2)^2 \Big)
\end{equation} where
\be \lambda(s,m_1^2,m_2^2) = \Big[s-(m_1+m_2)^2\big]\big[s-(m_1-m_2)^2\Big]~~;~~ s = k^2_0-k^2\,.   \label{lamb}\ee Combining this result with the relation (\ref{KMS}) we obtain
\be \rho(q_0,q) = \frac{\mathrm{sign}(q_0)}{8\pi [q^2_0 - q^2]}\, \Bigg\{ \Big[q^2_0-q^2 -(m_1+m_2)^2 \Big]\,\Big[q^2_0-q^2 -(m_1-m_2)^2 \Big]   \Bigg\}^{\frac{1}{2}}\,\Theta\Big(q^2_0-q^2 -(m_1+m_2)^2 \Big)\,.\label{rho2parfin}  \ee

\vspace{1mm}

\end{document}